\PassOptionsToPackage{table,xcdraw}{xcolor}
\documentclass[a4paper,fleqn]{cas-dc} 

\usepackage[numbers]{natbib}
\usepackage{amsmath,amssymb,amsfonts}
\usepackage{graphicx}
\usepackage{xcolor}

\makeatletter\long\def\@textcolor#1#2#3{\protect\leavevmode{\color#1{#2}#3}}\makeatother

\makeatletter\newcommand{\srcitem}[2]{\item[{[#1]}]\protected@edef\@currentlabel{[#1]}\label{#2}}\makeatother   

\usepackage{tikz}
\usetikzlibrary{arrows.meta,positioning,calc}
\usepackage{tcolorbox}
\tcbuselibrary{breakable}
\usepackage{pdflscape}
\usepackage{rotating}
\usepackage{longtable}
\usepackage{multirow}
\usepackage{enumitem}
\usepackage{fontawesome}
\usepackage{tablefootnote}
\usepackage{colortbl}
\usepackage[export]{adjustbox}
\usepackage{framed}
\usepackage{svg}
\usepackage{subfig}
\usepackage{rotating}
\usepackage{amssymb}
\usepackage{placeins}
\usepackage{url}  

\usepackage{hyperref}
\usepackage{xurl}
\usepackage{array}
\usepackage{float}
\usepackage{hyperref}
\usepackage{balance}
\usepackage{tabularx}
\usepackage{booktabs}
\usepackage{ragged2e}
\usepackage{longtable}
\usepackage{amssymb}
\usepackage{graphicx}

\usepackage{subfig}

\usepackage{xcolor}

\definecolor{lightgrayrow}{gray}{0.92}

\usepackage{multibib}
\newcites{WL}{Selected Peer-Reviewed Literature}
\newcites{GL}{Selected Grey Literature}
\makeatletter
\providecommand{\@biblabelWL}[1]{[WL#1]}
\providecommand{\@biblabelGL}[1]{[GL#1]}
\renewcommand{\@biblabelWL}[1]{[WL#1]}
\renewcommand{\@biblabelGL}[1]{[GL#1]}
\makeatother
\usepackage{enumitem}

\renewcommand{\arraystretch}{1.15}

\usepackage{setspace} 
\usepackage{algorithm}
\usepackage{algpseudocode}
\usepackage{xcolor}
\usepackage[utf8]{inputenc}
\usepackage[T1]{fontenc}

\usepackage{fancyhdr}

\fancypagestyle{first}{
    \fancyhf{}
    \fancyhead[L]{%
        \fontsize{10pt}{12pt}\selectfont
        \textit{Accepted in The Journal of Systems \& Software (JSS)}
    }
    
}

\begin{document}
\begin{sloppypar}

\let\WriteBookmarks\relax
\def\floatpagepagefraction{1}
\def\textpagefraction{.001}

\shorttitle{Multivocal Review of Vibe Coding}
\shortauthors{S.Siddeeq et al.}

\title[mode=title]{Vibe Coding in Software Development: A Multivocal Literature Review}

\author[tampere]{Shahbaz Siddeeq}[
    orcid=0009-0003-9030-8841
]
\ead{shahbaz.siddeeq@tuni.fi}
\credit{Conceptualization, Methodology, Investigation, Data curation, Formal analysis, Visualization, Writing of the original draft}

\author[tampere]{Muhammad Waseem}[
    orcid=0000-0001-7488-2577
]
\ead{muhammad.waseem@tuni.fi}
\credit{Validation, Investigation, Writing review and editing}

\author[tampere]{Kai-Kristian Kemell}[
    orcid=0000-0002-0225-4560
]
\ead{kai-kristian.kemell@tuni.fi}
\credit{Supervision, Writing review and editing}

\author[tampere]{Mika Saari}[
    orcid=0000-0001-7677-2355
]
\ead{mika.saari@tuni.fi}
\credit{Writing review and editing}

\author[tampere]{Jussi Rasku}[
    orcid=0000-0002-4401-8013
]
\ead{jussi.rasku@tuni.fi}
\credit{Writing review and editing}

\author[tampere]{Pekka Abrahamsson}[
    orcid=0000-0002-4360-2226
]
\ead{pekka.abrahamsson@tuni.fi}
\credit{Supervision, Funding acquisition, Writing review and editing}

\affiliation[tampere]{
  organization={Faculty of Information Technology and Communication Sciences, Tampere University},
  country={Finland}
}

\cortext[cor1]{Corresponding author: Shahbaz Siddeeq <\textit{shahbaz.siddeeq@tuni.fi}>}

\begin{abstract}
Vibe coding is an AI-assisted software development approach in which developers express intent in natural language and large language models generate code. To our knowledge, no review has synthesized academic and practitioner sources under a common protocol. We conducted a multivocal literature review of publications from January 2022 to 9 September 2026, identified through searches in October 2025 and September 2026. After assessing source quality and credibility, we analyzed 266 sources (76 peer-reviewed, 190 grey) across eight research questions: 146 were direct, 93 adjacent, and 27 background. Presence and outcome findings receive reporting-support and evidence-strength labels, respectively. The second author conducted twelve checks; pre-reconciliation agreement was low in several, but substantial for the final directness rule and moderate for the outcome fields, which were checked in full. Sources describe a generation-evaluation-revision loop, with developer work shifting from writing code toward directing, supervising, and validating. Productivity is reported in 109 of the 159 sources contributing to the outcome question, of which 22 provide assertions only; these counts indicate reporting, not effect size. Most comparable sources report short-term productivity gains and long-term maintainability costs, but both receive Weak evidence-strength labels because some comparable sources report opposing directions. Short-term quality and defects are labeled Weak. Long-term productivity, long-term quality, and short-term reproducibility are labeled Moderate. Prototyping is reported more often than production use, and user-interface work more often than data and machine-learning work, but task-domain evidence is sparse. Future work should evaluate safeguards, session-level dynamics, long-term maintainability, and production use.
\end{abstract}


\maketitle

\section{Introduction} \label{sec:intro}

Large Language Models (LLMs) are changing how developers specify, generate, evaluate, and maintain software. Early coding tools completed lines or blocks the developer had begun, while newer tools work through conversation: the developer describes a need in natural language, and the model returns code, tests, or fixes~\cite{russo2024navigating,ross2023programmer}. Some tools act as agents that plan a task, run commands, and change several files with little human input. Software development is in this way moving from code completion toward conversational, intent-driven, and agentic work~\cite{sauvola2024future}, and vibe coding is one form of this shift.

In vibe coding, a term Andrej Karpathy introduced in early 2025, the developer expresses intent in natural language, and one or more LLMs generate, revise, and refine code that the developer often does not read line by line~\cite{karpathy2025theres}. The developer still guides the work, through prompts, feedback, constraints, and validation rather than direct editing, so the focus moves from writing code toward specifying, supervising, evaluating, and validating.

\begin{figure}
    \centering
    \includegraphics[width=0.80\columnwidth]{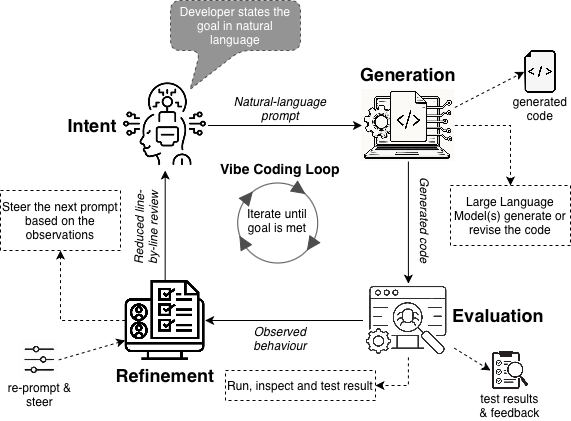}
    \caption{The vibe coding loop of \emph{intent}, \emph{generation}, \emph{evaluation}, and \emph{refinement}, repeated until the goal is met. Review of the generated code line by line is often reduced or skipped~\cite{karpathy2025theres}. The level of review is not part of the minimum definition of Section~\ref{sec:opdef}, and the sources differ on how much review the practice involves (Section~\ref{sec:rq1}).}
    \label{fig:vibe_loop}
\end{figure}

Vibe coding is usually described as a short loop of intent, generation, evaluation, and refinement, shown in Figure~\ref{fig:vibe_loop}. The developer evaluates each result by running, inspecting, or testing it, and uses what they observe, such as an error message or a failed test, to refine the next prompt until the result meets the goal. Vibe coding matters because it may reduce the effort needed to produce working software. A working version can come from a short description, which supports rapid prototyping, and the natural-language interface opens software creation to people who are not expert programmers. Early evidence from AI-assisted programming also points to productivity gains~\cite{peng2023impact}.

Vibe coding also raises software engineering risks. Code is often accepted on the basis of how it behaves, without a full reading, so mistakes can pass unnoticed. Models can return output that looks plausible but is incorrect, incomplete, insecure, or inconsistent with the intended requirements~\cite{mastropaolo2023robustness,akhoroz2025conversational}. Weak verification can leave maintainability problems, security weaknesses, and defects that surface later. Over a long session the context can drift and trust can exceed the evidence, and the origin of generated code is often unclear, which raises concerns about provenance, licensing, and reproducibility. These risks suggest that vibe coding is a new way of working and a software engineering research problem.

Evidence on vibe coding is spread across peer-reviewed studies, preprints, industry reports, blogs, vendor documentation, videos, and practitioner discussions~\cite{gaspar2025state,huang2025vibecoding,harkar2025vibecoding,ChowdhuryMann2025VibeCoding,edwards2025future}, which differ in depth, rigor, purpose, and credibility. This fragmentation makes it hard for practitioners to use vibe coding safely and for researchers to see what is known, how well it is supported, and what is open.

Existing reviews are useful starting points but are limited in scope, evidence coverage, or methodological transparency. Ge~et~al.~\cite{ge2025survey} do not appraise grey literature, Ray~\cite{ray2025review} does not document its search, and Fawzy~et~al.~\cite{fawzy2025vibe} follow a documented process but cover grey literature only. Michels~et~al.~\cite{michels2026vibe} cover productivity evidence and risk in a state-of-the-art review, which they describe as not systematic. We are not aware of a review that integrates peer-reviewed and grey literature under one documented protocol while jointly synthesizing definitions, workflows, developer-role changes, reported outcomes, risks, safeguards, tool ecosystems, and open research challenges. Section~\ref{sec:rw} compares these reviews in more detail.

This study brings the two streams together in a multivocal literature review (MLR), which adds grey literature to a systematic review and suits a topic where much of the activity is reported outside academic venues~\cite{garousi2019guidelines}. It gives an evidence-weighted account of how vibe coding is conceptualized, practiced, evaluated, and governed, which separates findings that many sources report from tentative ones. We applied selection and credibility criteria to both streams. We then synthesized 266 sources, 76 peer-reviewed and 190 grey, using descriptive mapping and thematic synthesis. They date from 2022 to September 2026. The searches were run in October 2025 and September 2026 under one protocol. For the analysis the corpus is divided into two publication windows: window~1 holds publications from January 2022 to October 2025 (113 sources), and window~2 those from November 2025 to 9 September 2026 (153). The main contributions of this study are as follows:

\begin{itemize}
\item We provide a documented MLR protocol that integrates peer-reviewed and grey literature on vibe coding, documenting the search, selection, quality assessment, and synthesis so that each step can be traced.
\item We synthesize the evidence on definitions, workflows, developer roles, outcomes, risks, safeguards, contexts, and tools, which brings the scattered material into one structured account.
\item We label each finding with a decision matrix. A finding that records a reported phenomenon receives a reporting-support label, which shows how many sources, and of what standing, report it. A finding on an outcome receives an evidence-strength label, which also takes into account whether the reported effects agree. The labels may help the reader judge how far each claim rests on the sources.
\item We propose a descriptive conceptual framework. It is stated as a process model, a taxonomy, and four hypothesis-generating propositions. We also give a research agenda that points to areas needing further empirical work.
\end{itemize}

The rest of the paper is organized as follows. Section~\ref{sec:methodology} describes the methodology, Section~\ref{sec:results} reports the results, Section~\ref{sec:discussion} discusses them and presents the conceptual framework, Section~\ref{sec:validity} covers validity threats, Section~\ref{sec:rw} compares earlier reviews, and Section~\ref{sec:conclusion} concludes.
\section{Methodology} \label{sec:methodology}

This study is an MLR following the guidelines of Garousi~et~al.~\cite{garousi2019guidelines}. These guidelines extend systematic review principles~\cite{kitchenham2007guidelines} with procedures for grey-literature search, credibility assessment, and combined synthesis. Much of the evidence on vibe coding, a recent practitioner-driven practice, appears outside peer-reviewed venues, in blogs, reports, guides, videos, and practitioner discussions. An MLR brings these sources together with peer-reviewed studies under one protocol.

\subsection{Goal and Research Questions}\label{sec:goal}

To synthesize current evidence on vibe coding, we formulated eight Research Questions (RQs) (Table~\ref{tab:rqs_rational}), following the Goal-Question-Metric (GQM) paradigm~\cite{basili1994gqm} as Garousi~et~al.~\cite{garousi2019guidelines} recommend for multivocal reviews. The metrics, reported descriptively rather than as combined statistics, are the number and share of contributing sources, the source type, the quality and credibility scores, and an evidence-strength label per finding.

\begin{table*}[!ht]
\centering
\caption{Research questions and their rationale}
\label{tab:rqs_rational}
\footnotesize
\renewcommand{\arraystretch}{1.1}
\begin{tabular}{|p{0.7cm}|p{6.4cm}|p{9.2cm}|}
\hline
\rowcolor[HTML]{EFEFEF}
\textbf{} & \textbf{Research Questions} & \textbf{Rationale} \\
\hline
\rowcolor[HTML]{DDDDDD}
\multicolumn{3}{|c|}{\textbf{Concept and Practices}} \\
\hline
\multirow{1}{=}{RQ1}
 & How is ``vibe coding'' defined and conceptualized in academic and industry literature? & The term is recent and not consistently defined. This RQ synthesizes how sources describe it, for example as prompt-based development, AI-assisted programming, agentic programming, or a development practice. \\
\hline
\multirow{1}{=}{RQ2}
 & What workflows characterize vibe coding and what tactics/tools are reported? & This RQ identifies recurring workflow patterns, such as iterative generation and testing loops, context provisioning, error-driven prompting, and retrieval grounding, with their tactics and tools. \\
\hline
\rowcolor[HTML]{DDDDDD}
\multicolumn{3}{|c|}{\textbf{Developer Role and Outcomes}} \\
\hline
\multirow{1}{=}{RQ3}
 & How does the developer role shift?
    \begin{itemize}
        \item RQ3.1: Changes in trust, flow, agency, cognitive load, and cognitive debt
        \item RQ3.2: Session-level dynamics such as momentum and context loss
        \item RQ3.3: Differences by audience and task type
    \end{itemize}
& RQ3 examines how developer work changes during interaction with LLM systems, including trust calibration, perceived agency, cognitive load, session-level dynamics, and differences across user groups and task types. \\
\hline
\multirow{1}{=}{RQ4}
 & What is reported about productivity, time-to-prototype, code and user-interface (UI) quality, defects, maintainability, and reproducibility? & RQ4 synthesizes reported outcomes. These include short-term effects such as productivity and prototyping speed, and longer-term defects, maintainability, and reproducibility. \\
\hline
\rowcolor[HTML]{DDDDDD}
\multicolumn{3}{|c|}{\textbf{Risks, Safeguards, and Context}} \\
\hline
\multirow{1}{=}{RQ5}
 & What risks and safeguards are reported for vibe-coded software?
    \begin{itemize}
        \item RQ5.1: Failure modes
        \item RQ5.2: Governance and guardrails
    \end{itemize}
& RQ5 examines reported risks and mitigation strategies. Failure modes include incorrect code generation, context-related errors, minimal review practices, and licensing or provenance concerns, and safeguards include review practices, validation pipelines, sandboxing, and provenance tracking. \\
\hline
\multirow{1}{=}{RQ6}
 & Where is vibe coding used, by whom, and with what constraints? & RQ6 examines adoption contexts, such as prototyping or production, user groups, and constraints such as privacy, cost, and domain sensitivity. \\
\hline
\rowcolor[HTML]{DDDDDD}
\multicolumn{3}{|c|}{\textbf{Tools and Open Challenges}} \\
\hline
\multirow{1}{=}{RQ7}
 & What tools and platforms support vibe coding? & RQ7 analyzes the tool ecosystem, including coding assistants, agentic development environments, and supporting infrastructure such as evaluation tools and testing integrations. \\
\hline
\multirow{1}{=}{RQ8}
 & What open challenges and future research directions are identified? & RQ8 synthesizes reported open challenges, including limitations in empirical evidence, evaluation practices, and representation of user groups or task domains. \\
\hline
\end{tabular}%
\end{table*}

\paragraph{Scope boundaries between related RQs:}
Five terms are kept distinct. A \emph{workflow} is a recurring sequence of activities in a development session. A \emph{tactic} is a single action a developer takes inside a workflow, such as supplying an error message to the model. A \emph{tool} is a named product or platform. A \emph{safeguard} is a technical or process control applied to generated code, such as a test suite or a sandbox. A \emph{governance mechanism} is an organizational rule about who may do what with generated code, such as a review policy or a disclosure requirement. Validation is coded under RQ2 when a source presents it as a step of the development loop, and under RQ5.2 when a source presents it as a control against risk. A source can carry both codes. RQ2 covers workflows and tactics, and RQ7 covers named tools. RQ4 covers product-level outcomes, and RQ5 covers process-level risks and safeguards. RQ3.3 covers differences by user and task, and RQ6 covers adoption context and constraints. The codebook gives worked examples of these boundaries.

\subsection{Operational definition and source directness}\label{sec:opdef}

We use the definition at two levels. The minimum definition, used to select and classify sources, is a software development practice in which (1) the developer expresses intent primarily in natural language and (2) one or more large language models generate and iteratively revise the code, with the developer steering through further natural-language turns. How closely the developer reviews the code is not part of the minimum definition. The stricter form, as first described by Karpathy~\cite{karpathy2025theres}, adds two features: (3) line-by-line review of the generated code is reduced or absent, and (4) the developer accepts or rejects the output mainly from its observed behavior. The stricter form is one end of a range of oversight that runs from little or no review to structured validation with tests and human review (Section~\ref{sec:rq1}). The term was coined in February 2025, so a source may show the practice without using the term.

Because the search terms also retrieve related work, every source is classified by its directness to the minimum definition. A source is \emph{direct} when two conditions both hold. First, its object of study is the human practice, that is, people (professional developers, students, end users, or the author in a first-person account) creating or changing software through natural-language interaction with LLMs. Second, it names vibe coding as the practice it studies in its title, keywords, or abstract, or, for an item without an abstract, in its opening text, or, without the term, its own description of the practice shows the stricter form, all four features explicitly. The stricter form is required in that case because features 1 and 2 alone also describe natural-language programming and AI-assisted programming in general. A source that names vibe coding and studies the human practice is direct at any level of review, because the sources differ on how much review the practice involves (Section~\ref{sec:rq1}). A source that names vibe coding but evaluates models, coding agents, benchmarks, or tools technically, or that uses the term only to frame such an evaluation, is \emph{adjacent}. A tool or system paper is direct only when it studies people using the tool in the practice. A conceptual, position, or review source whose subject is this practice is treated as studying it. More generally, a source is adjacent when it studies intent-driven code generation through conversation or agents (features 1 and 2) without the human practice as its object, for example benchmarks of models or agents on natural-language programming tasks, systems for agentic or conversational code generation evaluated technically, and studies of people doing natural-language-to-code work that neither name the practice nor show features 3 and 4. A source is \emph{background} when it studies AI-assisted programming in general without an intent-driven generation workflow as its object, for example code-completion assistants, usability or trust surveys of assistants, security of assistant output, reviews of LLM code generation, and taxonomies of human-AI collaboration. The rule, with its decision notes, is in the replication package.

The first author applied the rule to all 266 sources, from the title, keywords, abstract or opening text, and the extraction cells. The decision notes in the replication package record the boundary cases and how each was decided. Section~\ref{sec:reliability} reports the independent checks of directness.

Peer-reviewed sources are denoted WL and grey sources GL, with the numbers of the source lists at the end of the paper. Each list is numbered with the sources of publication window 1 first, and the replication package links each number to the record of that source in the data files. The corpus has 146 direct sources (35 peer-reviewed, 111 grey), 93 adjacent sources (24 peer-reviewed, 69 grey), and 27 background sources (17 peer-reviewed, 10 grey). Window 1 contributes 63 direct, 30 adjacent, and 20 background sources, and window 2 contributes 83, 63, and 7 (Section~\ref{sec:update2026}). Of the 146 direct sources, 41 are conceptual, position, or review sources about the practice, and 45 are experience reports. Of the 146, 120 name vibe coding in the title, 23 name it in the keywords, the abstract, or the opening text, and 3 rest on the four features alone. Nine sources that name the term in none of the title, keywords, or abstract were read in full. Three are grey items without an abstract whose opening text names the practice as their subject (\ref{GL105}, \ref{GL106}, \ref{GL203}), and they are direct. Three are direct because all four features are explicit (\ref{GL153}, \ref{GL155}, \ref{GL199}), and \ref{GL155} names the term only in its body. Three show features 1 and 2 only and are adjacent (\ref{GL146}, \ref{GL162}, \ref{GL168}). The classification of every source is in the replication package, and its distribution by stream and publication window is in Table~\ref{tab:corpus_profile} (Section~\ref{sec:study_context}).

Directness describes what a source is about, not the kind of evidence it gives. Of the 146 direct sources, 58 are empirical studies (54) or tool evaluations (4), and 88 are experience reports (45), position pieces (19), secondary sources (12), conceptual sources (10), or not classified (2) (Table~\ref{tab:corpus_profile}). A direct source therefore does not by itself provide empirical evidence on the effects of the practice, and this holds in particular for productivity and software quality (Section~\ref{sec:rq4}). The evidence category and the measurement field of each outcome carry that information, and criteria C2 and C3 of Section~\ref{sec:evidence_strength} count only empirical sources.

Direct and adjacent sources may support every research question. Background sources never count toward a label, for any research question. They are removed from the supporting set of every finding before the criteria of Section~\ref{sec:evidence_strength} are computed. They are still counted as reports in the ``Sources reporting'' column of each results table, and the ``Background'' column shows how many of them are background (in Table~\ref{tab:rq4_outcomes}, the Bg.\ column under Directness). Because of this rule, the labels computed on the 239 direct and adjacent sources are the main labels, so this subset is not shown again in the sensitivity tables. Section~\ref{sec:sensitivity} recomputes every finding on the 146 direct sources alone and on the 143 direct sources that name the term, and sheet S4 of the supplementary workbook in the replication package gives each label on these subsets.

\subsection{Review design}

Following MLR guidance~\cite{garousi2019guidelines}, peer-reviewed and grey literature were treated as two evidence streams, each with its own selection criteria and credibility checks. Four terms are used with fixed meanings. A \emph{record} is one hit returned by one database or channel, so a publication indexed in three databases produces three records. A \emph{publication} is one unique document after duplicate records are merged. A \emph{source} is a publication or grey item included in the corpus. A \emph{study} is an included source that reports empirical work. Counts of records and of unique publications are reported separately at every stage.

\begin{figure*}[!tbp]
    \centering
    \includegraphics[width=\textwidth, clip, trim=100 150 135 140]{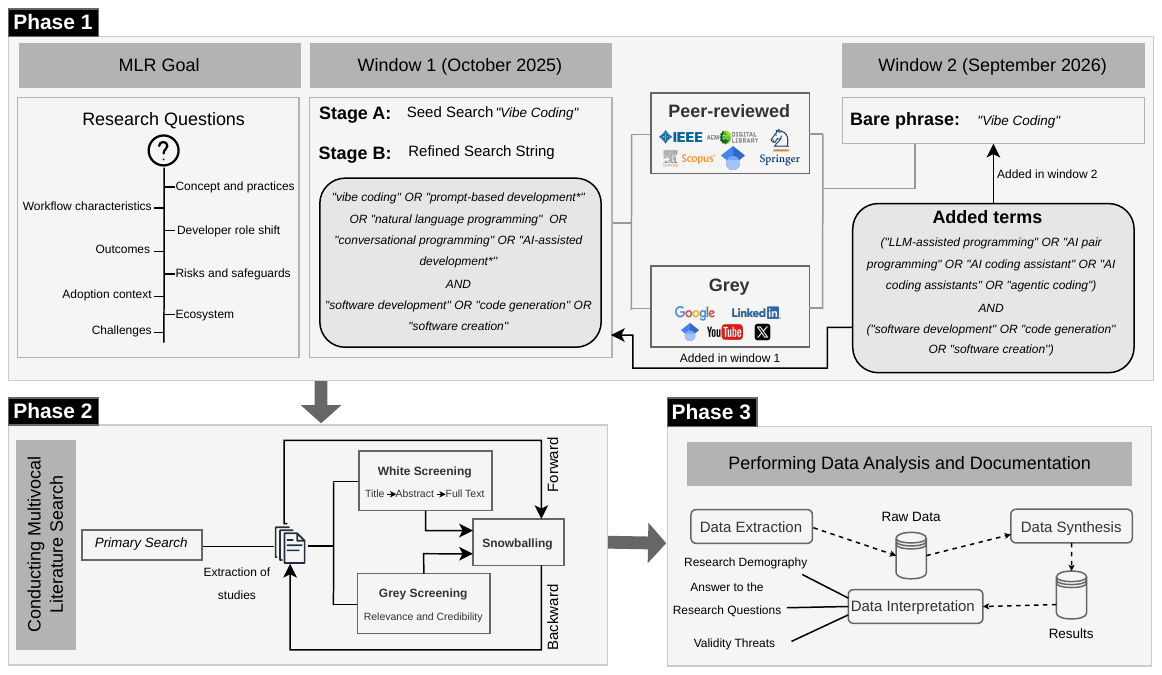}
    \caption{Overview of the review protocol. The searches were run in October 2025 and September 2026 and followed one selection, assessment, extraction, and synthesis procedure. The corpus is divided into two publication windows by publication date.}
    \label{fig:work_flow}
\end{figure*}

\begin{figure*}[!tbp]
    \centering
    \includegraphics[width=\textwidth,height=0.9\textheight,keepaspectratio]{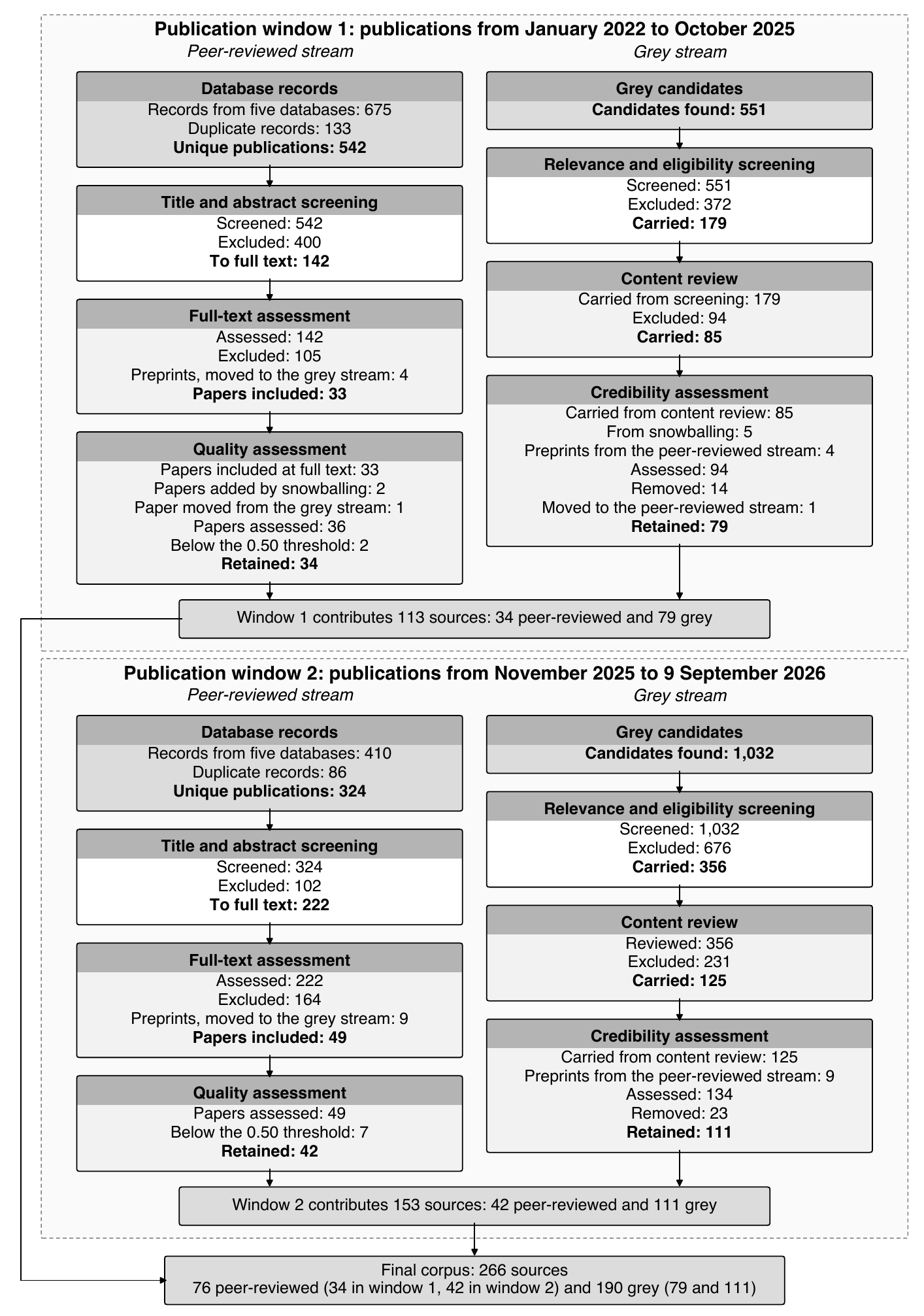}
    \caption{Flow of records for each publication window, for the peer-reviewed and grey streams, to the final corpus of 266 sources. Window 1 draws on the searches of October 2025 and September 2026, and window 2 on those of September 2026. Counts come from the decision files in the replication package. Duplicate records are explained in Section~\ref{sec:wl_selection} and Table~\ref{tab:db_records}. Grey candidates without a known date take the window of the query that returned them (Section~\ref{sec:grey_selection}).}
    \label{fig:round1_flow}
\end{figure*}

The review followed one protocol (Figure~\ref{fig:work_flow}). The searches were run in October 2025 and in September 2026. In October 2025 the first database search ran the Stage~B string in five databases, and five grey channels were searched. In September 2026 the second database search ran the bare phrase and four broader terms in the same five databases, and a logged grey search covered the grey channels. The searches of September 2026 covered all publications from January 2022, so they also found earlier work. All records went through the same selection criteria, quality and credibility checklists, extraction form, codebook, and synthesis, and a publication found by more than one search was counted once. The search strings, the treatment of background records in the peer-reviewed stream, and the grey search protocol differ between the searches of October 2025 and those of September 2026, as stated in Sections~\ref{sec:search_string_a_b} to~\ref{sec:update2026}. The corpus is then divided by publication date into two windows, window~1 to 31~October 2025 with 113 sources and window~2 from 1~November 2025 to 9~September 2026 with 153 sources (Section~\ref{sec:update2026}). The windows show how far each finding rests on the most recent work (Section~\ref{sec:sensitivity}). Figure~\ref{fig:round1_flow} shows the flow of records in each window, through both streams, to the corpus of 266 sources.

\subsubsection{Search strategy} \label{sec:search_string_a_b}

Because the term may not appear consistently in academic indexing, we developed the query of the first database search iteratively.

\paragraph{Stage A: Seed and scoping search}
A search for the exact phrase ``vibe coding'' returned 31 seed sources. Their recurring synonyms and related concepts (\emph{prompt-based development}, \emph{natural language programming}, \emph{conversational programming}, and \emph{AI-assisted development}) were added to the Stage~B search string. Seed sources that met all inclusion criteria were kept in the final corpus.

\paragraph{Stage B: Search string of the first database search}
The final search string was:

\begin{tcolorbox}[
    breakable,
    colback=gray!10,
    colframe=gray!60,
    title=\textbf{Search String},
    fonttitle=\bfseries,
    coltitle=black,
    arc=3mm,
    width=\columnwidth,
]
\begin{center}
\onehalfspacing
\textit{(
``vibe coding'' OR ``prompt-based development*'' OR ``natural language programming''
OR ``conversational programming'' OR ``AI-assisted development*''
)}\\[3pt]

\textbf{AND}\\[3pt]

\textit{(
``software development'' OR ``code generation'' OR ``software creation''
)}
\end{center}
\end{tcolorbox}

This follows guidance for emergent and inconsistently labeled concepts~\cite{kitchenham2007guidelines} and for grey-literature coverage~\cite{garousi2019guidelines}.

\paragraph{Search terms considered for the first database search}
During scoping we also considered four broader terms, ``LLM-assisted programming,'' ``AI pair programming,'' ``AI coding assistant,'' and ``agentic coding.'' We left them out of the first database search because they retrieve a much wider literature not focused on vibe coding. Section~\ref{sec:validity} discusses the recall trade-off.

\paragraph{Search strings of the second database search}
The second database search ran two strings in the five databases over 2022 to 2026, and each record was assigned to a window by its publication date. Q1 is the bare phrase \texttt{"vibe coding"}. Q2 pairs the four broader terms considered during scoping with the second block of the first database search: \texttt{("LLM-assisted programming" OR "AI pair programming" OR "AI coding assistant" OR "AI coding assistants" OR "agentic coding") AND ("software development" OR "code generation" OR "software creation")}. The plural of one term is written out because some databases do not allow a wildcard inside a quoted phrase. The Stage~B string was not run in the second database search. Section~\ref{sec:wl_selection} reports what the two strings returned for each window.

\paragraph{Search Sources}
In both searches we applied the search strings to five databases for the peer-reviewed stream: \emph{IEEE Xplore, ACM Digital Library, Scopus, SpringerLink, and Google Scholar}. A Google Scholar record was kept in the peer-reviewed stream only when its venue was a confirmed peer-reviewed journal, conference, workshop, or symposium. Other Google Scholar records were treated as grey literature or excluded, and when a work appeared as both a preprint and a published paper, we kept the published version. The grey stream was searched separately (Section~\ref{sec:grey_selection} and Table~\ref{tab:grey_search_2026}), following MLR guidance on using multiple grey channels~\cite{garousi2019guidelines}.

\paragraph{Inclusion and exclusion criteria for the peer-reviewed literature}

Table~\ref{tab:selection_criteria_white} lists the criteria and the stage at which each applies: title and abstract screening (T/A), full-text screening (F), or all stages (All).

\begin{table}[!htbp]
\centering
\caption{Inclusion and exclusion criteria for the peer-reviewed literature.}
\label{tab:selection_criteria_white}
\footnotesize
\renewcommand{\arraystretch}{1.1}
\begin{tabular}{|p{0.75cm}|>{\raggedright\arraybackslash}p{5.4cm}|p{0.95cm}|}
\hline
\rowcolor[HTML]{EFEFEF}
\textbf{ID} & \textbf{Criterion} & \textbf{Stage} \\
\hline
\multicolumn{3}{|c|}{\textbf{Inclusion criteria}} \\
\hline
I1 & Studies address vibe coding as defined in Section~\ref{sec:opdef}, or intent-driven or AI-assisted development in the context of software development. The directness of each retained study to the definition is classified as direct, adjacent, or background. & All \\
\hline
I2 & Studies evaluate or review tools or platforms for AI-driven software creation. & T/A \\
\hline
I3 & Studies are published in 2022 or later. & T/A \\
\hline
I4 & Studies are published in peer-reviewed journals or conferences relevant to software engineering. & T/A \\
\hline
I5 & Studies provide evidence-based contributions relevant to vibe coding, including empirical evaluations, design proposals, theoretical arguments, or position statements directly addressing vibe coding practices or outcomes. & T/A \\
\hline
I6 & Studies discuss developer experience or human-AI collaboration in coding. Such studies enter as adjacent or background evidence unless they meet the definition in Section~\ref{sec:opdef}. & T/A \\
\hline
\multicolumn{3}{|c|}{\textbf{Exclusion criteria}} \\
\hline
E1 & Not written in English. & T/A \\
\hline
E2 & Duplicate study or extension of an already included work. & T/A \\
\hline
E3 & Irrelevant to AI or software development. & All \\
\hline
E4 & Non-peer-reviewed source. & F \\
\hline
E5 & Not accessible via institutional access or public repositories. & T/A \\
\hline
E6 & Focuses on AI applications outside software development. & F \\
\hline
E7 & Published before 2022. & F \\
\hline
\end{tabular}
\end{table}

\paragraph{Grey literature eligibility (credibility and relevance)}

The eligibility criteria for grey sources are listed in Table~\ref{tab:gl_criteria}.

\begin{table}[!t]
\centering
\caption{Grey literature eligibility criteria.}
\label{tab:gl_criteria}
\footnotesize
\renewcommand{\arraystretch}{1.1}
\begin{tabular}{|p{0.95cm}|>{\raggedright\arraybackslash}p{2.1cm}|>{\raggedright\arraybackslash}p{4.05cm}|}
\hline
\rowcolor[HTML]{EFEFEF}
\textbf{ID} & \textbf{Criterion} & \textbf{Rule} \\
\hline
GL-I1 & Topical relevance &
Must address vibe coding or AI-assisted software development, including workflows, tools, or practices relevant to the RQs. \\
\hline
GL-I2 & Substantive technical content &
Must include concrete discussion of workflows, tools, risks, mitigations, or implementation detail rather than high-level opinion. \\
\hline
GL-I3 & Credibility threshold &
Must score $>$4 out of 15 on the 15-item credibility checklist of Garousi et al.~\cite{garousi2019guidelines} (Table~\ref{tab:qa_grey}). \\
\hline
\multicolumn{3}{|c|}{\textbf{Exclusion criteria}} \\
\hline
GL-E1 & Opinion-only or low substance &
Exclude sources dominated by commentary without substantive technical or methodological detail. \\
\hline
GL-E2 & Duplicate or replicative content &
Exclude near-duplicate sources or sources that primarily restate previously included material. \\
\hline
\end{tabular}
\end{table}

\subsubsection{Selection of peer-reviewed literature} \label{sec:wl_selection}

Both database searches applied the criteria of Table~\ref{tab:selection_criteria_white} at title and abstract screening, full-text assessment, and quality assessment. The records of the first database search were deduplicated after abstract screening, and the records of the second database search before screening. Table~\ref{tab:db_records} gives the records per database and stage, and Figure~\ref{fig:round1_flow} the flow of records of both searches.

\paragraph{First database search (October 2025)}
The first database search returned 540 records (463 unique publications).

Title screening against the criteria in Table~\ref{tab:selection_criteria_white} reduced the 540 records to 246, and abstract screening retained 90 as candidates for deduplication and full-text assessment. At the title stage the first author judged each title without consulting the abstract. A record was excluded when its title showed that it was not about AI or software development (E3), applied AI outside software development (E6), predated 2022 (E7), or did not concern AI-assisted or intent-driven development (I1, I6). No written rule for ambiguous titles was recorded at the time, and such titles were decided by judgment.

The counts above are database records. Duplicates were counted at every stage by normalized title, first author, and year, with a manual check of near matches. The 540 records are 463 unique publications (77 duplicate copies), the 246 records retained at title stage are 218 unique publications (28 copies), and the 90 records retained at abstract stage are 80 unique publications (10 copies). The title stage therefore excluded 245 unique publications. Each database export of the first database search was screened separately, in the order the databases were searched and under the same criteria, and duplicates were removed only after abstract screening. For 29 of the 67 publications retrieved from more than one database, the title was retained from one database and not from another. Each went forward from at least one database, so this affects the per-database counts in Table~\ref{tab:db_records} but not the set of publications screened next. A publication retrieved from several databases is attributed to the first database that returned it, in the search order given above.

Under the same criteria, 27 of the 80 unique publications passed full-text assessment and 53 were excluded. The second author screened all title-stage and full-text exclusions of the first database search (Section~\ref{sec:reliability}). After reconciliation, two title-stage exclusions were read in full and three full-text exclusions were reread. Of these five, three papers (\ref{WL74}, \ref{WL75}, \ref{WL76}) and one preprint (\ref{GL204}) were included, and one was excluded. In all, 82 unique publications of the first database search were read in full. Thirty papers and one preprint were included, and 51 publications were excluded.

Backward and forward snowballing~\cite{wohlin2014guidelines} on the reference lists and citing works of the 27 papers included at full-text assessment of the first database search yielded 47 candidate references. Of these, 31 were duplicates of papers already captured by the primary search, 9 addressed AI-assisted software development broadly without meeting I1 or I6, 3 were published before 2022 (I3/E7), and 2 failed the accessibility criterion (E5). The remaining 2 met all criteria and were added. The first database search and its snowballing together yielded 32 papers for quality assessment: the 27 papers included at full-text assessment, the 2 papers from snowballing, and \ref{WL74}, \ref{WL75}, and \ref{WL76}.

\begin{table*}[!t]
\centering
\footnotesize
\renewcommand{\arraystretch}{1.1}
\setlength{\tabcolsep}{4pt}
\caption{Records per database. First database search (October 2025): records retrieved and kept after each screening stage. Second database search (September 2026): records retrieved and unique records per publication window, each attributed to the first database in the order shown. The last row counts each publication once within a search. The two searches retrieved 5 publications in common, so together they give 866 unique publications (463 plus 408, less 5).}
\label{tab:db_records}
\begin{tabular}{|p{3.0cm}|p{1.3cm}|p{1.3cm}|p{1.6cm}|p{1.4cm}|p{1.6cm}|p{2.4cm}|p{2.4cm}|}
\hline
\rowcolor[HTML]{EFEFEF} & \multicolumn{4}{c|}{\textbf{First database search (October 2025)}} & \multicolumn{3}{c|}{\textbf{Second database search (September 2026)}} \\
\cline{2-8}
\rowcolor[HTML]{EFEFEF}\multirow{-2}{=}{\textbf{Database}} & \textbf{Retrieved} & \textbf{After title} & \textbf{After abstract} & \textbf{Included at full text} & \textbf{Retrieved} & \textbf{Unique, window 1} & \textbf{Unique, window 2} \\
\hline
IEEE Xplore & 148 & 52 & 21 & 11 & 67 & 9 & 51 \\
\hline
ACM Digital Library & 81 & 45 & 14 & 8 & 70 & 6 & 54 \\
\hline
Scopus & 191 & 92 & 25 & 6 & 193 & 20 & 81 \\
\hline
Springer & 18 & 7 & 6 & 1 & 54 & 11 & 35 \\
\hline
Google Scholar & 102 & 50 & 24 & 1 & 161 & 38 & 103 \\
\hline
\textbf{All records} & 540 & 246 & 90 & 27 & 545 & 84 & 324 \\
\hline
\textbf{Unique publications} & 463 & 218 & 80 & 27 & 408 & 84 & 324 \\
\hline
\end{tabular}
\end{table*}

\paragraph{Second database search (September 2026)}
The two strings of the second database search returned 545 records, 408 unique after cross-database deduplication (Table~\ref{tab:db_records}). Each record was assigned to a publication window by the date in its export: 84 unique records fall in window 1 and 324 in window 2. Of the 324, Q1 returned 250 and Q2 74. Title and abstract screening excluded 124 of the 408 records (22 of window 1 and 102 of window 2), and 284 went to full-text assessment (62 and 222). The second author screened the records of the second database search independently, and the reconciliation reversed six title-level exclusions of window 2, whose records are among those read in full (Section~\ref{sec:reliability}). Full-text assessment excluded 220 and passed 64. In window 1 it passed three papers (\ref{WL29}, \ref{WL30}, \ref{WL31}) and three preprints, and in window 2 it passed 49 papers and 9 preprints. Of the 58 of window 2, 43 came from Q1 and 12 from Q2 at the first screening, and three are among the six records whose title-level exclusion was reversed.

The largest groups of the 220 full-text exclusions of the second database search are background records (60, Section~\ref{sec:update2026}), records that mention the practice without contributing evidence about it (40), records whose full text could not be obtained through the institutional subscription and were excluded under E5 without an interlibrary-loan request (27), and records in which the practice only frames a study about something else (22). Whole proceedings volumes, books, and theses were excluded (10, 4, and 6 records), because the review synthesizes individual sources and the grey stream is reserved for practitioner sources. Thirty-three of the Q2 records of window 2 were excluded at full text as background.

\paragraph{Both database searches together}
Five publications were retrieved by both database searches, and each is counted once (Table~\ref{tab:db_records}). Figure~\ref{fig:round1_flow} counts each in the window of the record that went furthest in selection, four in window 1 and one in window 2. Two of them were read in full in both searches: \ref{WL27}, included by the first database search, and the preprint of \ref{GL1}. For window 1, Figure~\ref{fig:round1_flow} therefore shows 542 unique publications screened: the 463 of the first database search and the 84 of the second, less the four retrieved by both searches in this window and less the one whose second-search record falls in window 2, where it is counted. Of these, 142 were read in full: the 82 of the first database search and the 62 of the second, less the two read in both. The 33 papers included are 30 from the first database search and 3 from the second, and the 4 preprints are 1 and 3. Window 2 holds the 324 unique publications of the second database search alone, of which 222 were read in full and 49 papers and 9 preprints were included. Over both searches, 866 unique publications were screened, 364 were read in full, 269 were excluded, and 82 papers and 13 preprints were included (Figure~\ref{fig:round1_flow}). The 13 preprints went to the grey stream. With the 2 papers from snowballing and \ref{WL77}, a published paper found by the grey search, 85 papers went to quality assessment. The record of the five shared publications is in the replication package.

\paragraph{Quality assessment}
The first author assessed every retained peer-reviewed paper with the eleven-item checklist of Dyb{\aa} and Dings{\o}yr~\cite{dybaa2008empirical} (Table~\ref{tab:qa_white}), widely used in software engineering reviews~\cite{kitchenham2007guidelines,garousi2019guidelines}. Each item was scored \textit{Yes}~=~1, \textit{Partly}~=~0.5, or \textit{No}~=~0. An item was marked \textit{Not Applicable} only when it could not apply to the study type, for example participant recruitment in a non-empirical paper, and was then removed from that paper's denominator, so the score is the proportion of applicable items. Papers scoring below 0.5 were excluded from data extraction. Retained papers were grouped into three quality bands, \textit{High} (proportion $\geq$0.75), \textit{Medium} (0.60 to 0.74), and \textit{Low} (0.50 to 0.59), used only for evidence weighting.

Preprints were routed to the grey stream and scored with the fifteen-item checklist of Table~\ref{tab:qa_grey}, under the same thresholds. Of the 85 papers, nine scored below 0.50 after the second scoring was reconciled (Section~\ref{sec:reliability}), two from the first database search and its snowballing and seven from the second database search, and were excluded from data extraction. The 76 retained papers are 34 of window 1 and 42 of window 2. Of the 13 preprints, the TechRxiv copy of \ref{GL1} was removed as a duplicate, and none of the others fell at or below the grey exclusion floor.

The checklist score measures reporting quality, and for conceptual and secondary papers it is not a measure of empirical rigor. Sheet S1 of the supplementary workbook therefore gives, for every retained paper, its evidence category, study method, and number of applicable items beside its band and score. Outcome findings (RQ4) and the safeguard and failure findings (RQ5) are weighted only by empirical and tool-evaluation sources, through criterion C2 (Section~\ref{sec:evidence_strength}).

\begin{table}[!ht]
\centering
\caption{Quality assessment checklist for peer-reviewed papers,
adapted from Dyb{\aa} and Dings{\o}yr~\cite{dybaa2008empirical} and
organized into four evaluation stages taken from the Critical Appraisal Skills Programme. Scoring: Yes~=~1, Partly~=~0.5, No~=~0.}
\label{tab:qa_white}
\footnotesize
\renewcommand{\arraystretch}{1.1}
\begin{tabular}{|p{1.45cm}|p{0.8cm}|>{\raggedright\arraybackslash}p{4.85cm}|}
\hline
\rowcolor[HTML]{EFEFEF}
\textbf{Stage} & \textbf{QA$_s$} & \textbf{Quality assessment question} \\
\hline
\multirow{3}{=}{Reporting}
 & QA$_1$ & Is the paper based on research (not merely a ``lessons
   learned'' or expert-opinion report)? \\
\cline{2-3}
 & QA$_2$ & Is there a clear statement of the aims of the research? \\
\cline{2-3}
 & QA$_3$ & Is there an adequate description of the context in which
   the research was carried out? \\
\hline
\multirow{5}{=}{Rigor}
 & QA$_4$ & Was the research design appropriate to address the aims
   of the research? \\
\cline{2-3}
 & QA$_5$ & Was the recruitment or sampling strategy appropriate to
   the aims of the research? \\
\cline{2-3}
 & QA$_6$ & Was there a control group or comparison baseline with
   which to compare the approach (where appropriate)? \\
\cline{2-3}
 & QA$_7$ & Was the data collected in a way that addressed the
   research issue? \\
\cline{2-3}
 & QA$_8$ & Was the data analysis sufficiently rigorous? \\
\hline
Credibility
 & QA$_9$ & Has the relationship between researcher and participants
   been considered adequately? \\
\hline
\multirow{2}{=}{Relevance}
 & QA$_{10}$ & Is there a clear statement of findings? \\
\cline{2-3}
 & QA$_{11}$ & Is the study of value for research or practice? \\
\hline
\end{tabular}
\end{table}

\subsubsection{Selection of grey literature} \label{sec:grey_selection}

\paragraph{Five grey channels (October 2025)}
Google web search and Google Scholar were searched with four terms (vibe coding, AI-assisted programming, LLM-based development, and natural language programming), YouTube with the first two, and LinkedIn and X with vibe coding alone. The five channels gave 24 candidates: 14 from Google web search, 7 from Google Scholar and preprint servers, and one each from YouTube, LinkedIn, and X. Searches stopped when three consecutive result pages across all active channels yielded no new topically relevant source, the saturation rule recommended by Garousi et al.~\cite{garousi2019guidelines}.

Screening for topical relevance, credibility, and technical substance (Table~\ref{tab:gl_criteria}) retained 16 sources (9 from Google web search, 4 from Google Scholar and preprint servers, and the three single candidates) and excluded eight for limited technical content, opinion-dominant framing, or duplication. Backward and forward snowballing, tracing hyperlinks, named references, and cited resources within the 16 retained grey sources, produced 12 candidates. Of these, 5 were duplicates or near-duplicates of already-included material and 2 lacked substantive technical content (excluded by GL-E1), leaving 5 candidates for quality assessment. Two of the 5 scored $\leq$4 out of 15 (GL-I3) and were excluded, and the remaining 3 were added, giving 19 grey sources. With the preprint \ref{GL204} from the first database search (Section~\ref{sec:wl_selection}), 22 grey sources of the first database search and the five grey channels were assessed for credibility (the 16 retained at screening, the 5 snowballing candidates, and \ref{GL204}), and 20 were retained. The three preprints of window 1 from the second database search were also assessed. Two (\ref{GL21}, \ref{GL22}) were retained, and one was removed as the same article as \ref{GL1}. In all, 25 grey sources outside the logged grey search were assessed, 2 scored at or below the floor, 1 was a duplicate, and 22 were retained.

\paragraph{Logged grey search (September 2026)}
The logged grey search covered the grey channels under a logged protocol over both windows. Its retained sources are assigned to a window by their publication date. Google web searches were run in a logged-out browser with personalization disabled and language and region fixed (parameters \texttt{hl=en}, \texttt{gl=fi}, \texttt{pws=0}, ten results per page). We used the four terms of the five grey channels and eight facet queries, each combining ``vibe coding'' with one facet per research question, fixed before the run. The same stopping rule applied, with at most ten pages. Of the 24 Google queries run in the browser (G01 to G24 in the replication package), one kept its full item list of ten pages. Four lost their item-level list, and one of these four was interrupted by a verification challenge at page 7. A verification challenge blocked the other 19. Nine queries, the four incomplete ones and five of the blocked ones, were repeated as native captures of ten pages each. With one control query of the session settings, 34 Google queries are logged, and ten of them returned items (Section~\ref{sec:threat_selection}).

YouTube was searched logged out with the upload-date filter and two sort orders (relevance and the interface's popularity order), sixty results per run. LinkedIn was searched from a user account for posts in relevance and latest order, since the interface offered no separate articles filter. X was searched with date, language, and engagement operators for the first hundred results in top and latest order. arXiv was added as a named channel and queried through its interface. A general web-search service was used as a supplementary channel. Forty queries were sent through it. Thirty-two used the review terms, most with date operators for one of the two windows, and eight searched for the titles of known items. The service ranks results by its own relevance and offers no session controls. It returned 458 items, and 19 sources in the corpus come from it. Every inspected item, with its rank, page, date, and title-level decision, and the per-query log are in the replication package. Table~\ref{tab:grey_search_2026} gives the yield per channel.

The run produced 1{,}559 candidate items across six channels, of which 1{,}172 were carried forward on title, excerpt, and outlet type. Four eligibility tests were then applied in order: directness to the operational definition (Section~\ref{sec:opdef}), a promise of usable evidence (first-hand experience, a concrete workflow, a comparison with stated criteria, a measured or observed outcome, an incident, a safeguard practice, or a definitional contribution), a source identifiable from the record, and independence from other records. Explainer and glossary pages, vendor marketing, ranked tool lists without stated criteria, course advertisements, and news items that only relay a statement fail the second test. The 519 items left for full-content review include 27 items that the first author had kept at title level and that the eligibility tests had then dropped. The reconciliation of the independent screening check returned them to content review (Section~\ref{sec:reliability}). Reading the content excluded items whose full content could not be obtained (92, of which 52 web and social items had only a short page or a search excerpt and 40 videos had no usable transcript), items that restate another source in the corpus (62), items too weak to support a finding (55), and smaller groups. Of the 194 items then taken to credibility assessment on the fifteen-item checklist, 34 were removed and one moved to the peer-reviewed stream. Nineteen of the 34 duplicate a record already counted: 13 a record of the second database search, mostly as its preprint, five a source of the first database search or the five grey channels, and one another grey item. Among them is a 2022 web copy of \ref{WL12}, which is a published paper of the first database search. It was removed so that the study is counted once. Six were published after the cutoff, four scored at or below the exclusion floor of 4, one has a deleted page with no archived copy, and one was an April Fools' post found by credibility scoring. Three grey sources authored by members of the review team were removed so that no author's assessment of their own work enters the analysis. Their removal changes no label.

The search export carried no publication date for 87 of these grey items, 68 of them from the Google native channel. The authors read each date from the page text or metadata, or from an archived copy for five pages that are no longer online. 84 dates are precise to the day and 2 to the month. For one page without a date in its live or archived version, the first author established only the year (2026). Six of the 87 were published after the cutoff of 9 September 2026 and left under the post-cutoff rule of the grey screening instruction, all six Tier~3 and restricted. Twenty-nine were published before 1 November 2025. They were kept, because the search covered both windows, except two that duplicate peer-reviewed sources of the first database search and are among the duplicates above (a web copy of \ref{WL12} and a grey copy of \ref{WL74}). Of the 159 grey sources that this search retained, 57 are dated in window 1 and belong to window 1, and 102 are dated in window 2 and belong to window 2. For Figure~\ref{fig:round1_flow} its candidates were split by window in the same way. A candidate takes the window of its date where the date is known, and otherwise the window of the query that returned it, since most queries were run for one window. Of the 1{,}559 candidates, 527 fall in window 1 and 1{,}032 in window 2.

Under the stream rule of the grey screening instruction, a journal or conference paper belongs to the peer-reviewed stream even when the grey search finds it. One grey source, a published conference paper with no copy in the peer-reviewed stream, therefore moved to that stream as \ref{WL77}. Scored on the eleven-item checklist from its full text, it reached 6.5 of 10 applicable items (0.65, Medium band). QA9 was not applicable, because the authors are the only users of their prototype. \ref{WL77} was published in 2024 and belongs to window 1.

\begin{table}[!ht]
\centering
\caption{Credibility bands for grey sources, from the score on the 15-item checklist of Garousi et al.~\cite{garousi2019guidelines} (0 to 15, Table~\ref{tab:qa_grey}).}
\label{tab:gl_tiers}
\footnotesize
\renewcommand{\arraystretch}{1.1}
\setlength{\tabcolsep}{4pt}
\begin{tabular}{|p{1.25cm}|>{\raggedright\arraybackslash}p{1.55cm}|>{\raggedright\arraybackslash}p{1.9cm}|>{\raggedright\arraybackslash}p{2.6cm}|}
\hline
\rowcolor[HTML]{EFEFEF}
\textbf{Tier} & \textbf{Score} & \textbf{Interpretation} &
\textbf{Use in synthesis} \\
\hline
GL~Tier~1 & $\geq 12$ ($\geq 80\%$) & High credibility grey &
May support major themes and is still labeled as grey evidence. \\
\hline
GL~Tier~2 & $9 \leq s < 12$ ($60$ to $79\%$) & Moderate credibility grey &
Supports themes with caution, triangulated with peer-reviewed literature
where possible. \\
\hline
GL~Tier~3 & $4 < s < 9$ ($30$ to $57\%$) & Low credibility grey &
Include only if uniquely relevant, not used for strong
conclusions. \\
\hline
\multicolumn{4}{|p{8.15cm}|}{\textit{Exclusion floor (not a substantive tier):} sources scoring $\leq$4 ($<$27\%) were excluded from data extraction before tier assignment, consistent with MLR credibility guidance~\cite{garousi2019guidelines}.} \\
\hline
\end{tabular}
\end{table}

\paragraph{Both grey searches together}
The grey stream combines the 24 candidates of the five grey channels and the 1{,}559 candidates of the logged grey search (Figure~\ref{fig:round1_flow}). With the 13 preprints from the database searches, 228 grey sources were assessed for credibility. Of these, 37 were removed, one moved to the peer-reviewed stream, and 190 were retained. By publication date, 79 belong to window 1 (20 from the five grey channels and the first database search, 2 preprints of the second database search, and 57 from the logged grey search) and 111 to window 2 (102 from the logged grey search and 9 preprints).

\paragraph{Credibility assessment}
The first author scored every retained grey source on the 15-item checklist of Garousi et al.~\cite{garousi2019guidelines}, reproduced verbatim in Table~\ref{tab:qa_grey}. Sources scoring $\leq$4 out of 15 were excluded from data extraction (GL-I3), following MLR guidance~\cite{garousi2019guidelines}. Retained sources were grouped into three credibility bands by their score, for evidence weighting during synthesis (Table~\ref{tab:gl_tiers}).

This inclusion threshold is deliberately permissive so that definitional and discourse sources, such as the post in which the term was coined, remain in the corpus. Tier~3 sources, and sources in the social-media, video, vendor-documentation, and journalism categories, contribute to definitions (RQ1), practitioner discourse, and tool visibility (RQ7) only, and never to outcome, failure, or safeguard findings. Section~\ref{sec:sensitivity} recomputes every label without the Tier~3 sources and, separately, without those source categories.

After the reconciliation of the independent second scoring (Section~\ref{sec:reliability}), the 19 grey sources of the five grey channels are 3 in Tier~1 (\ref{GL1}, \ref{GL7}, \ref{GL10}), 10 in Tier~2, and 6 in Tier~3. The restriction rule applies to nine of them, the six Tier~3 sources and the three Tier~2 sources in the vendor-documentation and journalism categories. The 159 grey sources of the logged grey search are 67 in Tier~1, 51 in Tier~2, and 41 in Tier~3, and 64 of them are restricted. By window, 37 grey sources of window 1 and 36 of window 2 are restricted, 73 in all. Where peer-reviewed evidence reported the same phenomenon, findings resting on Tier~2 or Tier~3 sources are reported with the peer-reviewed citations (Section~\ref{sec:results}). The tier distribution by window is in Table~\ref{tab:corpus_profile}.

\begin{table*}[!b]
\centering
\footnotesize
\renewcommand{\arraystretch}{1.1}
\setlength{\tabcolsep}{4pt}
\caption{Logged grey search by channel, over both windows. A record found through several channels is counted under each of them, so the channel rows add up to more than the last row, which counts each record once. The per-query log is in the replication package.}
\label{tab:grey_search_2026}
\resizebox{\textwidth}{!}{%
\begin{tabular}{|p{4.4cm}|p{1.37cm}|p{1.76cm}|p{1.57cm}|p{2.16cm}|p{2.16cm}|p{2.25cm}|}
\hline
\rowcolor[HTML]{EFEFEF} & \multicolumn{2}{c|}{\textbf{Queries}} & & & & \\
\cline{2-3}
\rowcolor[HTML]{EFEFEF}\multirow{-2}{=}{\textbf{Channel}} & \textbf{Logged} & \textbf{With items} & \multirow{-2}{=}{\textbf{Raw items}} & \multirow{-2}{=}{\textbf{Entered screening}} & \multirow{-2}{=}{\textbf{Kept at title level}} & \multirow{-2}{=}{\textbf{Retained in corpus}} \\
\hline
Google web search & 34 & 10 & 981 & 862 & 593 & 80 \\
\hline
Web search service (supplementary) & 40 & 37 & 458 & 309 & 298 & 19 \\
\hline
X & 7 & 4 & 391 & 159 & 72 & 7 \\
\hline
YouTube & 8 & 8 & 385 & 269 & 252 & 6 \\
\hline
arXiv & 2 & 1 & 139 & 113 & 106 & 55 \\
\hline
LinkedIn & 6 & 2 & 120 & 75 & 62 & 1 \\
\hline
Google Scholar (grey portion) & 2 & 0 & 0 & 0 & 0 & 0 \\
\hline
\textbf{All, each record once} & 99 & 62 & 2,474 & 1,559 & 1,172 & 159 \\
\hline
\end{tabular}
}
\end{table*}

\subsubsection{Assignment of sources to publication windows}\label{sec:update2026}

Because the practice and its literature are recent, the corpus is divided into two publication windows, following guidance on updating secondary studies~\cite{wohlin2020guidelines,mendes2020when}. A source belongs to the window of its publication date, whichever search found it. Window 1 holds the sources published from 1~January 2022 to 31~October 2025, and window 2 those published from 1~November 2025 to 9~September 2026. Every record of the first database search and the five grey channels falls in window 1. A record of the second database search was assigned to a window by the publication date in its export, and a journal article that appeared online before its printed issue is dated by its online publication date, which the bibliographic record in the replication package carries (for example \ref{WL73}, online in June 2026 and in the December 2026 issue). A grey source was assigned by the date read from the page or its metadata (Section~\ref{sec:grey_selection}).

All searches used one set of selection criteria, checklists, extraction form, and codebook. The searches of October 2025 and September 2026 differ in three respects. The second database search ran the strings Q1 and Q2 in place of the Stage~B string (Section~\ref{sec:search_string_a_b}). The grey search followed a logged protocol over both windows (Section~\ref{sec:grey_selection}). In the peer-reviewed stream, background records (Section~\ref{sec:opdef}) were excluded at full text. Of the 11 background sources found by the second database search and the logged grey search, ten are grey and one, \ref{WL51} (window 2), is peer-reviewed. The 16 peer-reviewed background sources of window 1 all come from the first database search. The direct and adjacent strata therefore reach 9~September 2026 through both strings and the logged grey search.

A publication found by more than one search was counted once (Sections~\ref{sec:wl_selection} and~\ref{sec:grey_selection}). Window 1 holds 113 sources: 34 peer-reviewed (30 papers of the first database search and its snowballing, three papers of the second database search, and \ref{WL77}) and 79 grey (20 from the first database search and the five grey channels, two preprints from the second database search, and 57 from the logged grey search). Window 2 holds 153 sources, 42 peer-reviewed and 111 grey (102 from the logged grey search and nine preprints). The corpus therefore holds 266 sources, 76 peer-reviewed and 190 grey (Figure~\ref{fig:round1_flow}).

The replication package holds the strings and per-database syntax of both searches, the dates and result counts, the reason per excluded record, the scores of the excluded papers, the date evidence and method per item, the window of every source with the basis of its date, and the item scores with quoted evidence for \ref{WL74}, \ref{WL75}, \ref{WL76}, and \ref{WL77}. Section~\ref{sec:validity} discusses the limitations that follow from the differences between the searches, and Section~\ref{sec:sensitivity} reports every finding on each window alone as a sensitivity analysis.

\begin{table*}[!t]
\centering
\caption{Quality assessment criteria for grey literature, adopted
verbatim from Garousi et al.~\cite{garousi2019guidelines}. Each of the fifteen sub-questions scores 0, 0.5, or 1 as described in the last column, and the total (0 to 15) is mapped to tiers in Table~\ref{tab:gl_tiers}.}
\label{tab:qa_grey}
\footnotesize
\renewcommand{\arraystretch}{1.15}
\resizebox{\textwidth}{!}{%
\begin{tabular}{|>{\raggedright\arraybackslash}p{2.6cm}|>{\raggedright\arraybackslash}p{5.6cm}|>{\raggedright\arraybackslash}p{8.4cm}|}
\hline
\rowcolor[HTML]{EFEFEF}
\textbf{Criterion} & \textbf{Question} & \textbf{Score description} \\
\hline
\multirow{4}{=}{Authority of the producer} & Is the publishing organization reputable? & 1: Reputable and well-known organization; 0.5: Existing but not well-known organization; 0: Unknown or low-reputation organization \\
\cline{2-3}
 & Is an individual author associated with a reputable organization? & 1: Yes; 0: No \\
\cline{2-3}
 & Has the author published other work in the field? & 1: More than three works; 0.5: 1 to 2 works; 0: No other works \\
\cline{2-3}
 & Does the author have expertise in the area? & 1: Relevant professional role; 0: Not relevant \\
\hline
\multirow{3}{=}{Methodology} & Does the source have a clearly stated aim? & 1: Yes; 0: No \\
\cline{2-3}
 & Is the source supported by authoritative references? & 1: Reputable sources; 0.5: Non-highly reputable sources; 0: No references \\
\cline{2-3}
 & Does the work cover a specific question? & 1: Yes; 0.5: Not explicit; 0: No \\
\hline
\multirow{4}{=}{Objectivity} & Does the work seem to be balanced in presentation? & 1: Yes; 0.5: Partially; 0: No \\
\cline{2-3}
 & Is the statement objective or subjective? & 1: Objective; 0.5: Partially objective; 0: Subjective \\
\cline{2-3}
 & Are the conclusions free of bias? & 1: No vested interest; 0.5: Partial interest; 0: Strong interest \\
\cline{2-3}
 & Are the conclusions supported by the data? & 1: Yes; 0.5: Partially; 0: No \\
\hline
Date & Does the item have a clearly stated date? & 1: Yes; 0: No \\
\hline
Position w.r.t.\ related sources & Have related sources been discussed? & 1: Yes; 0: No \\
\hline
Novelty & Does it add something unique? & 1: Yes; 0.5: Partially; 0: No \\
\hline
Outlet type & Outlet control & 1: High credibility (books, reports, white papers); 0.5: Moderate credibility (news, videos, Stack Overflow); 0: Low credibility (blogs, tweets, emails) \\
\hline
\end{tabular}
}
\end{table*}

\subsection{Data extraction}\label{sec:extraction}

All 266 sources were extracted with the same form. The form is organized around the research questions (Table~\ref{tab:data_extraction}), and the first author filled it in. Each field holds a short analytical summary and, where possible, a supporting excerpt or section reference, so each item is traceable to its source. Missing or unclear information was marked as \emph{Not reported} or \emph{Unknown} rather than inferred.

Coding was hybrid~\cite{cruzes2011recommended}. Codes were derived deductively from the eight research questions, and further codes were added inductively while coding the sources first selected from the first database search and the five grey channels. The codebook was then frozen at 56 codes before the other sources were coded, so that the second author's coding check and the evidence-strength labels rest on one codebook. Each code is linked to one research question, so a source contributing to several questions does so through different codes and is not counted twice.

Content outside the codebook was noted during coding but not coded, and it is not reported as findings. Section~\ref{sec:threat_coding} lists these topics.

Beside the codes, three further fields were recorded. The first is the reporting mode of each workflow pattern (RQ2) and each safeguard (RQ5.2) a source carries, recorded for each source and item under a written rule: evaluated when the source reports an evaluation of the item, in use when it reports using the item without an evaluation, and recommended when it advises the item without reporting its use. A source that uses and recommends an item is therefore recorded as in use, and one that uses and evaluates it as evaluated. The mode was read from the extraction cell of the question and, for safeguards, also from the outcome cell, since evaluations of a safeguard are often reported with the outcomes of a tool. The second field is the direction, measurement, and horizon of each RQ4 outcome a source reports. The third is the record of safeguard evaluations (Table~\ref{tab:safeguard_evaluations}): for every source and coded safeguard whose mode is evaluated, the safeguard, what was evaluated, the design, and the reported result. Table~\ref{tab:data_extraction} lists them with the other fields.

The outcome fields are recorded separately for each outcome code a source carries, under a written instruction with seven decision notes. The record holds 438 pairs of source and outcome from 241 sources. Of these, 332 pairs come from the 175 sources that contribute to RQ4, and the labels use the 303 pairs of the 159 sources among them that are not background. The other 106 pairs come from 66 restricted sources, which do not contribute to outcome findings. A source that reports a gain on one outcome and a cost on another has each outcome recorded on its own, so neither is mixed. Mixed is kept for a gain and a loss, or a conditional result, within the same outcome. No effect is recorded when the source reports that the outcome was measured or observed and did not change. Not stated is recorded when the source names the outcome without a direction. No effect counts as a direction in criterion C4, and Not stated does not (Section~\ref{sec:evidence_strength}). Each RQ4 row of the evidence matrix therefore uses only what a source says about that outcome. The first author recorded the fields from the RQ4 extraction cells. The first author reviewed a stratified 30\% sample of 135 pairs (134 after the removal described in Section~\ref{sec:grey_selection}) against the instruction. The review differed from the recorded values in 39 pairs. Values were corrected in 19 of them, and in 13 further pairs the value was harmonized under the decision notes. This is part of the first author's recording and not a reliability check. Section~\ref{sec:reliability} reports the second author's check of these fields. The instruction and the record are in the replication package.

\subsection{Data synthesis}

The synthesis had two stages, descriptive mapping and thematic synthesis.

\paragraph{Descriptive mapping}

Sources were summarized by publication year, venue or source type, stream, and research question coverage.

\paragraph{Thematic synthesis}

We applied the codes to the extracted text, grouped them under RQ1 to RQ8, and compared peer-reviewed and grey evidence for each RQ. We then consolidated related codes into higher-level themes. Each theme is linked to its source identifiers and representative excerpts.

\begin{table*}[!t]
\centering
\caption{Data extraction form (grouped view). The field-level extraction for every source is in the replication package.}
\label{tab:data_extraction}
\small
\renewcommand{\arraystretch}{1.25}
\resizebox{\textwidth}{!}{%
\begin{tabular}{|p{3.8cm}|p{10.0cm}|p{2.1cm}|}
\hline
\rowcolor[HTML]{EFEFEF}
\textbf{Data item} & \textbf{Description} & \textbf{Linked RQ} \\
\hline
Study metadata & Title, authors, year, venue or outlet, source type (peer-reviewed or grey), and publication type & Demographic \\
\hline
Study method & Research method or type of contribution & Demographic \\
\hline
Conceptualization & How the source defines and frames vibe coding & RQ1 \\
\hline
Workflow and tactics & Interaction patterns, task distribution, and automation boundaries & RQ2 \\
\hline
Developer role & Trust, flow, agency, cognitive load, and cognitive debt & RQ3.1 \\
\hline
Session dynamics & Momentum, context loss, and disengagement during a session & RQ3.2 \\
\hline
Audience and task & Differences by user group and task type & RQ3.3 \\
\hline
Outcomes & Productivity, prototyping speed, code and UI quality, defects, maintainability, and reproducibility & RQ4 \\
\hline
Failure modes & Ways in which the human-AI process breaks down & RQ5.1 \\
\hline
Safeguards & Governance and guardrails & RQ5.2 \\
\hline
Adoption context & Where, by whom, and under what constraints vibe coding is used & RQ6 \\
\hline
Tools and platforms & Named tools and supporting infrastructure & RQ7 \\
\hline
Open challenges & Reported open challenges and future directions & RQ8 \\
\hline
Reporting mode & For each workflow pattern or safeguard a source carries: evaluated, in use, or recommended & RQ2, RQ5.2 \\
\hline
Outcome fields & Direction, measurement type, and horizon, recorded for each outcome a source reports (values in Section~\ref{sec:extraction}) & RQ4 \\
\hline
Safeguard evaluations & The safeguard, what was evaluated, the design, and the result, for each safeguard item recorded as evaluated & RQ5.2 \\
\hline
\end{tabular}
}
\end{table*}

\subsection{Labels of the findings}\label{sec:evidence_strength}

\begin{table}[!ht]
\centering
\caption{Decision matrix for the labels: reporting support for presence findings, evidence strength for outcome findings. Rows apply from top to bottom, and a finding takes the label of the first row whose conditions all hold. C1: independent sources. C2: empirical peer-reviewed sources in the High or Medium band. C3: empirical Tier~1 or Tier~2 grey sources. C4: consistency of the reported directions (outcome findings only). Restricted share: share of Tier~3 or restricted-category sources (Section~\ref{sec:evidence_strength}).}
\label{tab:evidence_matrix}
\footnotesize
\renewcommand{\arraystretch}{1.1}
\begin{tabular}{|p{1.4cm}|>{\raggedright\arraybackslash}p{6.15cm}|}
\hline
\rowcolor[HTML]{EFEFEF}
\textbf{Label} & \textbf{Rule (all conditions of a row must hold)} \\
\hline
Strong & C1 $\geq$ 8, C2 $\geq$ 3, restricted share $<$ 25\%, and C4 consistent (outcome findings) or not applicable (presence findings) \\
\hline
Moderate & C1 $\geq$ 5, (C2 $\geq$ 1 or C3 $\geq$ 2), and C4 consistent, minor disagreement, or not applicable \\
\hline
Weak & C1 of 3 or 4, or C1 $\geq$ 5 with C2 = 0 and C3 = 0, or C4 contested or not assessable \\
\hline
Emerging & C1 of 1 or 2 \\
\hline
\end{tabular}
\end{table}

Each finding (each row of a results table) receives a label from four recorded criteria. C1 is the number of independent sources. Two sources count once when they are two copies of one source, when they report one study or dataset, or when one restates or reports on the other (the dependency rule, in the replication package). Citing the same definition, post, or review is not a dependency. A dependency field in the extraction form of every peer-reviewed source and the credibility form of every grey source names the work each source builds on, and it has an entry for 160 of the 266 sources. Every corpus source named in such an entry, every pair with a shared author name or a near-identical title, and every pair judged dependent at an earlier reconciliation, 191 candidate pairs in all, was decided under the rule with a written reason. Ten pairs are dependent: two copies of one source, two companion papers on one study, and seven sources whose evidence restates another source's results. The other 181 are citations of a definition or framing (136), different work by one author (6), or unrelated sources with similar titles (39). The dependent pairs are merged within each finding, which lowers C1 in 32 of the 60 findings, and Section~\ref{sec:sensitivity} shows that no label changes without merging. Background sources are removed before C1 to C4 are computed. C2 is the number of peer-reviewed sources among them whose evidence category is empirical or tool evaluation, that report a study design, and that are in the High or Medium quality band. C3 is the number of Tier~1 or Tier~2 grey sources among them with an empirical basis (observation, interviews, mining, survey). C3 also records the share of the supporting sources that are Tier~3 or belong to the social-media, video, vendor-documentation, or journalism categories, counted before dependent sources are merged (the restricted share).

C4 is the consistency of direction among sources of comparable standing. Its rule depends on the type of finding.

\begin{table*}[!t]
\centering
\caption{Reliability checks by the second author that bear on the final corpus and labels. Agreement and Cohen's kappa ($\kappa$) are computed before reconciliation. Blind: made without sight of the first author's decision. N/D: not defined. The full account is in the replication package.}
\label{tab:reliability}
\footnotesize
\renewcommand{\arraystretch}{1.0}
\setlength{\tabcolsep}{2.5pt}
\begin{tabular}{|>{\raggedright\arraybackslash}p{4.7cm}|>{\raggedright\arraybackslash}p{2cm}|>{\raggedright\arraybackslash}p{0.8cm}|>{\raggedright\arraybackslash}p{1.1cm}|>{\raggedright\arraybackslash}p{2.4cm}|>{\raggedright\arraybackslash}p{2.4cm}|>{\raggedright\arraybackslash}p{2.5cm}|}
\hline
\rowcolor[HTML]{EFEFEF}
\textbf{Check} & \textbf{Coverage} & \textbf{Blind} & \textbf{N} & \textbf{Agreement} & \textbf{$\kappa$} & \textbf{Effect on labels} \\
\hline
First database search, exclusions at the title stage & all 245, plus 100 controls & partly & 345 & 0.74 & 0.15 & \ref{WL74} added \\
\hline
First database search, exclusions at full text & all 53 & no & 53 & 0.77 & N/D\textsuperscript{a} & \ref{WL75}, \ref{WL76}, \ref{GL204} added \\
\hline
Second database search, screening & kept or undecided, plus a sample of exclusions & yes & 304 & 0.71 & 0.36 & \ref{WL73}, \ref{GL31} added \\
\hline
Logged grey search, screening & carried forward, plus a sample of exclusions & yes & 1,249 & 0.62 & 0.28 & 7 items added \\
\hline
Directness, first form of the rule, sources found in October 2025 & all 47 & yes & 47 & 0.94 & 0.89 & none \\
\hline
Directness, first form of the rule, sources classified in 2026 & 30\% sample & yes & 67 & 0.69 & 0.37 & none \\
\hline
Directness, final rule & 30\% sample & yes & 80 & 0.88 & 0.79 & none \\
\hline
Quality, peer-reviewed (items / band) & 30\% sample & yes & 242 / 22 & 0.78 / 0.68 & 0.56 / 0.19 & none; one paper left \\
\hline
Credibility, grey (items / tier) & 30\% sample & yes & 900 / 60 & 0.69 / 0.68 & 0.50 / 0.53 & none \\
\hline
Coding, 56 codes & 25\% sample & yes & 3,808 & 0.94 & 0.83 & none \\
\hline
Supporting sets, 18 findings & all candidates & no & 891 & 0.85 & N/D\textsuperscript{b} & none \\
\hline
Outcome fields, per outcome (direction / measurement / horizon) & all 332 pairs & yes & 332 & 0.61 / 0.67 / 0.72 & 0.47 / 0.49 / 0.41 & none \\
\hline
\end{tabular}

\smallskip
\raggedright\footnotesize \textsuperscript{a}Every record carries the same first decision, so the informative result is the number marked for rereading. \textsuperscript{b}Every candidate is in the first author's set by construction, so the informative result is the number judged not to support the finding.
\end{table*}

For an outcome finding (RQ4), C4 is derived from the direction that the C2 and C3 sources record for the outcome of the row (Section~\ref{sec:extraction}). A source that reports several outcomes contributes to each row only what it says about that outcome. Positive, negative, mixed, and no effect count as directions. Sources whose direction is not stated are ignored. C4 is \emph{not assessable} when fewer than two of these sources record a direction. Otherwise, it is \emph{contested} when at least one records a positive and at least one a negative direction. Otherwise, it is a \emph{minor disagreement} when directions of more than one kind occur. Otherwise, it is \emph{consistent}, that is, all record the same direction. Dissent from Tier~3 or non-empirical sources is recorded as a flag and does not change C4. Under this rule, one comparable source that reports the opposite direction makes a row contested, even when most comparable sources report one direction. Section~\ref{sec:sensitivity} reports a variant in which a row is contested only when the smaller of the positive and negative counts is at least 20\% of the directions.

The findings of the other research questions, 51 in all, are presence findings. They record that a phenomenon is reported, not the direction of an effect. The codebook has no code for denying or contradicting a phenomenon, so presence findings have no direction to compare, and C4 is \emph{not applicable}. In Table~\ref{tab:evidence_matrix}, not applicable is accepted wherever consistent is accepted. The label of a presence finding therefore rests on C1, C2, C3, and the restricted share alone. For this reason, the two kinds of finding carry two kinds of label. A presence finding carries a \emph{reporting-support} label: Strong, Moderate, Weak, or Emerging reporting support. It says how many independent sources, and of what standing, report the phenomenon. It does not say that the phenomenon is effective or beneficial, or that the sources agree about an effect. An outcome finding carries an \emph{evidence-strength} label: Strong, Moderate, or Weak evidence. It also takes into account how consistent the reported directions are. The word label is used below for both kinds.

The labels follow Table~\ref{tab:evidence_matrix}, whose rows are applied from top to bottom. A finding receives the label of the first row whose conditions all hold, so Weak takes precedence over Emerging. The nine outcome findings are contested in four rows (short-term productivity, quality, and defects, and long-term maintainability), a minor disagreement in two (long-term productivity and long-term quality), consistent in one (short-term reproducibility), and not assessable in two (short-term maintainability and long-term defects). C4 is not applicable for the other 51 findings. The 56 codes give 51 presence findings and five outcome codes. Each outcome is split into a short-term and a long-term row, which would give ten rows, but no source reports a long-term reproducibility outcome, so RQ4 has nine rows and the review has 60 findings. Of the 51 presence findings, 42 have Strong, 8 Moderate, and 1 Weak reporting support. Of the nine outcome findings, 3 have Moderate and 6 Weak evidence, and none has Strong evidence. Over all 60 findings, the labels are 42 Strong, 11 Moderate, and 7 Weak. Section~\ref{sec:reliability} reports which parts of this procedure a second person checked.

\begin{table*}[!t]
\centering
\footnotesize
\renewcommand{\arraystretch}{1.1}
\setlength{\tabcolsep}{4pt}
\caption{Corpus profile by stream and publication window (266 sources). Window 1: publications to October 2025 (113 sources). Window 2: publications from November 2025 to 9 September 2026 (153 sources). Directness follows Section~\ref{sec:opdef}.}
\label{tab:corpus_profile}
\resizebox{\textwidth}{!}{%
\begin{tabular}{|p{2.6cm}|p{4.3cm}|p{1.33cm}|p{1.33cm}|p{1.33cm}|p{1.33cm}|p{1.33cm}|p{1.33cm}|}
\hline
\rowcolor[HTML]{EFEFEF} & & \multicolumn{3}{c|}{\textbf{Peer-reviewed}} & \multicolumn{3}{c|}{\textbf{Grey}} \\
\cline{3-8}
\rowcolor[HTML]{EFEFEF}\multirow{-2}{=}{\textbf{Dimension}} & \multirow{-2}{=}{\textbf{Category}} & \textbf{Window 1} & \textbf{Window 2} & \textbf{All} & \textbf{Window 1} & \textbf{Window 2} & \textbf{All} \\
\hline
Publication type or outlet & Web article, blog, or guide & 0 & 0 & 0 & 42 & 47 & 89 \\
\hline
 & Preprint or repository & 0 & 0 & 0 & 24 & 54 & 78 \\
\hline
 & Conference & 22 & 29 & 51 & 0 & 0 & 0 \\
\hline
 & Journal & 8 & 12 & 20 & 0 & 0 & 0 \\
\hline
 & Social media or forum & 0 & 0 & 0 & 7 & 6 & 13 \\
\hline
 & Video & 0 & 0 & 0 & 2 & 4 & 6 \\
\hline
 & Workshop & 4 & 1 & 5 & 0 & 0 & 0 \\
\hline
 & Journalism & 0 & 0 & 0 & 2 & 0 & 2 \\
\hline
 & Vendor or industry report & 0 & 0 & 0 & 2 & 0 & 2 \\
\hline
Evidence category & Empirical & 14 & 34 & 48 & 17 & 29 & 46 \\
\hline
 & Tool/System & 10 & 5 & 15 & 6 & 17 & 23 \\
\hline
 & Experience report & 0 & 0 & 0 & 31 & 25 & 56 \\
\hline
 & Secondary & 3 & 0 & 3 & 6 & 11 & 17 \\
\hline
 & Conceptual & 6 & 1 & 7 & 7 & 7 & 14 \\
\hline
 & Position & 1 & 2 & 3 & 10 & 22 & 32 \\
\hline
 & Not classified & 0 & 0 & 0 & 2 & 0 & 2 \\
\hline
Directness & Direct & 8 & 27 & 35 & 55 & 56 & 111 \\
\hline
 & Adjacent & 10 & 14 & 24 & 20 & 49 & 69 \\
\hline
 & Background & 16 & 1 & 17 & 4 & 6 & 10 \\
\hline
Quality band (peer-reviewed) & High & 21 & 29 & 50 & N/A & N/A & N/A \\
\hline
 & Medium & 10 & 7 & 17 & N/A & N/A & N/A \\
\hline
 & Low & 3 & 6 & 9 & N/A & N/A & N/A \\
\hline
Credibility tier (grey) & GL Tier 1 & N/A & N/A & N/A & 22 & 58 & 80 \\
\hline
 & GL Tier 2 & N/A & N/A & N/A & 31 & 32 & 63 \\
\hline
 & GL Tier 3 & N/A & N/A & N/A & 26 & 21 & 47 \\
\hline
\textbf{All} & & 34 & 42 & 76 & 79 & 111 & 190 \\
\hline
\end{tabular}
}
\par\smallskip\noindent\footnotesize{Grey preprints are scored on the grey credibility checklist. Research method categories are in Table~\ref{tab:corpus_method}. N/A: the category does not apply to the stream.}
\end{table*}

Outcome findings are split by the horizon recorded for the outcome (short-term, long-term) before labeling, and a source with both horizons enters both rows, so that support for short-term productivity is not read as support for long-term effectiveness. Opinion-based sources alone never exceed Weak, because C2 = 0 and C3 = 0 satisfy only the Weak and Emerging rows of Table~\ref{tab:evidence_matrix}, and repeated unverified practitioner claims count once. A finding with few but consistent empirical sources can reach a higher label than a finding with many sources whose reported directions conflict, so the labels are not a ranking by count. The one combination that satisfies no row, C1 $\geq$ 5 with C2 = 0 and C3 = 1, would be labeled Weak, and no finding of this review falls there. Sheet S3 of the supplementary workbook lists the criterion values and label for every finding, and Section~\ref{sec:sensitivity} reports how many labels change when each threshold is moved by one step. The labels are relative to the criteria and thresholds of Table~\ref{tab:evidence_matrix}. Other criteria or other thresholds would give other labels, and Section~\ref{sec:sensitivity} shows how many labels move with each one-step change. A label summarizes how much, by which sources, and how consistently a phenomenon or outcome is reported in this corpus. It is not a conclusion about whether vibe coding is effective.

\subsection{Reliability procedures}\label{sec:reliability}

The first author made the primary pass at every stage of both searches. During the selection of 2025, a sample of the decisions, scores, codes, and themes was reviewed, and disagreements were resolved through discussion. The size of that sample and the initial disagreements were not recorded. In September and October 2026 the second author made twelve checks that bear on the final corpus and labels, from the screening of the exclusions of the first database search to the RQ4 outcome fields (Table~\ref{tab:reliability}). Four earlier checks, whose results the labels do not use, are reported in the replication package. Four checks covered a full set, and most of the others covered a sample. A blind check of a 30\% sample of the per-outcome RQ4 record preceded the full check of that record, and its values are in the replication package. Each check followed a written instruction that was frozen before it. Most checks were blind, that is, made without sight of the first author's decisions, and Table~\ref{tab:reliability} marks the exceptions. Agreement was measured before reconciliation as percent agreement and Cohen's kappa. This section describes a kappa below 0.40 as low, from 0.40 to 0.60 as moderate, from 0.61 to 0.80 as substantial, and above 0.80 as near perfect~\cite{landis1977measurement}. The two authors then reconciled the disagreements, and both confirmed each reconciliation. The check-by-check account, with the samples, the timing, the instructions, the prevalence- and bias-adjusted kappa (PABAK), and every disagreement and its resolution, is in the replication package.

\begin{table*}[!t]
\centering
\footnotesize
\renewcommand{\arraystretch}{1.1}
\setlength{\tabcolsep}{4pt}
\caption{Research method by stream and publication window (266 sources). Method descriptions were grouped by a rule set given in the replication package.}
\label{tab:corpus_method}
\begin{tabular}{|p{5.5cm}|p{1.61cm}|p{1.61cm}|p{1.61cm}|p{1.61cm}|p{1.61cm}|p{1.61cm}|}
\hline
\rowcolor[HTML]{EFEFEF}
& \multicolumn{3}{c|}{\textbf{Peer-reviewed}} & \multicolumn{3}{c|}{\textbf{Grey}} \\
\cline{2-7}
\rowcolor[HTML]{EFEFEF}\multirow{-2}{=}{\textbf{Category}} & \textbf{Window 1} & \textbf{Window 2} & \textbf{All} & \textbf{Window 1} & \textbf{Window 2} & \textbf{All} \\
\hline
Benchmark & 1 & 3 & 4 & 4 & 10 & 14 \\
\hline
Case study & 3 & 9 & 12 & 0 & 4 & 4 \\
\hline
Design proposal or tool description & 0 & 4 & 4 & 0 & 3 & 3 \\
\hline
Experience report or practitioner guide & 0 & 0 & 0 & 24 & 22 & 46 \\
\hline
Experiment or user study & 4 & 1 & 5 & 3 & 2 & 5 \\
\hline
Interviews & 5 & 3 & 8 & 3 & 3 & 6 \\
\hline
Mixed methods & 2 & 10 & 12 & 3 & 18 & 21 \\
\hline
Not reported & 0 & 1 & 1 & 2 & 0 & 2 \\
\hline
Opinion, essay, or journalism & 0 & 0 & 0 & 6 & 18 & 24 \\
\hline
Other & 5 & 4 & 9 & 21 & 15 & 36 \\
\hline
Repository mining & 0 & 4 & 4 & 0 & 3 & 3 \\
\hline
Review or mapping & 2 & 0 & 2 & 6 & 5 & 11 \\
\hline
Static or automated analysis & 5 & 1 & 6 & 3 & 3 & 6 \\
\hline
Survey & 5 & 2 & 7 & 1 & 1 & 2 \\
\hline
Taxonomy or framework & 2 & 0 & 2 & 3 & 4 & 7 \\
\hline
\textbf{All} & 34 & 42 & 76 & 79 & 111 & 190 \\
\hline
\end{tabular}
\end{table*}

Before reconciliation, kappa in Table~\ref{tab:reliability} ranged from 0.15 to 0.89. Agreement was near perfect for coding (kappa 0.83) and for the first directness check of the sources from the first database search and the five grey channels (0.89). It was substantial for directness under the final rule (0.79). Five kappa values in the table were low: the title-stage exclusions of the first database search (0.15), the screening of the records of the second database search and the logged grey search (0.36 and 0.28), the directness of the sources classified in 2026 under the first form of the rule (0.37), and the quality band (0.19). For the per-outcome RQ4 fields, checked on all 332 pairs, agreement was moderate for direction (0.47), measurement type (0.49), and horizon (0.41). The two authors reconciled the 334 disagreements of that check, in 185 pairs, under the written rule. Thirty values changed, five directions, 16 measurement types, and nine horizons, and no label changed. Three of the four earlier checks in the replication package also had low kappa values. These were a check of C4 as a judgment on 18 findings (0.34), a first check of outcome direction at the source level, which was not blind (0.14), and a judgment of dependence between the 891 candidate sources of the 18 findings (0.07, with agreement 0.89), which the dependency rule of Section~\ref{sec:evidence_strength} has since replaced. The fourth, a blind check of the source-level outcome fields, gave kappa 0.70 for direction, 0.56 for measurement type, and 0.48 for horizon. The labels use the per-outcome record and the rule of Section~\ref{sec:evidence_strength}, so they do not use the results of these three checks. These values suggest that the two authors applied the screening rules differently, and that quality bands and the outcome fields involve judgment. Reconciliation changed which sources are in the corpus. Four records of window 1 entered (\ref{WL74}, \ref{WL75}, \ref{WL76}, \ref{GL204}), \ref{WL73}, \ref{GL31} and seven grey items entered, and one paper left. The labels are computed on the corpus after these changes. The reconciliations of the later checks, from directness to the per-outcome fields, changed no label.

The checks have limits. Most covered a sample, some were not blind, and the coding, evidence-set, C4, and outcome-field checks were made from the first author's extraction cells, so they do not measure agreement in extraction. Section~\ref{sec:sensitivity} tests how far the RQ4 labels depend on the reconciled values. Three checks made before the removal of the three grey sources authored by members of the review team (Section~\ref{sec:grey_selection}) included these sources, and their values are reported as computed then. The reporting mode of the RQ2 and RQ5.2 items, the record of safeguard evaluations (Section~\ref{sec:extraction}), and the dependency links were not checked by a second person, and no label depends on the dependency links (Section~\ref{sec:sensitivity}). Section~\ref{sec:threat_coding} discusses these limits.

\subsection{Reproducibility}

We followed a written protocol based on Garousi et al.~\cite{garousi2019guidelines}, with the criteria, checklists (Tables~\ref{tab:qa_white} and~\ref{tab:qa_grey}), and codebook described above. The data-extraction matrix, the codebook, both forms of the directness rule, and the item-level quality and credibility scores are released in the replication package, whose contents and deposit location are given in the Data availability statement.

\FloatBarrier

\begin{table*}[!t]
\centering
\footnotesize
\renewcommand{\arraystretch}{1.1}
\setlength{\tabcolsep}{3pt}
\caption{Contributing sources per research question, by stream and evidence category (266 sources). The later tables use these contributing sources as their denominator.}
\label{tab:rq_coverage}
\resizebox{\textwidth}{!}{%
\begin{tabular}{|>{\raggedright\arraybackslash}p{3.7cm}|p{1.05cm}|p{0.65cm}|p{1.75cm}|p{1.6cm}|p{0.8cm}|p{0.8cm}|p{0.8cm}|p{0.8cm}|p{0.8cm}|p{0.8cm}|p{0.8cm}|}
\hline
\rowcolor[HTML]{EFEFEF}
\textbf{Research question} & \textbf{Peer-reviewed} & \textbf{Grey} & \textbf{Contributing} & \textbf{Share of corpus} & \textbf{Emp.} & \textbf{Tool} & \textbf{Exp.} & \textbf{Sec.} & \textbf{Conc.} & \textbf{Pos.} & \textbf{N.c.} \\
\hline
RQ1 Definition & 65 & 174 & 239 & 90\% & 87 & 34 & 48 & 17 & 21 & 30 & 2 \\
\hline
RQ2 Workflows & 70 & 102 & 172 & 65\% & 77 & 33 & 22 & 14 & 17 & 9 & 0 \\
\hline
RQ3.1 Role and cognition & 66 & 89 & 155 & 58\% & 72 & 24 & 21 & 13 & 18 & 7 & 0 \\
\hline
RQ3.2 Session momentum & 39 & 57 & 96 & 36\% & 54 & 15 & 14 & 5 & 6 & 2 & 0 \\
\hline
RQ3.3 Audience and task differences & 60 & 84 & 144 & 54\% & 68 & 21 & 19 & 13 & 13 & 10 & 0 \\
\hline
RQ4 Outcomes & 66 & 109 & 175 & 66\% & 78 & 33 & 23 & 15 & 16 & 10 & 0 \\
\hline
RQ5.1 Failure modes & 55 & 96 & 151 & 57\% & 67 & 30 & 16 & 12 & 16 & 10 & 0 \\
\hline
RQ5.2 Safeguards & 67 & 100 & 167 & 63\% & 79 & 33 & 17 & 15 & 15 & 8 & 0 \\
\hline
RQ6 Usage contexts and constraints & 39 & 71 & 110 & 41\% & 45 & 17 & 17 & 11 & 12 & 8 & 0 \\
\hline
RQ7 Tools & 68 & 162 & 230 & 86\% & 86 & 33 & 48 & 18 & 18 & 25 & 2 \\
\hline
RQ8 Challenges & 44 & 79 & 123 & 46\% & 62 & 26 & 9 & 10 & 11 & 5 & 0 \\
\hline
\end{tabular}
}
\par\smallskip\noindent\footnotesize{Evidence category: Emp.: empirical; Tool: tool or system; Exp.: experience report; Sec.: secondary; Conc.: conceptual; Pos.: position; N.c.: not classified (no recorded evidence category).}
\end{table*}

\section{Results} \label{sec:results}
The per-question tables use these columns. \emph{Sources reporting} gives the number of sources coded with the finding (Count) and its percentage of the sources contributing to the question (Table~\ref{tab:rq_coverage}). \emph{Peer-reviewed} and \emph{Grey} split this count by stream. \emph{Background} gives how many of these sources are background sources (Section~\ref{sec:opdef}). Background sources count as reports in these columns, but they are set aside when the label is computed. \emph{Empirical sources} counts the comparable empirical sources by stream: the empirical peer-reviewed studies of High or Medium quality, and the Tier~1 or Tier~2 empirical grey sources. Restricted sources are Tier~3 grey sources and grey sources in the social-media, video, vendor-documentation, or journalism categories. They support only definitions, practitioner discourse, and tool visibility (Section~\ref{sec:grey_selection}). \emph{Reporting support} is the label of Section~\ref{sec:evidence_strength}, with criterion values in sheet S3 of the supplementary workbook in the replication package (the sheets named below are in the same workbook). It says how many independent sources, and of what standing, report the finding. The label and the comparable counts are computed on the evidence set of each finding, after background sources and non-supporting candidates are removed and dependent sources are merged, so they may differ from the number of sources reporting. The RQ4 table has its own columns, which Section~\ref{sec:rq4} describes, and its outcome findings carry an evidence-strength label. Its text gives counts and shares over the 159 RQ4 sources that count toward the labels, while Table~\ref{tab:rq4_outcomes} keeps all 175 contributing sources as its base. All counts describe reporting, not effect size.

\begin{figure}
    \centering
    \includegraphics[width=\columnwidth]{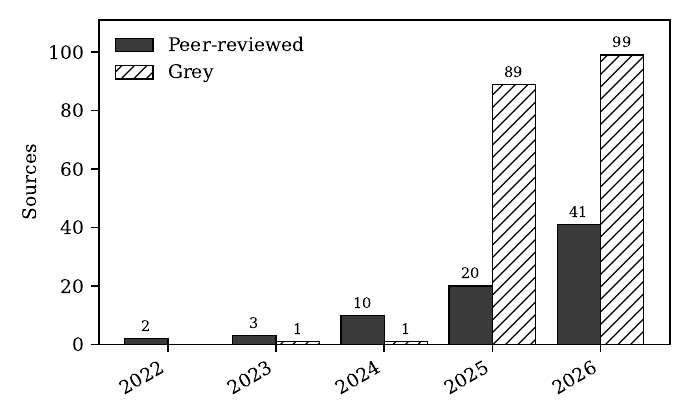}
    \caption{Distribution of the 266 sources by publication year, per stream, generated from the record.}
    \label{fig:Publication_Source_Trend}
\end{figure}

\subsection{Study Context}\label{sec:study_context}
The corpus holds 266 sources, 76 peer-reviewed and 190 grey. Publication window 1 contributes 113 sources (34 peer-reviewed, 79 grey), and publication window 2 contributes 153 (42 peer-reviewed, 111 grey). Tables~\ref{tab:corpus_profile} and~\ref{tab:corpus_method} give the corpus profile and the research methods by stream and publication window. Table~\ref{tab:rq_coverage} gives the sources contributing to each research question.

Two thirds of the peer-reviewed sources are conference papers (Table~\ref{tab:corpus_profile}). Most grey sources are web articles or guides (89) and preprints (78). The web articles and guides are spread over both windows (42 and 47), and most preprints (54 of 78) come from window 2.

Empirical sources, that is sources with reported data collection, are the largest evidence category (94 sources). Experience reports, first-hand accounts of practice without a study design, follow with 56 sources, all of them grey. The peer-reviewed stream is mostly empirical or tool-oriented, and the grey stream holds most of the position pieces. Two grey sources carry no evidence category in the record and are counted as not classified in Tables~\ref{tab:corpus_profile} and~\ref{tab:rq_coverage}.

Among the peer-reviewed sources, case studies and mixed-methods designs are the most frequent research methods, followed by interviews and surveys (Table~\ref{tab:corpus_method}). Among the grey sources, experience reports or practitioner guides and opinion pieces are frequent. Repository-mining studies appear only among the sources of window 2, in both streams. Most benchmarks (13 of 18) also come from window 2.

Table~\ref{tab:corpus_profile} gives directness by stream and publication window under the rule of Section~\ref{sec:opdef}. Most background sources (16 of 27) are window-1 peer-reviewed papers (Section~\ref{sec:update2026}).

\begin{table*}[!t]
\centering
\footnotesize
\renewcommand{\arraystretch}{1.1}
\setlength{\tabcolsep}{4pt}
\caption{RQ1: Conceptualizations of vibe coding. Percentages are relative to the 239 sources contributing to RQ1 (65 peer-reviewed, 174 grey). Column definitions are given at the start of Section~\ref{sec:results}. Restricted sources are counted. Background sources count as reports only.}
\label{tab:rq1_conceptualizations}
\resizebox{\textwidth}{!}{%
\begin{tabular}{|>{\raggedright\arraybackslash}p{4.64cm}|p{1.06cm}|p{0.87cm}|p{1.50cm}|p{0.77cm}|p{1.69cm}|p{1.50cm}|p{0.77cm}|p{1.79cm}|}
\hline
\rowcolor[HTML]{EFEFEF} & \multicolumn{2}{c|}{\textbf{Sources reporting}} & & & & \multicolumn{2}{c|}{\textbf{Empirical sources}} & \\
\cline{2-3}\cline{7-8}
\rowcolor[HTML]{EFEFEF}\multirow{-2}{=}{\textbf{Finding}} & \textbf{Count*} & \textbf{\%} & \multirow{-2}{=}{\textbf{Peer-reviewed}} & \multirow{-2}{=}{\textbf{Grey}} & \multirow{-2}{=}{\textbf{Background}} & \textbf{Peer-reviewed} & \textbf{Grey} & \multirow{-2}{=}{\textbf{Reporting support}} \\
\hline
Natural-language-to-code or prompt-based development & 164 & 69\% & 52 & 112 & 5 & 38 & 45 & Strong \\
\hline
Socio-technical practice & 139 & 58\% & 26 & 113 & 2 & 15 & 39 & Moderate \\
\hline
Agentic or semi-autonomous development & 80 & 33\% & 20 & 60 & 2 & 13 & 26 & Strong \\
\hline
AI-assisted development & 58 & 24\% & 20 & 38 & 5 & 10 & 13 & Strong \\
\hline
End-user or non-programmer development & 33 & 14\% & 8 & 25 & 0 & 7 & 8 & Moderate \\
\hline
AI-native IDE & 11 & 5\% & 3 & 8 & 1 & 3 & 3 & Strong \\
\hline
\end{tabular}
}
\par\smallskip\noindent\footnotesize{*A source can count under more than one finding.}
\end{table*}

Under the rule of Section~\ref{sec:grey_selection}, 73 grey sources (37 of window 1 and 36 of window 2) are restricted to definitions, practitioner discourse, and tool visibility. They do not support outcome or safeguard findings. The quality bands and credibility tiers are in Table~\ref{tab:corpus_profile}.

Figure~\ref{fig:Publication_Source_Trend} gives the distribution by year and stream. Only 17 sources, 15 peer-reviewed and 2 grey, are dated before 2025, and 140 sources are dated 2026.

Coverage is uneven (Table~\ref{tab:rq_coverage}). RQ1 (definitions) and RQ7 (tools and platforms) draw on 90\% and 86\% of the corpus, and RQ6 (contexts and users) and RQ3.2 (session momentum) on 41\% and 36\%. Grey sources outnumber peer-reviewed sources for every question, and empirical sources are the largest evidence category for every question. Experience reports and position pieces contribute mainly to RQ1 and RQ7, the two questions open to restricted sources.

These distributions suggest a recent topic whose evidence base is still uneven. Practitioner discussion seems to run ahead of peer-reviewed research. A large count may rest mainly on experience reports or restricted sources, so each finding should be read with its label.

\subsection{Vibe Coding Definition (RQ1)} \label{sec:rq1}

Sheet S2 of the supplementary workbook gives representative definitions of vibe coding in chronological order, as examples. They were not classified. The coded conceptualizations are in Table~\ref{tab:rq1_conceptualizations}.

RQ1 draws on 239 sources. It is one of the two questions open to restricted sources, and 65 of the 73 restricted sources contribute to it (Section~\ref{sec:grey_selection}). Table~\ref{tab:rq1_conceptualizations} gives six conceptualizations.

The most frequent framing is natural-language-to-code or prompt-based development, reported in 69\% of the contributing sources, with Strong reporting support. One study describes vibe coding as ``a variant of LLM-augmented programming that relies exclusively on iteratively writing prompts and evaluating program behavior''~\ref{WL36}. Many sources describe vibe coding as a socio-technical practice or culture, guided by feel, flow, and intuition, with minimal explicit planning (Moderate reporting support). Fewer sources frame vibe coding as agentic development under human supervision, as AI assistance in the copilot sense, as development by people without programming training, or through AI-native integrated development environments (IDEs).

The two streams differ in focus. Peer-reviewed sources lean toward the natural-language-to-code framing, while grey sources carry it and the socio-technical framing about equally (Table~\ref{tab:rq1_conceptualizations}). Grey sources also carry most of the agentic and end-user framings. This may suggest that the streams share the core of the term but differ on whether vibe coding is a technique for producing code or a way of working.

The RQ1 findings rest on how the sources define the practice, not on tests of it. Every framing has empirical sources of comparable standing in both streams, but the AI-native IDE framing has only six. The socio-technical and end-user framings have Moderate reporting support because 34\% and 36\% of their supporting sources are Tier~3 sources or belong to the restricted categories, above the 25\% limit for Strong. Without the restricted categories both would be Strong, and without the Tier~3 sources only the socio-technical framing would be (sheet S4 of the supplementary workbook). Their labels reflect the standing of the sources, not disagreement between them.

The degree of autonomy given to the model is not settled. A qualitative study reports that some developers equate vibe coding with fully agentic coding, whereas ``others frame it as software engineering supported by agents that still involves manual code editing and inspection''~\ref{WL34}. The agentic and AI-assisted framings also sit close to broader AI-assisted programming. When only direct sources are counted, the agentic, AI-assisted, and natural-language-to-code framings fall from Strong to Moderate, and the AI-native IDE framing falls to Weak (Section~\ref{sec:sensitivity}). These four labels also lie close to the restricted-share limit, and the AI-native IDE label rests on three comparable peer-reviewed studies, the smallest number that allows Strong (sheet S7).

\begin{table*}[!t]
\centering
\footnotesize
\renewcommand{\arraystretch}{1.1}
\setlength{\tabcolsep}{4pt}
\caption{RQ2: Workflow patterns reported for vibe coding. Percentages are relative to the 172 sources contributing to RQ2 (70 peer-reviewed, 102 grey). Column definitions are given at the start of Section~\ref{sec:results}. Restricted sources are not counted. Background sources count as reports only.}
\label{tab:rq2_workflows}
\resizebox{\textwidth}{!}{%
\begin{tabular}{|>{\raggedright\arraybackslash}p{4.64cm}|p{1.06cm}|p{0.87cm}|p{1.50cm}|p{0.77cm}|p{1.69cm}|p{1.50cm}|p{0.77cm}|p{1.79cm}|}
\hline
\rowcolor[HTML]{EFEFEF} & \multicolumn{2}{c|}{\textbf{Sources reporting}} & & & & \multicolumn{2}{c|}{\textbf{Empirical sources}} & \\
\cline{2-3}\cline{7-8}
\rowcolor[HTML]{EFEFEF}\multirow{-2}{=}{\textbf{Finding}} & \textbf{Count*} & \textbf{\%} & \multirow{-2}{=}{\textbf{Peer-reviewed}} & \multirow{-2}{=}{\textbf{Grey}} & \multirow{-2}{=}{\textbf{Background}} & \textbf{Peer-reviewed} & \textbf{Grey} & \multirow{-2}{=}{\textbf{Reporting support}} \\
\hline
\multicolumn{9}{|l|}{\textit{Interaction and refinement}} \\
\hline
Iterate-prompt-patch loop & 98 & 57\% & 34 & 64 & 2 & 22 & 33 & Strong \\
\hline
Chat-based iterative development loop & 59 & 34\% & 26 & 33 & 9 & 11 & 14 & Strong \\
\hline
Inline IDE-assisted completion & 7 & 4\% & 5 & 2 & 4 & 1 & 0 & Weak \\
\hline
\multicolumn{9}{|l|}{\textit{Validation and control}} \\
\hline
Validation and evaluation pipelines as a core loop & 82 & 48\% & 30 & 52 & 13 & 12 & 31 & Strong \\
\hline
\multicolumn{9}{|l|}{\textit{Context and grounding}} \\
\hline
Explicit context provisioning & 81 & 47\% & 25 & 56 & 6 & 12 & 31 & Strong \\
\hline
Retrieval and documentation grounding & 9 & 5\% & 5 & 4 & 1 & 3 & 1 & Strong \\
\hline
\multicolumn{9}{|l|}{\textit{Repair and collaboration}} \\
\hline
Patch and repair workflow & 30 & 17\% & 9 & 21 & 4 & 3 & 11 & Strong \\
\hline
Collaborative prompt practices & 7 & 4\% & 3 & 4 & 0 & 3 & 1 & Moderate \\
\hline
\end{tabular}
}
\par\smallskip\noindent\footnotesize{*A source can count under more than one finding.}
\end{table*}

\subsubsection{Synthesized working definition.} The sources describe vibe coding as more than code generation from prompts: a socio-technical practice in which natural language mediates developer intent, model generation, and human validation. From the definitions in sheet S2 of the supplementary workbook we synthesize the following working definition:

\textit{Vibe coding} is an intent-driven software development practice in which a developer specifies desired functionality, constraints, and context primarily through natural-language prompts, and one or more large language models generate, refine, and iteratively revise code artifacts in response. The developer's role shifts from direct code authoring toward expressing intent, evaluating generated outputs, providing corrective feedback, and steering successive iterations. In its stricter form, as originally framed by Karpathy, line-by-line review of the generated code is reduced or absent, and the developer accepts or rejects the output mainly from its observed behavior. In practice, the degree of developer oversight varies along a spectrum from minimal review (``pure'' vibe coding) to structured validation incorporating testing, linting, and human-in-the-loop review. The practice is characterized by rapid prototyping cycles, conversational interaction with AI systems, and a probabilistic generation process.

\begin{tcolorbox}[
    colback=gray!10,
    colframe=gray!60,
    title=\textbf{Takeaway 1},
    fonttitle=\bfseries,
    coltitle=black,
    arc=3mm,
    boxrule=0.8pt,
    width=\columnwidth,
]

\onehalfspacing
The sources describe vibe coding mainly as natural-language-to-code development and as a socio-technical practice guided by feel, flow, and intuition with little explicit planning. The autonomy given to the model varies across sources. The boundary with broader AI-assisted programming remains open, since several framings weaken when only direct sources are counted.
\end{tcolorbox}

\noindent This definition covers all six categories of Table~\ref{tab:rq1_conceptualizations}, so ``developer'' includes people without programming training. The spectrum of oversight covers the original Karpathy framing and the peer-reviewed definitions that keep human supervision and validation in the loop (for example \ref{WL42} and \ref{WL50}). In the terms of Section~\ref{sec:opdef}, every point of the spectrum meets the minimum definition used to classify sources, and only its minimal-review end meets the stricter form.

\begin{table*}[!t]
\centering
\footnotesize
\renewcommand{\arraystretch}{1.1}
\setlength{\tabcolsep}{4pt}
\caption{RQ3.1: Reported shifts in developer role and cognition. Percentages are relative to the 155 sources contributing to RQ3.1 (66 peer-reviewed, 89 grey). Column definitions are given at the start of Section~\ref{sec:results}. Restricted sources are not counted. Background sources count as reports only.}
\label{tab:rq3_role_shifts}
\resizebox{\textwidth}{!}{%
\begin{tabular}{|>{\raggedright\arraybackslash}p{4.64cm}|p{1.06cm}|p{0.87cm}|p{1.50cm}|p{0.77cm}|p{1.69cm}|p{1.50cm}|p{0.77cm}|p{1.79cm}|}
\hline
\rowcolor[HTML]{EFEFEF} & \multicolumn{2}{c|}{\textbf{Sources reporting}} & & & & \multicolumn{2}{c|}{\textbf{Empirical sources}} & \\
\cline{2-3}\cline{7-8}
\rowcolor[HTML]{EFEFEF}\multirow{-2}{=}{\textbf{Finding}} & \textbf{Count*} & \textbf{\%} & \multirow{-2}{=}{\textbf{Peer-reviewed}} & \multirow{-2}{=}{\textbf{Grey}} & \multirow{-2}{=}{\textbf{Background}} & \textbf{Peer-reviewed} & \textbf{Grey} & \multirow{-2}{=}{\textbf{Reporting support}} \\
\hline
Trust calibration and adaptive reliance & 99 & 64\% & 41 & 58 & 13 & 22 & 29 & Strong \\
\hline
Changes in cognitive load & 78 & 50\% & 31 & 47 & 7 & 19 & 22 & Strong \\
\hline
Shift from direct authoring toward supervision and steering roles & 58 & 37\% & 18 & 40 & 2 & 10 & 22 & Strong \\
\hline
Cognitive debt due to deferred understanding & 37 & 24\% & 15 & 22 & 5 & 9 & 6 & Strong \\
\hline
Flow effects & 25 & 16\% & 16 & 9 & 3 & 8 & 4 & Strong \\
\hline
\end{tabular}
}
\par\smallskip\noindent\footnotesize{*A source can count under more than one finding.}
\end{table*}

\subsection{Workflows and Tactics (RQ2)} \label{sec:rq2}

RQ2 draws on 172 sources. Table~\ref{tab:rq2_workflows} groups the eight workflow patterns in four families: interaction and refinement, validation and control, context and grounding, and repair and collaboration. Named tools are reported under RQ7 (Section~\ref{sec:rq7}).

The sources describe vibe coding as a short loop in which the developer prompts, runs the result, inspects it, and prompts again. The most frequent pattern, the iterate-prompt-patch loop of the interaction and refinement family, is reported in 57\% of the sources, 2 of them background (Strong reporting support). In it, errors or logs are fed back into successive prompts. Validation pipelines (82 sources, 13 of them background) and explicit context provisioning (81 sources, 6 of them background) follow, and both have Strong reporting support. These two patterns concern what developers give the model before generation and what they check after it. A repository mining study describes a case in which the continuous integration (CI) check was the main feedback: ``The vibe coder has to commit repeatedly to tune timeout parameters, effectively debugging via CI''~\ref{WL42}.

Three of these patterns are related but distinct. \emph{Validation and evaluation pipelines} are verification applied to generated code, such as automated tests, CI checks, linting, and runtime inspection. The \emph{iterate-prompt-patch loop} is the conversational debugging cycle across turns. The \emph{patch and repair workflow} is localized editing of an existing fragment, without full regeneration or a verification step.

Peer-reviewed and grey sources share the leading patterns but weight them differently (Table~\ref{tab:rq2_workflows}). A larger share of peer-reviewed sources report chat-based loops (37\% against 32\% of grey sources), inline completion, and retrieval grounding. A larger share of grey sources report the iterate-prompt-patch loop (63\% against 49\%), patch and repair, and context provisioning (55\% against 36\%). An interview study reports that technical users ``prefer tools like Cursor, which let them control which files the AI focuses on''~\ref{WL40}.

Most workflow items are described in use. The reporting mode is recorded for each source and pattern (Section~\ref{sec:extraction}). Of the 334 workflow items of the 157 contributing sources that are not background, 246 are described in use, 36 are recommended, 44 are evaluated empirically, and 8 carry no reported mode. Twenty-six sources evaluate at least one pattern, and 31 of the 44 evaluated items come from grey sources. The iterate-prompt-patch loop is evaluated most often (12 of its 96 items), and validation pipelines next (10 of 69).

The smaller patterns are less secure. Inline IDE-assisted completion is reported in seven sources, but four of them are background sources. It rests on three independent sources and has Weak reporting support. Collaborative prompt practices (seven sources, none background) have Moderate reporting support. They rest on seven independent sources, one fewer than Strong requires. Retrieval and documentation grounding (nine sources, one background) has Strong reporting support, but it rests on eight independent sources and three comparable peer-reviewed studies, the smallest numbers that allow Strong. A one-step move of either threshold would make it Moderate (sheet S7). Patch and repair also has Strong reporting support with three comparable peer-reviewed studies, at the threshold for Strong. When only direct sources are counted, retrieval grounding falls to Weak, inline completion to Emerging, and patch and repair to Moderate (Section~\ref{sec:sensitivity}). Only one of the seven inline-completion sources and four of the nine retrieval sources are direct. These patterns may therefore belong more to AI-assisted programming in general than to vibe coding in particular.

The sources describe vibe coding with more checking than its original framing implies. Of the 157 contributing sources that are not background, 69 report validation pipelines and 75 report context provisioning. The evidence does not show which workflow leads to better outcomes, which are reported under RQ4 (Section~\ref{sec:rq4}).

\begin{tcolorbox}[
    colback=gray!10,
    colframe=gray!60,
    title=\textbf{Takeaway 2},
    fonttitle=\bfseries,
    coltitle=black,
    arc=3mm,
    boxrule=0.8pt,
    width=\columnwidth,
]

\onehalfspacing
Vibe coding is described as an iterative loop of prompting, running, inspecting, and repairing, steered by the context given to the model and the checks applied after generation. Most workflow items are described in use and 44 of the 334 items are evaluated empirically, so the labels show how widely a pattern is reported and not its relative effectiveness.
\end{tcolorbox}

\begin{table*}[!t]
\centering
\footnotesize
\renewcommand{\arraystretch}{1.1}
\setlength{\tabcolsep}{4pt}
\caption{RQ3.2: Session dynamics and context momentum. 96 sources contribute to RQ3.2 (39 peer-reviewed, 57 grey), and counts are given without percentages. Column definitions are given at the start of Section~\ref{sec:results}. Restricted sources are not counted. Background sources count as reports only.}
\label{tab:rq3_session_dynamics}
\begin{tabular}{|>{\raggedright\arraybackslash}p{4.39cm}|p{2.32cm}|p{1.53cm}|p{0.79cm}|p{1.73cm}|p{1.53cm}|p{0.79cm}|p{1.83cm}|}
\hline
\rowcolor[HTML]{EFEFEF} & \textbf{Sources reporting} & & & & \multicolumn{2}{c|}{\textbf{Empirical sources}} & \\
\cline{2-2}\cline{6-7}
\rowcolor[HTML]{EFEFEF}\multirow{-2}{=}{\textbf{Finding}} & \textbf{Count*} & \multirow{-2}{=}{\textbf{Peer-reviewed}} & \multirow{-2}{=}{\textbf{Grey}} & \multirow{-2}{=}{\textbf{Background}} & \textbf{Peer-reviewed} & \textbf{Grey} & \multirow{-2}{=}{\textbf{Reporting support}} \\
\hline
Repetitive prompt-patch cycles without convergence & 49 & 19 & 30 & 1 & 14 & 14 & Strong \\
\hline
Context window or state-retention failures & 46 & 19 & 27 & 4 & 12 & 13 & Strong \\
\hline
Session momentum breaks due to failure, misalignment, or tool friction & 29 & 12 & 17 & 3 & 8 & 7 & Strong \\
\hline
\end{tabular}
\par\smallskip\noindent\footnotesize{*A source can count under more than one finding.}
\end{table*}

\subsection{Developer Role Shift (RQ3)} \label{sec:rq3}

This RQ reports one developer-role shift seen from three angles: the cognitive and behavioral constructs that accompany it (RQ3.1), the within-session interaction dynamics (RQ3.2), and the differences by audience and task (RQ3.3).

\subsubsection{Trust, Cognitive Load, and Cognitive Debt}

RQ3.1 draws on 155 sources (Table~\ref{tab:rq3_role_shifts}). The sources describe a move from writing code toward steering, curating, and reviewing generated output. They report it together with changes in trust, cognitive load, flow, and understanding. All five constructs have Strong reporting support. Of the 155 sources, 72 are in the empirical evidence category. Trust calibration and adaptive reliance is the most frequent (64\%, 13 of the 99 sources background). Its code records when developers rely on generated output and how they verify it. A workshop study records participants for whom ``The AI had all the power; we were increasingly `directed' to adapt to the tool''~\ref{WL49}.

Both streams report all five constructs, with empirical sources of comparable standing in each (Table~\ref{tab:rq3_role_shifts}). They report trust and cognitive load in similar shares. Flow effects are reported in a larger share of peer-reviewed sources, and the explicit role shift in a larger share of grey sources.

Individual sources report both relief and strain. A qualitative study quotes one practitioner for whom letting the tool deal with syntax ``frees up brain space for the more important work'', and another who is ``more mentally exhausted at the end of the day these days'' because of constant code review~\ref{GL34}. One source describes flow as smoother iteration on routine or exploratory tasks~\ref{GL1}. Since the codebook records no direction for these constructs, the review cannot weigh relief against strain.

Cognitive debt, the build-up of deferred understanding when developers accept code they have not fully understood, is reported in a quarter of the sources (Strong reporting support). \ref{WL15} discusses how rapid iteration may postpone reasoning about design decisions, and \ref{GL10} warns that early success can create an illusion of correctness. The labels of all five constructs hold when only direct sources are counted (Section~\ref{sec:sensitivity}). The role shift is widely reported, but the coded evidence does not record how cognitive debt develops over time.

\subsubsection{Session Dynamics and Context Momentum}

RQ3.2 draws on 96 sources (Table~\ref{tab:rq3_session_dynamics}). The sources describe sessions that lose direction. Developers repeat prompt and patch turns without convergence, the model loses earlier context, and momentum breaks after failures, misalignment, or tool friction. All three phenomena have Strong reporting support. An interview study quotes a practitioner: ``It forgets what the outcome was supposed to be and my earlier instructions, so I have to keep pulling it back on track''~\ref{WL50}.

The streams converge on all three phenomena, with empirical sources of comparable standing in each. Of the 96 sources, 54 are empirical studies. Context window failures are reported in a slightly larger share of peer-reviewed sources, and looping cycles in a slightly larger share of grey sources (Table~\ref{tab:rq3_session_dynamics}). One study reports that developers fork the conversation once the context grows too long~\ref{WL34}. \ref{GL7} describes context momentum, in which ``early prompts and the outcomes they yield can steer the project onto a particular trajectory that may become challenging to diverge from later in the session''.

The codes overlap in part. Of the 96 sources, 26 carry two or more of the three codes, 17 also carry flow effects (RQ3.1), and 25 also carry cumulative context errors (RQ5.1, Section~\ref{sec:rq5_1}). The evidence records that the phenomena occur, not how often within a session or under which conditions. Their labels hold when only direct sources are counted. Whether this depends on the tools of the period was not coded.

\subsubsection{Differences by Audience and Task}

RQ3.3 draws on 144 sources. Table~\ref{tab:rq33_audience_task} separates audience orientation (Panel~A) from task or application emphasis (Panel~B). The audience panel rests on many sources, while the task panel is thin.

Vibe coding is described for two audiences. Novice or non-programmer orientation is reported in 78\% of the sources and expert acceleration in 53\%. Both have Strong reporting support, with empirical sources of comparable standing in both streams, and both are divided similarly between the streams. One novice-oriented study describes vibe coding as a way into software creation for founders and other people without programming training~\ref{WL41}. One expert-oriented source describes faster routine implementation while the developer keeps responsibility for design and verification~\ref{GL1}.

\begin{figure*}[!htbp]
\centering
\subfloat[Audience emphasis against reported outcomes.\label{fig:heatmap_audience}]{\includegraphics[width=0.48\textwidth]{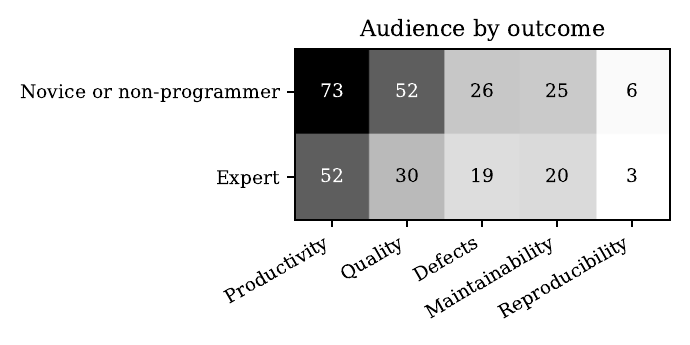}}
\hfill
\subfloat[Task domain against reported outcomes. ML: machine learning.\label{fig:heatmap_task}]{\includegraphics[width=0.48\textwidth]{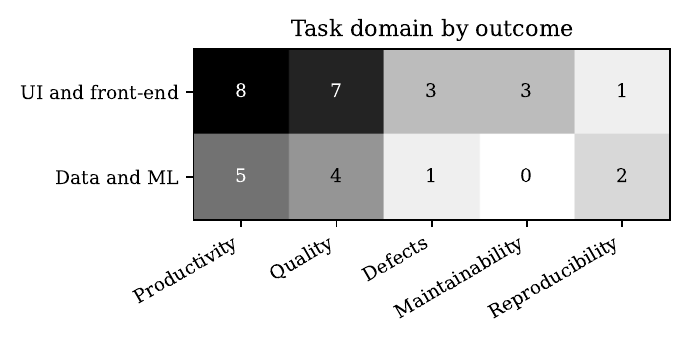}}
\caption{Cross-tabulation of the RQ3.3 codes against the outcome codes of RQ4, over the 159 RQ4 sources that are not background. A cell is the number of sources that carry both codes. The counts describe where outcomes are reported for each audience and task, not how large the outcomes are.}
\label{fig:rq33_rq4_heatmaps}
\end{figure*}

The codes record the audience a source addresses, not a comparison of groups, so the review reports no finding on differences between novices and experts. Crossed with the outcome codes, the novice and expert sources report the same outcomes in a similar order (Figure~\ref{fig:rq33_rq4_heatmaps} and Section~\ref{sec:rq4}).

\begin{table*}[!t]
\centering
\footnotesize
\renewcommand{\arraystretch}{1.1}
\setlength{\tabcolsep}{4pt}
\caption{RQ3.3: Audience and task emphasis. Percentages are relative to the 144 sources contributing to RQ3.3 (60 peer-reviewed, 84 grey). Column definitions are given at the start of Section~\ref{sec:results}. Restricted sources are not counted. Background sources count as reports only.}
\label{tab:rq33_audience_task}
\resizebox{\textwidth}{!}{%
\begin{tabular}{|>{\raggedright\arraybackslash}p{4.64cm}|p{1.06cm}|p{0.87cm}|p{1.50cm}|p{0.77cm}|p{1.69cm}|p{1.50cm}|p{0.77cm}|p{1.79cm}|}
\hline
\rowcolor[HTML]{EFEFEF} & \multicolumn{2}{c|}{\textbf{Sources reporting}} & & & & \multicolumn{2}{c|}{\textbf{Empirical sources}} & \\
\cline{2-3}\cline{7-8}
\rowcolor[HTML]{EFEFEF}\multirow{-2}{=}{\textbf{Finding}} & \textbf{Count*} & \textbf{\%} & \multirow{-2}{=}{\textbf{Peer-reviewed}} & \multirow{-2}{=}{\textbf{Grey}} & \multirow{-2}{=}{\textbf{Background}} & \textbf{Peer-reviewed} & \textbf{Grey} & \multirow{-2}{=}{\textbf{Reporting support}} \\
\hline
\multicolumn{9}{|l|}{\textit{Panel A. Audience orientation}} \\
\hline
Novice or non-programmer orientation & 112 & 78\% & 49 & 63 & 10 & 29 & 28 & Strong \\
\hline
Expert acceleration & 76 & 53\% & 31 & 45 & 12 & 14 & 15 & Strong \\
\hline
\multicolumn{9}{|l|}{\textit{Panel B. Task or application emphasis}} \\
\hline
UI and front-end tasks & 17 & 12\% & 8 & 9 & 5 & 2 & 5 & Moderate \\
\hline
Data and machine learning tasks & 6 & 4\% & 3 & 3 & 0 & 1 & 2 & Moderate \\
\hline
\end{tabular}
}
\par\smallskip\noindent\footnotesize{*A source can count under more than one finding.}
\end{table*}

The task panel rests on little direct empirical evidence. UI and front-end tasks and data and machine learning tasks both have Moderate reporting support (Table~\ref{tab:rq33_audience_task}). Five of the UI sources are background sources, and the other 12 include two comparable peer-reviewed studies, one fewer than Strong requires. The data label rests on five independent sources, the threshold for Moderate, with one comparable peer-reviewed study and two comparable grey sources. When only direct sources are counted, the data label falls to Weak and the UI label stays Moderate (Section~\ref{sec:sensitivity}).

The task evidence is too thin to say where vibe coding works better (Section~\ref{sec:discussion_contexts_tools}).

\begin{tcolorbox}[
    colback=gray!10,
    colframe=gray!60,
    title=\textbf{Takeaway 3},
    fonttitle=\bfseries,
    coltitle=black,
    arc=3mm,
    boxrule=0.8pt,
    width=\columnwidth,
]

\onehalfspacing
The developer role shifts from writing code toward supervising, validating, and steering generated output, with reported changes in trust, cognitive load, flow, and understanding. Within sessions, the sources report loops that do not converge, loss of earlier context, and broken momentum. The practice is described for novices and non-programmers and for experienced developers seeking acceleration, while the evidence by task type is thin.
\end{tcolorbox}

\subsection{Productivity, Quality, and Maintainability (RQ4)} \label{sec:rq4}

RQ4 draws on 175 sources. Of these, 16 are background sources, which count as reports in Table~\ref{tab:rq4_outcomes} but not toward the labels, so 159 sources count toward RQ4. The table pairs the two horizons for each outcome and gives directness and measurement in place of stream. Direction, measurement, and horizon are recorded for each outcome a source reports, so each row uses only what a source says about that outcome. Direction is positive, negative, mixed, no effect, or not stated, and Section~\ref{sec:extraction} defines these values. A source that reports both horizons for an outcome appears in both rows, and a source with no recorded horizon appears in neither.

\begin{table*}[!t]
\centering
\footnotesize
\renewcommand{\arraystretch}{1.1}
\setlength{\tabcolsep}{3pt}
\caption{RQ4: Reported outcomes of vibe coding, split by horizon (175 contributing sources: 66 peer-reviewed, 109 grey; percentages over all 175). Direction counts the comparable empirical sources with a recorded direction for that outcome. Counts describe reporting, not effect size.}
\label{tab:rq4_outcomes}
\begin{tabular}{|>{\raggedright\arraybackslash}p{2.13cm}|p{1.34cm}|p{0.67cm}|p{0.67cm}|p{0.67cm}|p{0.67cm}|p{0.67cm}|p{0.67cm}|p{0.67cm}|p{0.67cm}|p{0.67cm}|p{0.67cm}|p{0.67cm}|p{0.67cm}|p{1.34cm}|}
\hline
\rowcolor[HTML]{EFEFEF} & & \multicolumn{2}{>{\columncolor[HTML]{EFEFEF}\centering\arraybackslash}p{1.55cm}|}{\textbf{Sources reporting}} & \multicolumn{3}{c|}{\textbf{Directness}} & \multicolumn{4}{c|}{\textbf{Direction}} & \multicolumn{3}{c|}{\textbf{Measurement}} & \\
\cline{3-4}\cline{5-7}\cline{8-11}\cline{12-14}
\rowcolor[HTML]{EFEFEF}\multirow{-2}{=}{\textbf{Outcome}} & \multirow{-2}{=}{\textbf{Horizon}} & \textbf{n*} & \textbf{\%} & \textbf{Dir.} & \textbf{Adj.} & \textbf{Bg.} & \textbf{Pos.} & \textbf{Neg.} & \textbf{Mix.} & \textbf{N.e.} & \textbf{Meas.} & \textbf{Self} & \textbf{Ass.} & \multirow{-2}{=}{\textbf{Evidence strength}} \\
\hline
Productivity & short-term & 117 & 67\% & 69 & 36 & 12 & 34 & 3 & 14 & 4 & 45 & 46 & 26 & Weak \\
\cline{2-15}
 & long-term & 9 & 5\% & 5 & 3 & 1 & 0 & 1 & 5 & 0 & 5 & 4 & 0 & Moderate \\
\hline
Code or UI quality & short-term & 93 & 53\% & 42 & 46 & 5 & 13 & 15 & 33 & 0 & 55 & 28 & 10 & Weak \\
\cline{2-15}
 & long-term & 9 & 5\% & 4 & 4 & 1 & 0 & 1 & 0 & 2 & 4 & 2 & 3 & Moderate \\
\hline
Defect risks and error patterns & short-term & 47 & 27\% & 15 & 28 & 4 & 2 & 15 & 9 & 0 & 25 & 10 & 12 & Weak \\
\cline{2-15}
 & long-term & 5 & 3\% & 4 & 1 & 0 & 0 & 0 & 0 & 0 & 1 & 3 & 1 & Weak \\
\hline
Maintainability implications & short-term & 3 & 2\% & 2 & 1 & 0 & 0 & 1 & 0 & 0 & 2 & 1 & 0 & Weak \\
\cline{2-15}
 & long-term & 39 & 22\% & 21 & 14 & 4 & 2 & 15 & 1 & 1 & 11 & 11 & 17 & Weak \\
\hline
Reproducibility and consistency & short-term & 7 & 4\% & 5 & 2 & 0 & 0 & 3 & 0 & 0 & 3 & 0 & 4 & Moderate \\
\hline
\end{tabular}
\par\smallskip\noindent\footnotesize{n: sources reporting the outcome (*a source can count under more than one outcome). Dir.: direct; Adj.: adjacent; Bg.: background; Pos.: positive; Neg.: negative; Mix.: mixed; N.e.: no effect; Meas.: measured; Self: self-reported; Ass.: asserted. A source that reports both horizons appears in both rows, and a source with no recorded horizon appears in neither. Comparable sources with no recorded direction are not counted under Direction. Background sources count as reports but not toward the label.}
\end{table*}

Productivity is the most reported outcome. The shares in this section are over the 159 sources that count toward RQ4, whereas Table~\ref{tab:rq4_outcomes} counts all 175 RQ4 sources, background included, and splits each outcome by horizon. Productivity appears in 109 of the 159 sources (69\%), and 22 of them only assert it, without data or reported perception. In the table its short-term row shows 117. Code or UI quality follows in 97 sources (61\%). Maintainability appears in 40 sources (25\%), 19 of them by assertion only. Most reports concern the short term. Productivity, for example, is reported for the short term in 105 of these sources and for the long term in 8. Maintainability is the exception. It is recorded as long-term unless the maintenance work falls within the immediate task, and 35 of its 40 sources carry the long-term horizon. The two streams measure at a similar rate. Of the 54 peer-reviewed sources, 30 (56\%) record at least one measured outcome, and of the 105 grey sources, 54 (51\%) do.

Long-term reports rest less often on measurement. Of the 267 short-term reports, 130 (49\%) are measured and 52 (19\%) asserted. Of its 62 long-term reports, 21 (34\%) are measured and 21 (34\%) asserted, 17 of them on maintainability. The values are the reconciled ones after the second author's full check (Section~\ref{sec:reliability}), and these shares may still be read as approximate.

Two sources show how the per-outcome record separates outcomes. A case study measures a large time saving on the immediate task, a positive short-term productivity outcome, and adds that the saving ``does not establish quality equivalence between the two outputs''~\ref{WL62}. A repository study reports ``static-analysis warnings and cognitive complexity rising by roughly 18\% and 39\%, indicating sustained agent-induced technical debt even when velocity advantages fade''~\ref{WL33}. It is recorded as negative on long-term quality and maintainability and as mixed on productivity. In all, 28 of the 159 sources that count toward RQ4 record a positive direction on one outcome and a negative direction on another. The record does not code whether one outcome affects the other.

Among the sources of comparable standing, the predominant direction differs by outcome (Table~\ref{tab:rq4_outcomes}). For short-term productivity, 34 of the 55 comparable directions are positive, 14 mixed, 4 no effect, and 3 negative. For long-term maintainability, 15 of 19 are negative. For short-term defects, 15 of 26 are negative and 9 mixed. Short-term quality is mostly mixed, with 33 mixed, 15 negative, and 13 positive directions. The long-term rows for productivity and quality are small, with six and three comparable directions. Seven comparable directions are no effect, four on short-term productivity, two on long-term quality, and one on long-term maintainability.

The consistency criterion follows a written rule (Section~\ref{sec:evidence_strength}). A row is contested as soon as one comparable source reports a positive direction and another a negative one, whatever the balance between them. Four rows are contested and labeled Weak: short-term productivity, short-term quality, short-term defects, and long-term maintainability. Short-term productivity is contested by three negative directions (\ref{WL36}, \ref{WL42}, \ref{GL59}) against 34 positive ones. Long-term maintainability is contested by two positive directions (\ref{GL29}, \ref{GL179}) against 15 negative ones. For these two rows, the Weak label reflects that opposing results exist, not an even split. Two rows show a minor disagreement, with directions of more than one kind but not both positive and negative, and are labeled Moderate. These are long-term productivity (five mixed, one negative) and long-term quality (one negative, two no effect). Short-term reproducibility is consistent, with three negative directions. It is labeled Moderate because it rests on seven independent sources and one comparable peer-reviewed study, below the thresholds for Strong. These three Moderate rows rest on six or fewer comparable directions, so their agreement may partly reflect their size. Two rows are Weak because consistency is not assessable. Short-term maintainability has one comparable direction, and no comparable source records a direction for long-term defects. Reproducibility is reported for the short term only. Section~\ref{sec:sensitivity} reports a less strict variant of the rule for contested rows. These labels state how the comparable sources stand under the rule of Section~\ref{sec:evidence_strength}. A Weak label means that the reported directions conflict or that too few comparable sources record a direction. It does not mean that vibe coding is ineffective, and a Moderate label does not mean that it is effective.

\begin{table*}[!t]
\centering
\footnotesize
\renewcommand{\arraystretch}{1.1}
\setlength{\tabcolsep}{4pt}
\caption{RQ5.1: Failure modes. Percentages are relative to the 151 sources contributing to RQ5.1 (55 peer-reviewed, 96 grey). Column definitions are given at the start of Section~\ref{sec:results}. Restricted sources are not counted. Background sources count as reports only.}
\label{tab:rq5_failures}
\begin{tabular}{|>{\raggedright\arraybackslash}p{4.64cm}|p{1.06cm}|p{0.87cm}|p{1.50cm}|p{0.77cm}|p{1.69cm}|p{1.50cm}|p{0.77cm}|p{1.79cm}|}
\hline
\rowcolor[HTML]{EFEFEF} & \multicolumn{2}{c|}{\textbf{Sources reporting}} & & & & \multicolumn{2}{c|}{\textbf{Empirical sources}} & \\
\cline{2-3}\cline{7-8}
\rowcolor[HTML]{EFEFEF}\multirow{-2}{=}{\textbf{Finding}} & \textbf{Count*} & \textbf{\%} & \multirow{-2}{=}{\textbf{Peer-reviewed}} & \multirow{-2}{=}{\textbf{Grey}} & \multirow{-2}{=}{\textbf{Background}} & \textbf{Peer-reviewed} & \textbf{Grey} & \multirow{-2}{=}{\textbf{Reporting support}} \\
\hline
Hallucinations and incorrect generations & 109 & 72\% & 45 & 64 & 9 & 28 & 36 & Strong \\
\hline
Cumulative context errors and iteration drift & 34 & 23\% & 8 & 26 & 0 & 5 & 15 & Strong \\
\hline
Minimal-review commits and insufficient inspection practices & 31 & 21\% & 10 & 21 & 2 & 5 & 9 & Strong \\
\hline
Brittle prompts and prompt sensitivity & 27 & 18\% & 5 & 22 & 0 & 5 & 14 & Strong \\
\hline
Licensing and provenance uncertainty & 9 & 6\% & 3 & 6 & 2 & 1 & 2 & Moderate \\
\hline
\end{tabular}
\par\smallskip\noindent\footnotesize{*A source can count under more than one finding.}
\end{table*}

In both streams, negative is the most frequent comparable direction for short-term defects and long-term maintainability. The streams differ on short-term productivity, where 21 of the 28 grey comparable directions are positive against 13 of the 23 peer-reviewed ones, counting positive, negative, and mixed directions.

Directness and measurement also differ. Of the 159 sources that count toward RQ4, 87 are direct and 72 adjacent. A measured outcome is recorded for 34 of the 87 direct sources and for 50 of the 72 adjacent sources. For short-term productivity, 21 of the 69 direct sources measure the outcome, 35 report it as perception, and 13 assert it, whereas 20 of the 36 adjacent sources measure it. For short-term quality, 18 of the 42 direct sources and 34 of the 46 adjacent sources measure it. The measured evidence on productivity and quality therefore comes largely from adjacent sources, which evaluate models, agents, or tools rather than the human practice (Section~\ref{sec:opdef}), while the direct sources more often report perceptions and claims. A source that addresses vibe coding directly is thus not by itself empirical evidence on its effects.

Figure~\ref{fig:rq33_rq4_heatmaps} crosses the RQ3.3 codes with the outcome codes. Of the 159 RQ4 sources that are not background, 95 carry the novice code, 61 the expert code, and 39 both. Productivity is the most frequent outcome in both groups (73 and 52 sources), followed by quality (52 and 30), and the two distributions are similar in shape. The task dimension is thin. The UI and front-end code appears with an outcome in at most 8 sources per outcome and the data and machine-learning code in at most 5, too few for a comparison.

\begin{tcolorbox}[
    colback=gray!10,
    colframe=gray!60,
    title=\textbf{Takeaway 4},
    fonttitle=\bfseries,
    coltitle=black,
    arc=3mm,
    boxrule=0.8pt,
    width=\columnwidth,
]

\onehalfspacing
Productivity is the most frequently reported outcome, mostly for the short term. Comparable sources mostly report short-term productivity gains, more short-term defects, and long-term maintainability costs, and mixed results on short-term quality, with the opposite direction also reported in each of these rows. Long-term outcomes are reported far less often, and the findings do not show whether vibe coding improves or harms these outcomes on balance.
\end{tcolorbox}

\subsection{Risks and Safeguards (RQ5)} \label{sec:rq5}

\subsubsection{Failure Modes} \label{sec:rq5_1}

RQ5.1 draws on 151 sources. Of the 151 sources, 67 are in the empirical evidence category. Hallucinated or incorrect generation is the most frequently reported failure mode, in 109 sources (72\%, 9 of them background), and three workflow-related modes follow: cumulative context errors and iteration drift, minimal-review commits, and brittle prompts. All four have Strong reporting support, and licensing and provenance uncertainty has Moderate reporting support (Table~\ref{tab:rq5_failures}).

The two streams differ in focus. Hallucination is reported in 45 of the 55 peer-reviewed sources and is their main concern. Grey sources report the workflow and prompt related modes more often, for example context errors in 26 grey and 8 peer-reviewed sources. One peer-reviewed study warns that LLMs ``frequently generate erroneous code segments or hallucinate, presenting seemingly correct answers that are actually wrong''~\ref{WL24}. A participant in an interview study describes the minimal-review pattern as ``If you just press accept accept accept, that's simply not good. At some point you can't use the product anymore''~\ref{WL40}.

Each mode is supported by empirical sources of comparable standing in both streams. For the three workflow-related modes, however, these are mostly grey, with five peer-reviewed sources each. The codes record each mode separately, not how the modes relate. The findings record that a failure is reported. They do not show how often it occurs in use or how severe it is.

The labels depend partly on the subset. Licensing uncertainty falls from Moderate to Emerging when only direct sources are counted. On window 1 alone, hallucination stays Strong and the three workflow-related modes become Moderate. All four stay Strong on window 2 alone (Section~\ref{sec:sensitivity}). The Strong labels of the workflow-related modes may therefore depend more on the sources of window 2.

\subsubsection{Governance and Guardrails} \label{sec:rq5_2}

RQ5.2 draws on 167 sources (Table~\ref{tab:rq_coverage}). Validation and testing pipelines (120 sources, 72\%, 11 of them background) and human-in-the-loop review (115, 69\%, 7 of them background) are the most frequently reported safeguards. Provenance tracking, guidance on when to avoid vibe coding, and sandboxing are reported less often. All five have Strong reporting support (Table~\ref{tab:rq5_safeguards}). The two streams report the safeguards in broadly similar proportions.

\begin{table*}[!t]
\centering
\footnotesize
\renewcommand{\arraystretch}{1.1}
\setlength{\tabcolsep}{4pt}
\caption{RQ5.2: Safeguards and guardrails. Percentages are relative to the 167 sources contributing to RQ5.2 (67 peer-reviewed, 100 grey). Column definitions are given at the start of Section~\ref{sec:results}. Restricted sources are not counted. Background sources count as reports only. The evaluated safeguards are summarized in Table~\ref{tab:safeguard_evaluations}.}
\label{tab:rq5_safeguards}

\begin{tabular}{|>{\raggedright\arraybackslash}p{4.64cm}|p{1.06cm}|p{0.87cm}|p{1.50cm}|p{0.77cm}|p{1.69cm}|p{1.50cm}|p{0.77cm}|p{1.79cm}|}
\hline
\rowcolor[HTML]{EFEFEF} & \multicolumn{2}{c|}{\textbf{Sources reporting}} & & & & \multicolumn{2}{c|}{\textbf{Empirical sources}} & \\
\cline{2-3}\cline{7-8}
\rowcolor[HTML]{EFEFEF}\multirow{-2}{=}{\textbf{Finding}} & \textbf{Count*} & \textbf{\%} & \multirow{-2}{=}{\textbf{Peer-reviewed}} & \multirow{-2}{=}{\textbf{Grey}} & \multirow{-2}{=}{\textbf{Background}} & \textbf{Peer-reviewed} & \textbf{Grey} & \multirow{-2}{=}{\textbf{Reporting support}} \\
\hline
Validation and testing pipelines & 120 & 72\% & 46 & 74 & 11 & 29 & 42 & Strong \\
\hline
Human-in-the-loop review & 115 & 69\% & 50 & 65 & 7 & 31 & 36 & Strong \\
\hline
Provenance tracking and traceability mechanisms & 25 & 15\% & 9 & 16 & 2 & 7 & 7 & Strong \\
\hline
Explicit guidance on when to avoid vibe coding & 20 & 12\% & 10 & 10 & 0 & 9 & 4 & Strong \\
\hline
Sandboxing and constrained execution environments & 14 & 8\% & 7 & 7 & 2 & 3 & 2 & Strong \\
\hline
\end{tabular}
\par\smallskip\noindent\footnotesize{*A source can count under more than one finding.}
\end{table*}

Most safeguard items are recommended rather than evaluated. The reporting mode is recorded for each source and safeguard (Section~\ref{sec:extraction}). Of the 272 safeguard items of the 151 contributing sources that are not background, 149 are recommended, 85 are described in use, and 38 are evaluated empirically. Sixty-four sources only recommend the safeguards they report, and 40 only describe them in use. Evaluation is concentrated in the grey stream, which holds 30 of the 38 evaluated items, many of them tool or system reports. A repository study concludes that AI adoption ``will need to be paired with strong quality safeguards''~\ref{WL33}. An auto-ethnographic study notes that the community has tried ``formalised prompts, versioning, sandboxing and Model Context Protocol (MCP) servers, but these remain only partial solutions''~\ref{WL68}.

Where safeguards are evaluated, the results are mixed. Table~\ref{tab:safeguard_evaluations} counts the 38 evaluated instances of coded safeguards from 32 sources by safeguard and reported result. The instances, with their design, are listed in sheet S6 of the supplementary workbook. The record covers every source and coded safeguard whose reporting mode is evaluated. Validation pipelines account for 22 instances, human review for 15, and sandboxing for 1. No instance concerns provenance tracking or guidance on when to avoid vibe coding. Of the 38 instances, 16 are beneficial, 8 are mixed, 4 are limiting (\ref{WL64}, \ref{GL55}, \ref{GL81}, \ref{GL99}), and 10 state no separate result. One source (\ref{GL55}) reports a limit on one safeguard and a benefit on another, and six sources (\ref{WL2}, \ref{GL39}, \ref{GL70}, \ref{GL88}, \ref{GL120}, \ref{GL153}) state no separate result for any safeguard they evaluate. All four limits concern validation pipelines: a static analyzer that did not flag a recursion flaw, a single scanner that misrepresented the security posture, generated tests that degraded repair, and passing functional tests that left security issues in place. Several beneficial results concern a whole tool in which the safeguard is one part, so these evaluations do not isolate the safeguard.

\begin{table*}[!t]
\centering
\footnotesize
\renewcommand{\arraystretch}{1.1}
\setlength{\tabcolsep}{4pt}
\caption{Empirically evaluated safeguards: instances, sources, and reported result (38 instances from 32 sources). Beneficial: reduced the risk or improved the target. Mixed: a benefit with a cost, or a partial effect. Limiting: failed or left the problem in place. Not stated: no separate result. The instances are listed in sheet S6 of the supplementary workbook.}
\label{tab:safeguard_evaluations}
\resizebox{\textwidth}{!}{%
\begin{tabular}{|>{\raggedright\arraybackslash}p{6.20cm}|>{\centering\arraybackslash}p{1.50cm}|>{\centering\arraybackslash}p{1.50cm}|>{\centering\arraybackslash}p{1.50cm}|>{\centering\arraybackslash}p{1.50cm}|>{\centering\arraybackslash}p{1.50cm}|>{\centering\arraybackslash}p{1.50cm}|}
\hline
\rowcolor[HTML]{EFEFEF}
\textbf{Safeguard} & \textbf{Instances} & \textbf{Sources} & \textbf{Beneficial} & \textbf{Mixed} & \textbf{Limiting} & \textbf{Not stated} \\
\hline
Validation and testing pipelines & 22 & 22 & 7 & 7 & 4 & 4 \\
\hline
Human-in-the-loop review & 15 & 15 & 8 & 1 & 0 & 6 \\
\hline
Sandboxing and constrained execution & 1 & 1 & 1 & 0 & 0 & 0 \\
\hline
\textbf{All} & 38 & 32 & 16 & 8 & 4 & 10 \\
\hline
\end{tabular}
}
\end{table*}

Sandboxing is the least certain label. It rests on three comparable peer-reviewed studies, at the threshold for Strong, and falls to Moderate when only direct sources are counted (Section~\ref{sec:sensitivity}).

\begin{tcolorbox}[
    colback=gray!10,
    colframe=gray!60,
    title=\textbf{Takeaway 5},
    fonttitle=\bfseries,
    coltitle=black,
    arc=3mm,
    boxrule=0.8pt,
    width=\columnwidth,
]

\onehalfspacing
Hallucinated or incorrect generation is the most frequently reported failure mode, followed by context errors, minimal review, and brittle prompts, which grey sources report more often. Validation pipelines and human review are the most frequently reported safeguards, but most safeguard items are recommended rather than evaluated, and the 32 sources that evaluate one report beneficial, mixed, and limiting results, so the effectiveness of each safeguard remains uncertain.
\end{tcolorbox}

\subsection{Usage Contexts and Constraints (RQ6)} \label{sec:rq6}

RQ6 draws on 110 sources, and Table~\ref{tab:rq6_contexts} lists its four codes. Restricted grey sources (Section~\ref{sec:grey_selection}) do not count toward its findings. The sources place vibe coding mainly in prototyping and early-stage work (56 sources), and less often in production use (43 sources). Both findings have Strong reporting support.

Seventeen sources carry both context codes. One peer-reviewed study reports that ``developers agreed that vibe coding could speed up prototyping but argued that it could not do the work of engineers''~\ref{WL41}. The constraints the sources name are privacy or data and code exposure (31 sources) and cost or usage limits (26 sources), both with Strong reporting support.

Grey sources outnumber peer-reviewed sources in every row, most widely for cost constraints (6 peer-reviewed, 20 grey). For the two context rows, the peer-reviewed stream contributes more empirical sources of comparable standing than the grey stream, so these findings do not rest on practitioner accounts alone. The cost finding is the most uncertain. It rests mainly on grey sources and sits at the edge of its label, so a one-step move of the threshold for comparable peer-reviewed sources would make it Moderate (Section~\ref{sec:sensitivity}).

\begin{table*}[!t]
\centering
\footnotesize
\renewcommand{\arraystretch}{1.1}
\setlength{\tabcolsep}{4pt}
\caption{RQ6: Usage contexts and constraints. 110 sources contribute to RQ6 (39 peer-reviewed, 71 grey), and counts are given without percentages. Column definitions are given at the start of Section~\ref{sec:results}. Restricted sources are not counted. Background sources count as reports only.}
\label{tab:rq6_contexts}
\resizebox{\textwidth}{!}{%
\begin{tabular}{|>{\raggedright\arraybackslash}p{4.39cm}|p{2.32cm}|p{1.53cm}|p{0.79cm}|p{1.73cm}|p{1.53cm}|p{0.79cm}|p{1.83cm}|}
\hline
\rowcolor[HTML]{EFEFEF} & \textbf{Sources reporting} & & & & \multicolumn{2}{c|}{\textbf{Empirical sources}} & \\
\cline{2-2}\cline{6-7}
\rowcolor[HTML]{EFEFEF}\multirow{-2}{=}{\textbf{Finding}} & \textbf{Count*} & \multirow{-2}{=}{\textbf{Peer-reviewed}} & \multirow{-2}{=}{\textbf{Grey}} & \multirow{-2}{=}{\textbf{Background}} & \textbf{Peer-reviewed} & \textbf{Grey} & \multirow{-2}{=}{\textbf{Reporting support}} \\
\hline
Use in prototyping or early-stage development contexts & 56 & 24 & 32 & 1 & 17 & 13 & Strong \\
\hline
Use in production or operational deployment contexts & 43 & 17 & 26 & 2 & 15 & 7 & Strong \\
\hline
Privacy or exposure of data and code & 31 & 15 & 16 & 3 & 6 & 7 & Strong \\
\hline
Cost or usage-limit constraints & 26 & 6 & 20 & 2 & 3 & 8 & Strong \\
\hline
\end{tabular}
}
\par\smallskip\noindent\footnotesize{*A source can count under more than one finding.}
\end{table*}

\begin{tcolorbox}[
    colback=gray!10,
    colframe=gray!60,
    title=\textbf{Takeaway 6},
    fonttitle=\bfseries,
    coltitle=black,
    arc=3mm,
    boxrule=0.8pt,
    width=\columnwidth,
]

\onehalfspacing
The sources discuss vibe coding in prototyping more often than in production. Privacy and cost are the reported constraints. The cost finding rests mainly on grey sources and sits at the edge of its label.
\end{tcolorbox}

\subsection{Tools and Platforms (RQ7)} \label{sec:rq7}

RQ7 draws on the 230 sources that name at least one coded tool. Claude and Anthropic tools (59\%) and ChatGPT and OpenAI models (49\%) are named most often, followed by Cursor, GitHub Copilot, and coding agents and agent modes (Table~\ref{tab:tools_coverage_rq7}). Unlike RQ6, RQ7 counts the restricted grey sources, because tool visibility is one of the things they may support, and their share of each finding enters the label.

The streams differ in which tools they name. GitHub Copilot has the largest peer-reviewed count, while Claude, Cursor, and coding agents are named mostly in grey sources. One peer-reviewed empirical study notes that ``Developers now embed AI assistants like Copilot and Cursor into their development workflows''~\ref{WL42}.

\begin{table*}[!t]
\centering
\footnotesize
\renewcommand{\arraystretch}{1.1}
\setlength{\tabcolsep}{4pt}
\caption{RQ7: Tools and platforms named. Percentages are relative to the 230 sources contributing to RQ7 (68 peer-reviewed, 162 grey). Column definitions are given at the start of Section~\ref{sec:results}. Restricted sources are counted. Background sources count as reports only.}
\label{tab:tools_coverage_rq7}
\resizebox{\textwidth}{!}{%
\begin{tabular}{|>{\raggedright\arraybackslash}p{4.64cm}|p{1.06cm}|p{0.87cm}|p{1.50cm}|p{0.77cm}|p{1.69cm}|p{1.50cm}|p{0.77cm}|p{1.79cm}|}
\hline
\rowcolor[HTML]{EFEFEF} & \multicolumn{2}{c|}{\textbf{Sources reporting}} & & & & \multicolumn{2}{c|}{\textbf{Empirical sources}} & \\
\cline{2-3}\cline{7-8}
\rowcolor[HTML]{EFEFEF}\multirow{-2}{=}{\textbf{Finding}} & \textbf{Count*} & \textbf{\%} & \multirow{-2}{=}{\textbf{Peer-reviewed}} & \multirow{-2}{=}{\textbf{Grey}} & \multirow{-2}{=}{\textbf{Background}} & \textbf{Peer-reviewed} & \textbf{Grey} & \multirow{-2}{=}{\textbf{Reporting support}} \\
\hline
Claude and Anthropic tools & 135 & 59\% & 31 & 104 & 4 & 25 & 39 & Moderate \\
\hline
ChatGPT and OpenAI models & 113 & 49\% & 40 & 73 & 12 & 23 & 35 & Strong \\
\hline
Cursor & 104 & 45\% & 25 & 79 & 7 & 20 & 23 & Moderate \\
\hline
GitHub Copilot & 92 & 40\% & 41 & 51 & 19 & 19 & 19 & Strong \\
\hline
Coding agents and agent modes & 78 & 34\% & 15 & 63 & 4 & 8 & 27 & Strong \\
\hline
Evaluation frameworks and benchmarks & 15 & 7\% & 4 & 11 & 0 & 4 & 10 & Strong \\
\hline
\end{tabular}
}
\par\smallskip\noindent\footnotesize{*A source can count under more than one finding.}
\end{table*}

Four of the six findings have Strong reporting support. Claude and Cursor have Moderate reporting support, although Claude has the largest count, because 27\% of the Claude sources and 29\% of the Cursor sources are Tier~3 or belong to the restricted categories, above the 25\% limit for Strong (Section~\ref{sec:evidence_strength}). Both become Strong without the Tier~3 sources or the restricted categories (Section~\ref{sec:sensitivity}). Coding agents and agent modes stay just inside the limit, with a restricted share of 24\%. The labels of these three findings therefore depend on how much weight the less credible grey sources receive.

Two Strong findings are less certain. The finding on evaluation and benchmarks rests mostly on adjacent sources (14 of 15) and falls to Emerging when only direct sources are counted. The finding on coding agents falls to Moderate on direct sources alone and lies just inside the restricted-share limit (Section~\ref{sec:sensitivity}).

\begin{tcolorbox}[
    colback=gray!10,
    colframe=gray!60,
    title=\textbf{Takeaway 7},
    fonttitle=\bfseries,
    coltitle=black,
    arc=3mm,
    boxrule=0.8pt,
    width=\columnwidth,
]

\onehalfspacing
Claude, ChatGPT, Cursor, and GitHub Copilot are the most visible tools. Copilot is named most in peer-reviewed sources, and the others mostly in grey sources. The RQ7 codes record that a tool is named, not whether a source evaluates it, so the findings describe visibility, not performance, and the order may change quickly.
\end{tcolorbox}

\subsection{Open Challenges and Future Directions (RQ8)} \label{sec:rq8}

RQ8 draws on the 123 sources that report an open challenge. Of the 123 sources, 62 are in the empirical evidence category. The most frequent challenge is the lack of standardized evaluation and benchmarks, in 85 sources (69\%, 6 of them background). Trust, misuse, and reliability challenges, context management, security, and governance follow, and all five findings have Strong reporting support (Table~\ref{tab:rq8_challenges}). We relate each challenge to an earlier research question. The evaluation gap relates to the outcome measures of RQ4 (Section~\ref{sec:rq4}). Trust, misuse, and reliability challenges relate to the role shift of RQ3 (Section~\ref{sec:rq3}). Context management relates to the workflows of RQ2 (Section~\ref{sec:rq2}) and the session-level failures of RQ3.2. Security and governance build on the risks and safeguards of RQ5.

One peer-reviewed study states that ``There is currently no standardized benchmark for vibe coding efficiency in the literature, which limits direct quantitative comparison with prior work''~\ref{WL62}. Another names open questions ``particularly around trust calibration, evaluation without ground truth, and the risk of becoming black-box operators''~\ref{WL44}.

Grey sources outnumber peer-reviewed sources in four of the five rows. Security is the exception, reported in 10 peer-reviewed and 8 grey sources, for example \ref{WL40}, \ref{WL56}, \ref{GL12}, and \ref{GL16}. The RQ8 codes record that a challenge is named, not whether a source studies it. Context management, governance, and security fall from Strong to Moderate when only direct sources are counted. Governance and context management are the least certain. Each rests on three comparable peer-reviewed studies, so a one-step move of that threshold would make both Moderate (Section~\ref{sec:sensitivity}).

In the coded record, long-term outcomes are reported far less often than short-term ones (Section~\ref{sec:rq4}), and reproducibility is reported in only seven sources. The four propositions of the conceptual framework (Section~\ref{sec:discussion}, P1 to P4) are the testable part of this agenda. P1 to P3 name associations reported together in sources, and P4 is our reading. The coded record does not show that any of them has been tested.

\begin{tcolorbox}[
    colback=gray!10,
    colframe=gray!60,
    title=\textbf{Takeaway 8},
    fonttitle=\bfseries,
    coltitle=black,
    arc=3mm,
    boxrule=0.8pt,
    width=\columnwidth,
]

\onehalfspacing
The lack of standardized evaluation and benchmarks is the most frequently reported open challenge, followed by trust, misuse, and reliability challenges, context management, security, and governance. Long-term outcomes and reproducibility are the least reported parts of this agenda.
\end{tcolorbox}

\begin{table*}[!t]
\centering
\footnotesize
\renewcommand{\arraystretch}{1.1}
\setlength{\tabcolsep}{4pt}
\caption{RQ8: Open challenges. Percentages are relative to the 123 sources contributing to RQ8 (44 peer-reviewed, 79 grey). Column definitions are given at the start of Section~\ref{sec:results}. Restricted sources are not counted. Background sources count as reports only.}
\label{tab:rq8_challenges}
\resizebox{\textwidth}{!}{%
\begin{tabular}{|>{\raggedright\arraybackslash}p{4.64cm}|p{1.06cm}|p{0.87cm}|p{1.50cm}|p{0.77cm}|p{1.69cm}|p{1.50cm}|p{0.77cm}|p{1.79cm}|}
\hline
\rowcolor[HTML]{EFEFEF} & \multicolumn{2}{c|}{\textbf{Sources reporting}} & & & & \multicolumn{2}{c|}{\textbf{Empirical sources}} & \\
\cline{2-3}\cline{7-8}
\rowcolor[HTML]{EFEFEF}\multirow{-2}{=}{\textbf{Finding}} & \textbf{Count*} & \textbf{\%} & \multirow{-2}{=}{\textbf{Peer-reviewed}} & \multirow{-2}{=}{\textbf{Grey}} & \multirow{-2}{=}{\textbf{Background}} & \textbf{Peer-reviewed} & \textbf{Grey} & \multirow{-2}{=}{\textbf{Reporting support}} \\
\hline
Lack of standardized evaluation and benchmarks & 85 & 69\% & 27 & 58 & 6 & 20 & 34 & Strong \\
\hline
Trust, misuse, and reliability challenges & 31 & 25\% & 15 & 16 & 5 & 7 & 10 & Strong \\
\hline
Context management limitations & 23 & 19\% & 5 & 18 & 2 & 3 & 10 & Strong \\
\hline
Security concerns including unsafe code, vulnerabilities, and misuse risks & 18 & 15\% & 10 & 8 & 4 & 4 & 6 & Strong \\
\hline
Governance, disclosure, and organizational integration challenges & 12 & 10\% & 5 & 7 & 0 & 3 & 4 & Strong \\
\hline
\end{tabular}
}
\par\smallskip\noindent\footnotesize{*A source can count under more than one finding.}
\end{table*}

\subsection{Sensitivity analyses}\label{sec:sensitivity}

The label of each of the 60 findings in Sections~\ref{sec:rq1} to~\ref{sec:rq8} was recomputed on the six subsets of Table~\ref{tab:sensitivity_direct}. Each recomputation used the matrix of Section~\ref{sec:evidence_strength} and set background sources aside, as in the main analysis. The direct subset rests on the directness rule of Section~\ref{sec:opdef}, applied to all 266 sources. The three sources that are direct on the four features alone are too few for a subset of their own, so the last subset holds the 143 direct sources that name the term. On the full corpus, 42 of the 51 presence findings have Strong, 8 Moderate, and 1 Weak reporting support. Of the 9 outcome findings, 3 have Moderate and 6 Weak evidence. In this section, Strong, Moderate, Weak, and Emerging therefore mean reporting support for a presence finding and evidence strength for an outcome finding. The per-finding values and labels are in sheets S3 and S4 of the supplementary workbook, and the threshold check is in sheet S7.

On the 146 direct sources alone, 40 labels are unchanged, 18 become weaker, and 2 become stronger. Of the 18 weaker labels, 14 move one step, mostly from Strong to Moderate reporting support. AI-native IDE (RQ1) and retrieval grounding (RQ2) fall from Strong to Weak reporting support, licensing uncertainty (RQ5.1) from Moderate to Emerging reporting support, and evaluation frameworks (RQ7) from Strong to Emerging reporting support. Two RQ4 findings become stronger. Short-term defects and long-term maintainability move from Weak to Moderate evidence. Their positive comparable directions come from adjacent grey sources (\ref{GL151} and \ref{GL172} for defects, \ref{GL29} and \ref{GL179} for maintainability), so on direct sources alone these rows are no longer contested. The drops on the direct subset therefore come from setting aside the adjacent sources. The 143 direct sources that name the term give the same changes, so the three sources classified direct on the four features alone change no label.

Removing the Tier~3 grey sources (219 sources remain) makes no label weaker and 3 findings stronger, each from Moderate to Strong reporting support: socio-technical practice (RQ1), Claude and Anthropic tools (RQ7), and Cursor (RQ7). Removing the restricted categories (193 remain) also makes no label weaker. It makes the same three findings stronger, and end-user or non-programmer development (RQ1) as well, each from Moderate to Strong reporting support.

Counting every source as independent, without merging the dependent pairs of Section~\ref{sec:evidence_strength}, changes no label (Table~\ref{tab:sensitivity_direct}). Merging lowers C1 in 32 findings, by one to six sources, and no finding sits close enough to a C1 threshold for that to matter. The labels therefore do not depend on the dependency decisions.

A sensitivity analysis by publication window shows the largest movement (Table~\ref{tab:sensitivity_direct}). On the 113 sources of window 1 alone, 25 labels are unchanged, 32 become weaker, and 3 become stronger. The presence findings have 17 Strong, 27 Moderate, 3 Weak, and 4 Emerging reporting support, and the outcome findings 4 Moderate and 5 Weak evidence. Of the 32 weaker labels, 23 move from Strong to Moderate reporting support, among them use in production (RQ6). Avoiding vibe coding for high-risk work (RQ5.2) falls furthest, from Strong to Emerging reporting support. Long-term productivity and long-term quality fall from Moderate to Weak evidence, since only one comparable source of window 1 records a direction for each. The three stronger findings are short-term productivity, short-term defects, and long-term maintainability, which all move from Weak to Moderate evidence. On window 1 alone their comparable sources record no opposing directions. For productivity they record 10 positive, 2 no-effect, and 1 mixed direction, for defects 5 negative and 1 mixed, and for maintainability 5 negative. On the 153 sources of window 2 alone, 46 labels are unchanged, 12 become weaker, and 2 become stronger. The presence findings have 36 Strong, 11 Moderate, 2 Weak, and 1 Emerging reporting support, and the outcome findings 2 Moderate and 6 Weak evidence. Two findings have no supporting source, retrieval grounding (RQ2) and short-term reproducibility (RQ4). The two stronger findings are Claude and Anthropic tools and Cursor (RQ7), which move from Moderate to Strong reporting support.

Window 1 alone thus supports fewer findings at the same level than window 2 alone. Window 1 is also smaller than window 2 (113 against 153 sources). Two findings, retrieval grounding and short-term reproducibility, rest only on sources of window 1. The counts do not show whether the difference between the windows reflects a change in the literature over time, the searches behind each window, or the size of each window.

Threshold sensitivity was checked by moving each numeric threshold of the matrix one step in each direction, with the other thresholds fixed (sheet S7). Three of these ten moves change no label. Raising the number of independent sources needed for Strong from eight to nine changes one label, retrieval grounding (RQ2), and lowering it to seven gives collaborative prompt practices (RQ2) Strong reporting support. Raising the restricted-share limit from 25\% to 30\% gives Claude and Cursor (RQ7) Strong reporting support. Raising the number of comparable peer-reviewed studies needed for Strong from three to four changes the most labels, seven of the 60. The findings that move sit at the edge of a threshold, such as the findings with Strong reporting support that rest on three comparable peer-reviewed studies, the smallest number that allows Strong. The C4 rule was checked in the same way. If a row were contested only when the minority of positive against negative directions made up at least 20\% of its comparable directions, three rows would have Moderate instead of Weak evidence: short-term productivity, short-term defects, and long-term maintainability. Short-term quality would stay contested, since its 13 positive directions are 21\% of its 61 comparable directions. The per-threshold record is in the replication package.

\begin{table*}[!t]
\centering
\footnotesize
\renewcommand{\arraystretch}{1.1}
\setlength{\tabcolsep}{4pt}
\caption{Labels (reporting support or evidence strength) of the 60 findings, recomputed on each subset: sources in the subset, and labels unchanged, weaker, or stronger. The label of every finding on every subset is in sheet S4 of the supplementary workbook.}
\label{tab:sensitivity_direct}
\resizebox{\textwidth}{!}{%
\begin{tabular}{|>{\raggedright\arraybackslash}p{9.2cm}|>{\centering\arraybackslash}p{1.5cm}|>{\centering\arraybackslash}p{1.5cm}|>{\centering\arraybackslash}p{1.5cm}|>{\centering\arraybackslash}p{1.5cm}|}
\hline
\rowcolor[HTML]{EFEFEF}
\textbf{Subset} & \textbf{Sources} & \textbf{Unchanged} & \textbf{Weaker} & \textbf{Stronger} \\
\hline
Direct only & 146 & 40 & 18 & 2 \\
\hline
Without Tier~3 & 219 & 57 & 0 & 3 \\
\hline
Without restricted categories & 193 & 56 & 0 & 4 \\
\hline
Publication window 1 only (publications to October 2025) & 113 & 25 & 32 & 3 \\
\hline
Publication window 2 only (publications from November 2025) & 153 & 46 & 12 & 2 \\
\hline
No merging of dependent sources (C1 before merging) & 266 & 60 & 0 & 0 \\
\hline
Direct sources that name the term in the title, keywords, abstract, or opening text & 143 & 40 & 18 & 2 \\
\hline
\end{tabular}
}
\end{table*}

The RQ4 labels were also tested against the reconciled outcome fields, since agreement on them before reconciliation was moderate (Section~\ref{sec:reliability}). The measurement type does not enter any label. Recording every Not stated as No effect (26 pairs) changes no label, and recording every No effect as Not stated (7 pairs) gives long-term quality Weak instead of Moderate evidence. Recording every pair without a horizon as short-term (19 pairs) changes no label. Recording every pair with both horizons as long-term only changes no label, and recording them as short-term only gives long-term productivity Emerging and long-term quality Weak instead of Moderate evidence. For direction, the smallest number of comparable sources whose direction would have to change for a label to change was found for each outcome row. The four contested rows need two or more changes, and short-term quality needs more than four. The three rows with Moderate evidence rest on few comparable sources, and one change would alter each of them: long-term productivity, long-term quality, and short-term reproducibility. These three labels are therefore the least stable. The record of these tests is in the replication package.

\section{Discussion} \label{sec:discussion}

\subsection{Vibe Coding as an Iterative Control System}
\label{sec:discussion_control_system}

This review frames vibe coding as an \emph{iterative control system}, a feedback-driven process in which the developer specifies intent, evaluates AI-generated output, and refines the next prompt. The term is used conceptually, not as a formal control-theoretic model. The definitions synthesized in RQ1 converge on natural-language intent as the primary interface. The iterate-prompt-patch loop is the workflow pattern reported most often, followed by explicit context provisioning and validation pipelines (Section~\ref{sec:rq2}), so the loop and its checks are both frequent in the literature. The role shifts documented in RQ3 show developers moving from direct code authoring toward supervision and verification.

Validation and testing pipelines and human-in-the-loop review are reported by 109 and 108 of the 151 non-background RQ5.2 sources, which suggests that the sources treat generated output as something to check before integration. On this reading, developer effort moves from writing code to reviewing and validating generated outputs (RQ3.1). These counts show how often the sources report the two safeguards, not whether the safeguards work.

The two streams describe the loop with a different focus. Grey sources more often describe prompt-and-repair iteration and context provisioning, while chat-based loops and retrieval grounding appear at similar or slightly higher shares in peer-reviewed sources (Section~\ref{sec:rq2}). Validation pipelines appear in both streams. Of the 32 sources that evaluate a coded safeguard, 25 are grey, many of them tool or system reports (Section~\ref{sec:rq5_2}). The line between the streams is therefore not a line between evaluated and anecdotal evidence.

Prior reviews adopt narrower lenses (Section~\ref{sec:rw}). This review frames vibe coding as a socio-technical loop that spans intent, generation, validation, trust calibration, and governance, and the two streams weight the stages of the loop differently, so combining them gives a fuller picture.

Some sources pair a systematic loop, with tests and structured prompting, with workable results. Others pair limited review with hallucinations, technical debt, or cognitive debt. Propositions P2 and P3 below give sources that illustrate each pairing. The co-occurrence was not counted.

\subsection{Developer Role Shift and Cognitive Burden}
\label{sec:discussion_role_shift}

In the sources, developers contribute less by writing code and more by steering, supplying context, and reviewing generated output (RQ2, RQ3.1).

The sources do not describe the role shift as uniformly beneficial. The RQ3.1 code for cognitive load records changes in load, reduced or increased, that come from the burden of prompting and review. The code for cognitive debt records understanding that is deferred and later rework (RQ3.1).

Studies of AI-assisted programming report related effects: productivity gains from GitHub Copilot~\cite{peng2023impact}, the same assistant framed as both an asset and a liability~\cite{dakhel2023github}, the model as a conversational assistant to the programmer~\cite{ross2023programmer, akhoroz2025conversational}, and teams navigating generative-AI adoption~\cite{russo2024navigating}. The coded record does not compare vibe coding with other AI-assisted programming. In the sources of this review, the developer is described as a supervisor, validator, and prompt-based controller (RQ3.1). The role-shift findings keep their labels when only the direct sources are counted (Section~\ref{sec:sensitivity}). This suggests that they do not rest mainly on the broader literature on AI-assisted programming.

The sources address novices and experts (RQ3.3). Of the 129 non-background RQ3.3 sources, 102 are coded for novice or non-programmer orientation and 64 for expert acceleration. The codes record the audience a source addresses (Section~\ref{sec:rq3}), so the review reports no finding on differences between novices and experts.

For practitioners, this reframing suggests that verification and review may matter as much as authoring. This reading rests on how often the sources report these activities, not on tests of their effect.

\subsection{Conditional Outcomes and the Verification Gap}
\label{sec:discussion_outcomes_verification}

Productivity is the most frequently reported outcome (RQ4). Among comparable sources, short-term productivity is mostly reported as a gain and long-term maintainability mostly as a cost, but each row has opposing sources and is labeled Weak evidence (Section~\ref{sec:rq4}). The rows rest on different sources. Within single sources, 17 of the 32 non-background sources that report both outcomes record a productivity gain with a maintainability cost. Long-horizon outcomes are examined less often than short-term ones. Agreement on the outcome fields before reconciliation was moderate (Section~\ref{sec:reliability}), so these patterns may be best read as provisional. The direct sources contribute most of the reports on these outcomes but a minority of the measured ones (Section~\ref{sec:rq4}), so a source that addresses vibe coding directly is not by itself empirical evidence of its effect on productivity or quality.

We use the term verification gap for our reading that code may be generated faster than it can be verified. No code records this relation, so it is not a finding of the review. RQ5.1 records hallucinated or incorrect outputs, cumulative context errors, and changes integrated with limited review. The safeguards reported most often in both streams are validation and testing pipelines and human-in-the-loop review (RQ5.2). Of the 151 non-background RQ5.2 sources, 33 evaluate at least one coded safeguard (Table~\ref{tab:safeguard_evaluations}). The evidence is therefore stronger at identifying risks than at showing which mitigations work.

Earlier studies report productivity gains~\cite{peng2023impact}, quality concerns about generated code~\cite{mastropaolo2023robustness, dakhel2023github}, and a speed-quality trade-off~\cite{fawzy2025vibe}. Sources that report productivity gains often also report evaluation and validation practices. This co-occurrence does not establish that verification causes the gains or identifies the conditions under which the gains occur. Evaluation and validation practices, here validation pipelines as a workflow or a safeguard and human-in-the-loop review, are reported by most RQ4 sources whether or not they report gains: by 140 of the 159 non-background sources contributing to RQ4, by 60 of the 68 of them whose recorded productivity direction is positive, and by 80 of the other 91. The co-occurrence may therefore reflect how common these practices are in the literature.

For practitioners, the sources recommend validation and testing pipelines and human-in-the-loop review (RQ5.2), and 20 sources give guidance on when to avoid vibe coding. Both have Strong reporting support, which is not a test of whether they reduce risk. Scaling verification to the intended use was not coded. The included sources evaluate review and testing only within specific tools and settings (Table~\ref{tab:safeguard_evaluations}). For researchers, the priority is causal evidence on which safeguards reduce the reported risks (Section~\ref{sec:rq8}).

\subsection{Contexts, Tools, and Adoption Boundaries}
\label{sec:discussion_contexts_tools}

Prototyping is reported more often than production use, but task-domain evidence remains sparse. Among the two explicitly coded task domains, UI and front-end work appears more often than data and machine-learning work, but it is represented in only 17 of the 144 sources contributing to RQ3.3, 5 of them background. These counts describe literature coverage, not comparative effectiveness. Both task-domain findings rest on few empirical sources of comparable standing, and no source compares project types within one design. In RQ6, 35 of the 41 non-background sources that report use in production also report validation and testing pipelines or human-in-the-loop review (RQ5.2), as do 55 of the other 63 non-background RQ6 sources. These codes are reported by most sources, so the co-occurrence says little about production use in particular. Privacy and cost constraints are named as factors in adoption.

The literature reports different tool categories and associated workflow descriptions, but the available evidence does not establish that tool choice moderates workflow behavior or outcomes.

Unlike the tool taxonomy of Ray~\cite{ray2025review}, this review reads tools and contexts as possible moderators of the whole loop (Section~\ref{sec:discussion_control_system}), stated as the author-derived proposition P4. No included source tests this role.

These patterns are early indications. On window 1 alone, use in production and privacy constraints fall from Strong to Moderate reporting support, and on window 2 alone cost constraints do so (Section~\ref{sec:sensitivity}). For practitioners, the sources advise caution for production, safety-critical, and security-sensitive work (RQ5.2). This advice has not been evaluated in the included sources. Studies of production, data-intensive, and safety-critical settings could show whether what is reported for prototyping also holds there.

\subsection{Open Challenges and Research Agenda}
\label{sec:discussion_agenda}

The main open question concerns the evaluation loop. The sources describe it and its failure modes. The review did not code whether sources compare parts of the loop or how long a session they observe, so this record does not show which parts matter most for outcomes.

Earlier challenge lists come from one stream or from an undocumented base (Section~\ref{sec:rw}), and this review adds a view across both streams.

Several priorities follow. First, session-level dynamics (RQ3.2) are the smallest question by contributing sources. Full-session interaction traces could help measure them. Second, few sources evaluate a safeguard empirically, and controlled and field studies comparing safeguards may show which reduce risk. Third, task-domain evidence is sparse (Section~\ref{sec:discussion_contexts_tools}). Fourth, clearer operational definitions (RQ1) would support comparability across studies. Fifth, long-horizon outcomes (RQ4) are under-studied relative to short-term measures, and longer development cycles would be needed to observe them. Finally, four topics were noted during coding but lie outside the codebook and were not coded (Section~\ref{sec:threat_coding}). They are not findings of this review, and their systematic coding is left to future work.

\subsection{Tensions in the Evidence Base}

The review suggests five recurring tensions across the research questions:

\begin{itemize}
\item Productivity is the most frequently reported outcome (RQ4), and long-term maintainability is reported mostly as a cost (Table~\ref{tab:rq4_outcomes}). Cognitive debt (RQ3.1) is reported by 12 of the 40 non-background sources that report maintainability. Whether speed gains cost maintainability through deferred understanding was not coded.
\item Vibe coding is framed as lowering entry barriers for novices (RQ3.3), and hallucinations and incorrect generations are among the failure modes reported most often (RQ5.1). Whether novices detect incorrect output less often was not coded.
\item Agentic or semi-autonomous framings (RQ1) point toward delegation, whereas the most frequently reported safeguards, human-in-the-loop review and validation pipelines (RQ5.2), point back toward human control. Where the balance lies is not tested in the included sources.
\item Prototyping is reported more often than production deployment (RQ6), and no included source evaluates the move from prototype to production, so this move remains underexplored.
\item Explicit context provisioning is a frequent tactic (RQ2). Of the 75 non-background sources that report it, 34 also report context-window or state-retention failures (RQ3.2) or cumulative context errors (RQ5.1), against 21 of the other 82 non-background RQ2 sources. Whether the interaction that supplies context also degrades it was not coded.
\end{itemize}

Each tension is reported where the speed of generation may outpace the rigor of the surrounding evaluation loop (Section~\ref{sec:discussion_control_system}). This reading is ours and has not been tested.

\subsection{Conceptual Framework}\label{sec:framework}

\begin{figure*}[!htbp]
\centering
\includegraphics[width=\textwidth, clip, trim=0 400 0 0]{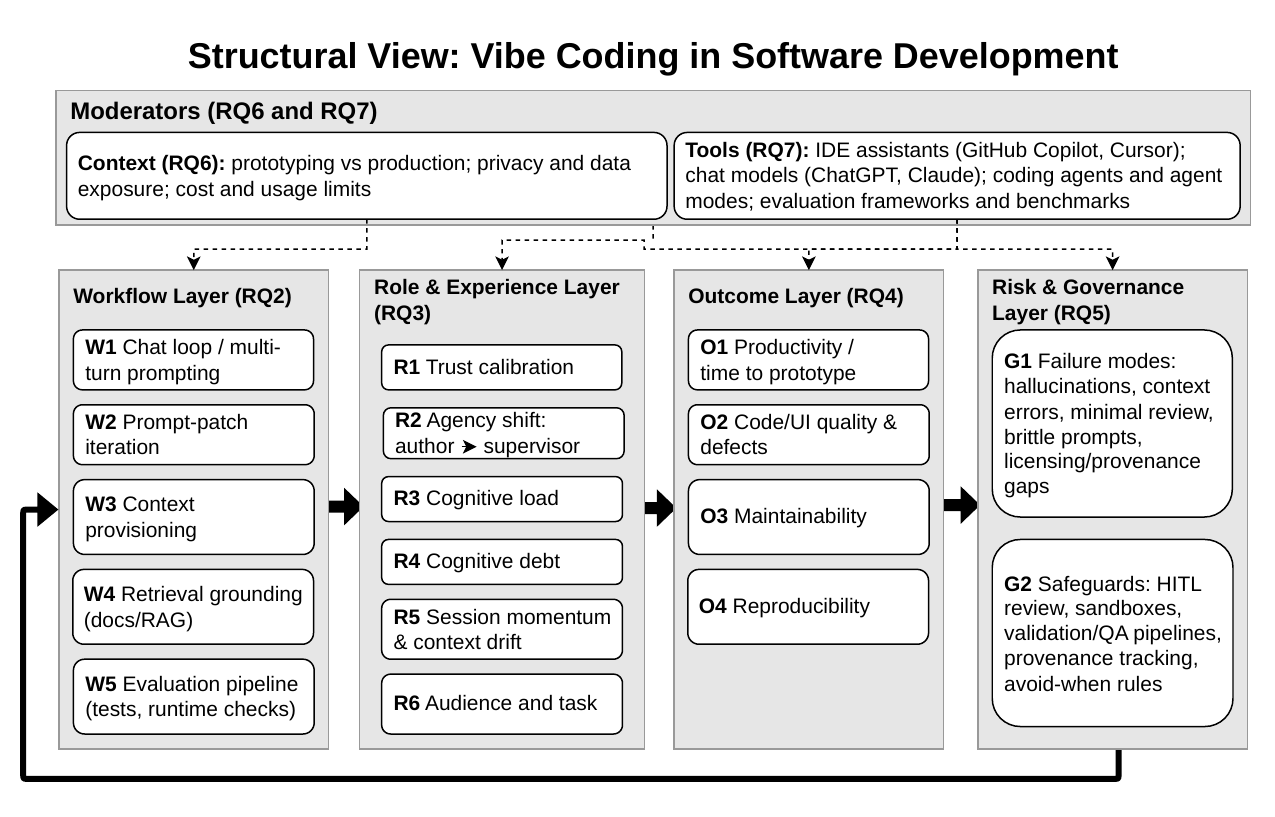}
\caption{Conceptual framework: the structural view of four layers, namely workflow (RQ2), role and experience (RQ3), outcomes (RQ4), and risk and governance (RQ5). P4 proposes that context (RQ6) and tools (RQ7) moderate them. The moderation is an author-derived proposition and has not been tested. Links between layers are propositions, not causal claims. The loop of a single session is shown in Figure~\ref{fig:vibe_loop}. HITL: human-in-the-loop. RAG: retrieval-augmented generation. QA: quality assurance of the generated code (not the quality-assessment items of Table~\ref{tab:qa_white}).}
\label{fig:conceptual_framework}
\end{figure*}

The framework (Figure~\ref{fig:conceptual_framework}) is descriptive, not a causal model or a governance framework (for the latter, see Section~\ref{sec:discussion_implications}). The structural view is a descriptive taxonomy of four layers. The propositions between the layers are hypotheses for testing. Figure~\ref{fig:vibe_loop} shows the loop of a session (Section~\ref{sec:discussion_control_system}), and Figure~\ref{fig:conceptual_framework} the four layers with the moderators proposed in P4.

The four layers are a chain of proposed relationships, not a static stack. The workflow layer (RQ2) holds the interaction patterns. The role and experience layer (RQ3) holds trust calibration, the shift to supervision, cognitive load, and cognitive debt. The outcome layer (RQ4) holds productivity, defects, maintainability, and reproducibility. The risk and governance layer (RQ5) holds the failure modes and safeguards.

\paragraph{How the layers interact} The sources report the constructs together, and no included source manipulates one layer to observe its effect on another. We therefore state the interactions as four propositions, with sources that illustrate them. The co-occurrences were not counted, no included source tests P1 to P3, and P4 is our reading of the evidence.
\begin{itemize}
\item \emph{P1, workflow design and cognitive load.} The interaction pattern chosen in the workflow layer (RQ2) is associated with the amount of prompt refinement, context repair, and output verification the developer carries in the role and experience layer (RQ3). Illustrated by \ref{WL7}, \ref{GL2}, and \ref{GL3}.
\item \emph{P2, developer validation and outcomes.} How closely the developer supervises and validates the loop (RQ3) is associated with the quality and maintainability outcomes reported in the outcome layer (RQ4). Illustrated by \ref{GL10} and \ref{GL49}.
\item \emph{P3, weak evaluation and risk.} Limited review and superficial verification in the workflow and role layers are associated with the failure modes in the risk and governance layer (RQ5.1), minimal-review commits and cumulative context errors, and with trust calibration (RQ3.1). Illustrated by \ref{GL7}, \ref{GL57}, and \ref{WL15}.
\item \emph{P4, moderation by tools and context.} The tool ecosystem (RQ7) and the development context and users (RQ6) may moderate the three relationships above. The same workflow may then produce different load, outcomes, and risk, depending on the tools used and on the context of use and its constraints. The RQ6 and RQ7 evidence describes tools and contexts separately (\ref{WL27}, \ref{WL42}, and \ref{GL1}).
\end{itemize}
We offer the four propositions for testing in future studies. 

\paragraph{Moderators (RQ6, RQ7)}
The moderators named in P4 are the context of use, prototyping or production, and privacy or cost constraints (RQ6), and the tools used (RQ7). The RQ7 codes record named tools and coding agents, not tool categories. Developer expertise and the contrasts between novices and experts and between UI and data or machine-learning work belong to RQ3.3 (Section~\ref{sec:discussion_role_shift}). No included source tests the role of these moderators (Section~\ref{sec:discussion_contexts_tools}).


\subsection{Implications for Practitioners}\label{sec:discussion_implications}

For practitioners, the sources suggest treating vibe coding as an engineering workflow with explicit evaluation practices and not as a shortcut. These are patterns in what the sources report, not evaluated recommendations. The paragraphs below state, for common adoption decisions, what the corpus reports and the level of that evidence. The evidence is evaluated when an included study tests the practice, described when sources report the practice in use, and recommended when sources advise it without evaluation. Some decisions were not systematically coded, and for these the text points to the nearest coded findings.

Which projects suit vibe coding is described in the sources but not evaluated. Prototyping is reported more often than production use, and task-domain evidence is sparse (Section~\ref{sec:discussion_contexts_tools}, Tables~\ref{tab:rq33_audience_task} and~\ref{tab:rq6_contexts}). The coded finding closest to suitability is explicit guidance on when to avoid vibe coding, given by 20 sources (RQ5.2).

For verification, the evidence is mostly recommended practice. Validation and testing pipelines and human-in-the-loop review are the safeguards named most often (Section~\ref{sec:discussion_control_system}), and their Strong reporting support is not a test of their effect. Which checks form a minimum baseline was not systematically coded, and no source tests whether such a baseline is sufficient.

Some safeguards have been evaluated. Table~\ref{tab:safeguard_evaluations} counts 38 evaluated instances from 32 sources, most of them validation pipelines and human review, with beneficial, mixed, and limiting results (Section~\ref{sec:rq5_2}). No source in this record evaluates provenance tracking or guidance on when to avoid vibe coding, and several beneficial results are for a whole tool in which the safeguard is one component, so the effect of a safeguard on its own remains uncertain.

The balance between short-term productivity and the cost of verification and maintenance is described but not tested. Productivity and maintainability are reported together by 32 of the 159 non-background sources contributing to RQ4. Of them, 17 record a positive productivity direction together with a negative maintainability direction. Both outcomes are measured in 6 of the 32. In these 6, productivity is positive in 4, mixed in 1, and recorded as no effect in 1, and maintainability is negative in 5 and positive in 1. Agreement on the outcome fields before reconciliation was moderate (Section~\ref{sec:reliability}), so these six may be read with caution. Short-term productivity has Weak evidence and long-term productivity Moderate evidence, and maintainability has Weak evidence in both horizons (Table~\ref{tab:rq4_outcomes}). No source tests how to balance the two.

Four decisions were not systematically coded, so the review gives no rule for them. They are when a prototype may be promoted to production, governance by the risk level of the software, the allocation of responsibility for generated code among developers, reviewers, managers, and tool providers, and different requirements for professional and non-professional developers. The nearest coded findings are guidance on when to avoid vibe coding and human-in-the-loop review (RQ5.2), use in production (RQ6), and governance, disclosure, and organizational integration as an open challenge (12 sources, RQ8). No included source evaluates a promotion rule, a governance scheme by risk level, or different requirements for the two groups.

Context-window or state-retention failures (RQ3.2) or cumulative context errors (RQ5.1) are reported by 58 non-background sources. The workflow codes closest to a response are explicit context provisioning and retrieval and documentation grounding (RQ2), and 34 of the 58 also report context provisioning. Whether these practices reduce breakdowns has not been tested.

\subsection{Implications for Research}\label{sec:discussion_research}

For research, the priorities are set out in Section~\ref{sec:rq8}. This points toward studying vibe coding as an end-to-end workflow, with verification effort, context management, trust calibration, and long-term maintainability measured alongside correctness and productivity.

\section{Threats to validity} \label{sec:validity}

This section discusses threats to validity following guidance for secondary studies \cite{kitchenham2007guidelines} and multivocal reviews \cite{garousi2019guidelines}.

\subsection{Study search}\label{sec:threat_search}

The search string may not capture all relevant work. The first database search paired vibe-coding terms with a limited set of software-development terms and excluded broader AI and LLM programming terms (Section~\ref{sec:search_string_a_b}), which favors precision over recall, so work that uses only broader terminology may have been missed. The strings also differ between the searches (Section~\ref{sec:update2026}). Window~1 holds sources found by both sets of strings, and window~2 only sources found by the strings of the second database search, so window-2 work that uses only the four terms of the first database search other than ``vibe coding'' may be missed. The search strings were not validated against a known set of relevant papers, so recall was not measured.

Background records were excluded at full text only in the second database search (Section~\ref{sec:update2026}), so the peer-reviewed background sources, which appear only as reports in the tables, mostly reflect the literature to October 2025. Publications dated after 9 September 2026 were not added, and six grey items whose dates were confirmed in October 2026 fell after the cutoff and were removed (Section~\ref{sec:grey_selection}). For one grey source only the year of publication is established. Web search results and platform rankings change over time, particularly on Google and YouTube. Re-executing the search later may therefore yield a different candidate set, even though the search dates, channels, and selection procedures are documented \cite{garousi2019guidelines}.

\subsection{Study selection}\label{sec:threat_selection}

Study selection involves subjective judgment, particularly for grey literature. Duplicates within and between the two database searches are identified by the matching rule of Section~\ref{sec:wl_selection}, and all counts use that rule. The five grey channels were not logged at the level of pages inspected or session controls. The logged grey search, which covered both windows, has five limitations (Section~\ref{sec:grey_selection}). A verification challenge blocked most of the Google queries run in the browser, so only ten of the 34 logged Google queries returned items. About half of the grey candidates carried a result text beside the title, so more of the other half went to full-content review. LinkedIn and X were searched from a user account, so personalization cannot be excluded. YouTube's popularity order stood in for a view-count order that the interface no longer offers. The supplementary web-search channel ranks results by its own relevance and offers no session controls, so its 40 queries may not return the same items if repeated. Explicit criteria (Table~\ref{tab:selection_criteria_white}), staged screening with recorded counts, and the credibility checklist (Table~\ref{tab:qa_grey}) reduce this subjectivity \cite{garousi2019guidelines}, but some remains in the credibility assessment of grey sources.

Items judged not applicable are left out of the quality score, so a paper that is not empirical can reach the High band. Such a paper does not count toward criterion C2, so it does not raise a label (Section~\ref{sec:evidence_strength}).

\subsection{Data extraction}\label{sec:threat_extraction}

Data extraction may introduce error or bias in how evidence is recorded. A structured template organized around RQ1 to RQ8 (Table~\ref{tab:data_extraction}) and retained excerpts linked to Paper\_IDs keep each synthesized claim traceable to its sources, although some interpretive variability remains.

\subsection{Coding and theme classification}\label{sec:threat_coding}

Assigning codes and defining themes involves interpretive judgment. The first author performed the main review steps. For the first database search, a sample of the screening, scoring, and coding decisions was reviewed during the 2025 selection. The size of that sample and the disagreements were not recorded. All of the second author's checks were made in September and October 2026. The second author's screening of the records that the first author excluded from the first database search was therefore made about a year after the first author's screening. Only the two screening checks of the second database search and the logged grey search were made while that screening was in progress. The checks of directness, quality, credibility, coding, evidence sets, C4, and outcome fields were made after the decisions they check. Directness was not assessed when the first database search was screened, and both authors classified those sources in September 2026. The timing of each check is given in the replication package.

Several limitations remain. The second author is a member of the author team and knew the included corpus, so the second author's screening of the records that the first author excluded at the title stage of the first database search was blinded only by interleaving retained records as controls. The second author's screening of the full-text exclusions of the first database search was not blind, because every record was a known full-text exclusion. Agreement before reconciliation was low at several stages (Section~\ref{sec:reliability}, Table~\ref{tab:reliability}), and the selection, classification, and evidence sets carry this uncertainty. The checks of coding, evidence sets, C4, and the outcome fields were made from the first author's extraction cells, not from the full texts. They measure agreement in applying codes and rules to the same extracted text, not agreement in extraction. The check of supporting sets was not blind, since every candidate shown to the second author was in the first author's set. Most checks covered a sample, and reconciliation settled the sampled records only. After reconciliation, four of the excluded records were included. This screening cannot show that no other relevant record was excluded. In the screening checks of the second database search and the logged grey search the first author recorded Unsure for many records and the second author did not, so agreement computed with Unsure set aside overstates agreement (replication package). The second author's check of directness under the final rule covered a 30\% sample (kappa 0.79), so the classes of the other sources rest on the first author's application of the rule (Section~\ref{sec:reliability}). The second author's earlier check of C4 as a judgment on 18 findings (Section~\ref{sec:reliability} and the replication package) does not test the C4 rule used for the labels. Under this rule, C4 is derived from the recorded directions for the outcome findings and is not applicable to the other findings (Section~\ref{sec:evidence_strength}). Three recorded fields were not checked independently and rest on the first author alone: how a source reports the RQ2 and RQ5.2 items, the record of safeguard evaluations (Table~\ref{tab:safeguard_evaluations}), and the dependency links under the final rule. Three grey sources authored by members of the review team were removed so that no author's assessment of their own work enters the analysis. Their removal changes no label.

A documented codebook of 56 codes, each linked to one research question, and traceability from each finding to its sources and excerpts reduce ambiguity, though some subjectivity remains in qualitative coding. Coding was hybrid. Codes were derived deductively from the research questions and added inductively while coding the sources of the first database search and the five grey channels. The codebook was then frozen at 56 codes before the other sources were coded, so that the independent coding pass and the labels rest on one codebook. Content outside the codebook was noted during coding but not coded. It is listed in the replication package and is not reported as findings, so the research questions are answered for the coded categories only. Four topics in these notes may merit future investigation with an extended codebook and the same reliability procedure and evidence-strength matrix: the security risk of generated code, tools outside the codebook, specification-driven and planning practice, and skill erosion or learning effects. Security concerns are coded as an open challenge under RQ8, but the security risk of generated code is not coded as a failure mode under RQ5.1.

\subsection{Synthesis}\label{sec:threat_synthesis}
Outcomes are measured with diverse methods, so the review relies on descriptive mapping and thematic synthesis, not meta-analysis, and vote-counting across studies may overinterpret. Theme boundaries may vary with the analyst, particularly for constructs that are not consistently operationalized, so another team may derive somewhat different themes from the same corpus. The corpus holds 20 secondary sources, among them three of the four reviews compared in Section~\ref{sec:rw}. The dependency rule counts a source and the sources it summarizes once, so that their evidence does not enter a finding twice.

\subsection{External validity}\label{sec:threat_external}
The included sources are concentrated in 2025 and 2026 and frequently discuss a limited set of widely used tools and development environments, so findings may not generalize to future model versions, other tool ecosystems, or less-represented contexts such as safety-critical or regulated domains. Task-domain coverage is uneven (Section~\ref{sec:rq3}). Model capabilities, context limits, agentic toolchains, and IDE integrations change quickly, so conclusions drawn from a snapshot that closes in September 2026 may become partially outdated.

The review remains narrower than broader narrative reviews of AI-assisted software development because of its focused scope and selection protocol. The low snowballing yield (2 peer-reviewed and 3 grey additions) may reflect a sparsely interconnected literature. Relevant sources, particularly grey items not indexed by our channels, may still have been missed, and later updates of the review may run the search protocol again and expand it.

\subsection{Construct validity}\label{sec:threat_construct}
``Vibe coding'' is a recent practitioner-driven term whose meaning varies across sources. Some describe intent-driven prompt programming, others agentic development environments, and others AI-assisted development in general, which risks construct drift in synthesis. The search included related terminology (prompt-based development, conversational programming, AI-assisted development), definitions and conceptual framings were coded under RQ1, and the two streams were analyzed separately during parts of the synthesis. Socio-technical constructs such as trust calibration, cognitive load, and cognitive debt are not consistently defined or measured across primary sources, so we reported where evidence was sparse or conceptually heterogeneous (e.g., session-level dynamics in RQ3.2). Every source is classified as direct, adjacent, or background to the operational definition (Section~\ref{sec:opdef}), and every label is recomputed on the direct sources alone (Section~\ref{sec:sensitivity}). The directness class records the subject of a source, not the kind of evidence it gives, so the recomputation on the direct sources tests whether a finding rests on sources about the practice, not whether it rests on empirical evidence (Section~\ref{sec:opdef}). On a 30\% sample, agreement on the class under the final rule was substantial (kappa 0.79, Section~\ref{sec:reliability}). The limits of the directness checks are stated in Section~\ref{sec:threat_coding}.

\subsection{Conclusion validity}\label{sec:threat_conclusion}
Outcomes reported in RQ4, including productivity, quality, and maintainability, rest on heterogeneous evidence, so we read outcome claims with the evidence stream and the consistency across sources in view. Counts and percentages are the frequency with which concepts appear in the reviewed sources, not estimates of prevalence, effect size, or practical impact. A practice reported by many sources is therefore not shown to be effective. The labels of Section~\ref{sec:evidence_strength} depend in the same way on the criteria and thresholds adopted there, and they would change under other criteria. They describe the state of the reporting in the reviewed sources and are not conclusions about the effectiveness of vibe coding. Publication and availability biases may favor positive or novel findings in both streams, and grey literature may be promotional, anecdotal, or tool-driven, which may overrepresent early-adopter experience.

Three limitations arise from the searches of September 2026. First, 27 of the 284 records at the full-text assessment of the second database search were excluded under E5 because the institutional subscription did not cover them, without an interlibrary-loan request (Section~\ref{sec:wl_selection}), and 92 grey items whose full content could not be obtained were excluded on a similar ground (Section~\ref{sec:grey_selection}). Second, the grey eligibility rule of the logged grey search is stricter than that of the five grey channels, since an item has to promise usable evidence. The grey sources from the two grey searches are therefore composed differently, and window~1 holds sources from both. Counts that mix them should be read with that in mind. Third, 190 of the 266 sources are grey, so the corpus is weighted toward practitioner material. The 73 restricted grey sources, 37 from window~1 and 36 from window~2, may support definitions, practitioner discourse, and tool visibility only, never an outcome or safeguard-effectiveness claim. The dependency rule of Section~\ref{sec:evidence_strength} was applied by the first author alone and was not checked by a second person, and without any merging no label changes (Section~\ref{sec:sensitivity}). Every finding is also computed for each publication window alone, and window~1 alone supports fewer findings at the same level (Section~\ref{sec:sensitivity}).

The outcome fields of RQ4 add a limitation. The labels use the direction, measurement, and horizon recorded for each outcome a source reports (Section~\ref{sec:extraction}), and the first author recorded them. The second author's blind check of all 332 pairs gave moderate agreement before reconciliation (Section~\ref{sec:reliability}). The two authors reconciled every disagreement under the written rule from the extraction cells, so the reconciled values rest on their reading of those cells, not of the full texts, and the counts of measured, self-reported, and asserted reports in Table~\ref{tab:rq4_outcomes} may be best read with this in mind. The split between No effect and Not stated changes one label, long-term quality, and the three Moderate outcome rows rest on few comparable sources, so one changed direction would alter each of them (Section~\ref{sec:sensitivity}). Under the C4 rule, one comparable source that reports the opposite direction makes a row contested (Section~\ref{sec:evidence_strength}). Four of the nine outcome rows are contested and two are not assessable, and all six are labeled Weak evidence, although in some contested rows most comparable sources report one direction (Table~\ref{tab:rq4_outcomes}). If a row were contested only when the smaller of the positive and negative counts is at least 20\% of the directions, three of the four contested rows would be Moderate evidence (Section~\ref{sec:sensitivity}).

Three properties of the recorded fields limit what the labels and counts can show. For a presence finding C4 is not applicable, so its label is one of reporting support. Strong reporting support shows how widely and by which sources the phenomenon is reported. It does not show that a reported practice or safeguard works. Nor does it show agreement about an effect, because the codebook has no code for contradicting a phenomenon (Section~\ref{sec:evidence_strength}). The RQ4 direction is recorded per outcome without the size of the effect, so a gain together with a loss on the same outcome appears as mixed. The reporting mode of RQ2 and RQ5.2 is recorded for each source and item from the extraction cells of the question and, for safeguards, the outcome cells, so an evaluation reported elsewhere in a source is not captured (Section~\ref{sec:extraction}). Two further limits apply. No label is computed on peer-reviewed sources alone, because every subset of the sensitivity analysis keeps grey sources (Section~\ref{sec:sensitivity}). The corpus is described by outlet category (Table~\ref{tab:corpus_profile}) but not summarized by venues.

\section{Related Work}
\label{sec:rw}

Four prior reviews address vibe coding, each with its own scope and method.

Ge~et~al.~\cite{ge2025survey} give a formal, infrastructure-oriented treatment of vibe coding. They model the practice as a constrained Markov decision process over a human-project-agent triad, and they derive five development models: unconstrained automation, iterative conversational collaboration, planning-driven, test-driven, and context-enhanced. The review covers code LLMs, agent architectures, execution environments, and feedback mechanisms in depth, describes itself as the first ``comprehensive and systematic'' review of the topic, and reports drawing on over 1{,}000 papers. However, it documents no search string, no inclusion or exclusion criteria, no quality appraisal, and no replication package, so its selection process is difficult to reproduce independently. It does not incorporate grey literature under an appraisal protocol, and compared with this MLR it gives limited structured attention to empirical outcomes, usage contexts, practitioner experience, and developer human factors.

Ray~\cite{ray2025review} offers the broadest narrative survey, drawing on around 245 references. It maps over twenty tools across five categories: browser-based environments, IDEs and editors, plugins and extensions, command-line agents, and task-management systems. It proposes a taxonomy of four interaction modalities, namely full delegation, guided delegation, active pairing, and expert consultation, placed on a delegation-pairing continuum, and it identifies twelve critical challenges and fourteen research directions. However, it documents no systematic protocol, no search string, no inclusion or exclusion criteria, no quality appraisal, and no replication package, and it blends peer-reviewed and grey sources without credibility weighting. Compared with this MLR, it gives limited structured attention to developer role and cognitive aspects and offers no synthesis of empirical outcomes.

Fawzy~et~al.~\cite{fawzy2025vibe} conduct a systematic Grey Literature Review~(GLR) of vibe coding. They analyze 101 practitioner sources against a 0 to 15 credibility rubric, extract 518 firsthand behavioral units through thematic analysis, and archive a replication package. Their findings report a speed-quality trade-off: 62\% of practitioners are motivated by speed, 68\% perceive the resulting code as fast but flawed, and 36\% skip quality assurance. Compared with this MLR, the review is scoped to practitioner grey literature with no peer-reviewed evidence, offers no tool taxonomy and no formal risk, safeguard, or governance analysis, measures outcomes through perceptions only, and treats developer-role change narratively. It addresses four research questions, whereas this MLR addresses eight. The review was published in the ICSE-SEIP 2026 proceedings after October 2025.

Michels~et~al.~\cite{michels2026vibe} present a state-of-the-art review of vibe coding organized around practice, performance, productivity, and risk, and state that the review is not a systematic one. The corpus was assembled by following tech journalism, vendor and practitioner publications, and social-media discussion, together with coverage of the scientific literature. The review cites 123 sources, of which 15 are peer-reviewed papers and 13 are preprints. It organizes sources by trustworthiness tier, archives ephemeral sources, inventories models and tools by form factor, summarizes benchmark results, decomposes capability by task type, tabulates productivity claims from 2022 to 2026, documents security incidents, code-quality telemetry, copyright exposure, and skill atrophy, and closes with four open research questions. Compared with this MLR, it documents no search string, no inclusion or exclusion criteria, and no replication package, so its source selection is difficult to reproduce. It treats the conceptualization of vibe coding, workflow patterns and tactics, and usage contexts only in passing, and it does not compare findings between the peer-reviewed and grey evidence streams.

This review grades preprints as grey literature in its own corpus. Three of the four reviews are preprints: Ge~et~al.\ (arXiv 2510.12399), Ray (TechRxiv), and Michels~et~al.\ (arXiv 2608.20446). Fawzy~et~al.\ is the one peer-reviewed review of the four, and the version our search retrieved is its preprint. Three of the four are also sources of this corpus, namely Ray~\ref{GL1}, Fawzy~et~al.~\ref{GL10}, and Michels~et~al.~\ref{GL30}, each classified as grey, secondary, and Tier~1, and recorded in the dependency field so that the sources they summarize are not counted twice (Section~\ref{sec:evidence_strength}). Ge~et~al.\ is not in the corpus. The arXiv channel of the logged grey search returned it (item GLR-0412 in the replication package), and it was read in full and excluded at content review as a background review with no evidence of its own.

The four reviews differ in method, evidence sources, and focus. Only Fawzy~et~al.\ apply a documented protocol and a public replication package, and only to grey literature. None of the four combines peer-reviewed and grey literature under one documented systematic selection protocol. This MLR addresses that gap. It integrates both evidence streams under a single protocol with a documented search string, inclusion and exclusion criteria, quality and grey-credibility tiers, and a replication package, and applies it across eight research questions so that findings can be weighted by evidence quality.

\definecolor{mlrrow}{RGB}{232, 245, 255}
\definecolor{mlrrowgrey}{RGB}{240,240,240}

Table~\ref{tab:rq_coverage_comparison} compares the reviews and links each coverage row to the section of this paper that documents the finding. In the \emph{Review type} row, NR stands for narrative review and SoTA for state-of-the-art review, the label Michels~et~al.\ use for their own review.

\begin{table}[!ht]
\centering
\caption{Comparison of this MLR with existing vibe coding reviews. The top two rows summarize corpus size and review type. The remaining rows summarize per-area coverage. \checkmark{} means addressed substantively, $\triangle$ means partially addressed or addressed only narratively, none means not addressed, and not documented means that no selection process is documented. Ge et al.\ reports drawing on over 1{,}000 papers, Ray cites approximately 245 references, and Michels et al.\ 123, none of them the product of a documented selection protocol.}
\label{tab:rq_coverage_comparison}
\small
\renewcommand{\arraystretch}{1.25}
\resizebox{\columnwidth}{!}{%
\begin{tabular}{|p{4.2cm}|p{1.3cm}|p{1.3cm}|p{1.4cm}|p{1.5cm}|p{1.5cm}|}
\hline
\rowcolor{mlrrowgrey}
\textbf{Contributions / Themes} & \textbf{Ge et al.\ \cite{ge2025survey}} & \textbf{Ray \cite{ray2025review}} & \textbf{Fawzy et al.\ \cite{fawzy2025vibe}} & \textbf{Michels et al.\ \cite{michels2026vibe}} & \textbf{Our MLR} \\
\hline
Number of selected sources & not documented & not documented & 101 & not documented & 266 \\
\hline
Review type & NR & NR & GLR & SoTA & MLR \\
\hline
Conceptualization of vibe coding (Section~\ref{sec:rq1}) & $\triangle$ & $\triangle$ & $\triangle$ & $\triangle$ & \checkmark \\
\hline
Workflow patterns and tactics (Section~\ref{sec:rq2}) & \checkmark & \checkmark & \checkmark & $\triangle$ & \checkmark \\
\hline
Developer role shift and cognitive aspects (Section~\ref{sec:rq3}) & none & none & $\triangle$ & \checkmark & \checkmark \\
\hline
Reported outcomes: productivity, quality, defects, maintainability, reproducibility (Section~\ref{sec:rq4}) & none & $\triangle$ & \checkmark & \checkmark & \checkmark \\
\hline
Failure modes (Section~\ref{sec:rq5_1}) & $\triangle$ & $\triangle$ & $\triangle$ & \checkmark & \checkmark \\
\hline
Safeguards and governance (Section~\ref{sec:rq5_2}) & $\triangle$ & $\triangle$ & $\triangle$ & \checkmark & \checkmark \\
\hline
Usage contexts and operational constraints (Section~\ref{sec:rq6}) & none & $\triangle$ & $\triangle$ & $\triangle$ & \checkmark \\
\hline
Tool ecosystem and platform taxonomy (Section~\ref{sec:rq7}) & none & \checkmark & none & \checkmark & \checkmark \\
\hline
Open challenges and future research directions (Section~\ref{sec:rq8}) & $\triangle$ & \checkmark & $\triangle$ & \checkmark & \checkmark \\
\hline
\end{tabular}%
}
\end{table}

Table~\ref{tab:rq_coverage_comparison} shows three patterns. First, among the reviews compared here, only this MLR draws its corpus from a documented systematic selection process applied to both streams. Second, \emph{workflow patterns and tactics} is the area most often addressed substantively by prior work. Ge~et~al., Ray, and Fawzy~et~al.\ cover it, and Michels~et~al.\ touch it through the history of tool form factors. Third, \emph{developer role shift and cognitive aspects} is not addressed by Ge~et~al.\ or Ray. Fawzy~et~al.\ touch it through practitioner accounts, and Michels~et~al.\ cover its cognitive side through evidence on skill atrophy and code comprehension. No prior review substantively covers the full set of areas this MLR synthesizes, which motivates the cross-stream integration adopted here.

\FloatBarrier

\section{Conclusion} \label{sec:conclusion}

This MLR synthesized 266 sources, 76 peer-reviewed and 190 grey, published from January 2022 to 9~September 2026 and found by searches in October 2025 and September 2026. Of these, 146 are direct sources on vibe coding as defined in Section~\ref{sec:opdef}, 93 are adjacent, and 27 are background. Directness records the subject of a source, not its evidence, and 58 of the 146 direct sources are empirical studies or tool evaluations. On a 30\% sample, the second author's agreement on these classes was substantial (kappa 0.79). When only the direct sources are counted, 40 of the 60 labels hold.

Of the 60 findings, 42 are labeled Strong, 11 Moderate, and 7 Weak. For the 51 presence findings, the label is reporting support (42 Strong, 8 Moderate, 1 Weak). It shows how widely, and by which sources, a phenomenon is reported, not whether a practice works. The 9 outcome findings carry evidence-strength labels (3 Moderate, 6 Weak). All labels depend on the criteria and thresholds of Section~\ref{sec:evidence_strength}, and they describe the reporting in the corpus, not the effectiveness of vibe coding.

Findings with Strong reporting support include the following. The sources describe vibe coding as an intent-driven and iterative mode in which natural language steers generation, evaluation, and revision (RQ1, RQ2). Developer work is reported to shift from authoring toward directing, supervising, and validating generated output, with trust calibration and cognitive load as recurring themes (RQ3.1). Repeated prompt cycles without convergence, context loss, and momentum breaks are reported as session-level failures (RQ3.2). Incorrect generations are the most frequently reported failure mode (RQ5.1). Human-in-the-loop review and validation pipelines are the most frequently reported safeguards, and 20 sources give guidance on when to avoid the practice (RQ5.2). Use is reported in prototyping and in production (RQ6), and the lack of standardized evaluation, trust, context management, security, and governance are named as open challenges (RQ8).

Cumulative context errors, prompt brittleness, patch-and-repair workflows, and the visibility of individual tools rest mainly on grey literature (RQ5.1, RQ2, RQ7). Two tool findings, and the description of vibe coding as a practice of feel, flow, and intuition (RQ1), have Moderate reporting support because a large share of their sources is restricted. Licensing and provenance uncertainty also has Moderate reporting support.

The outcome findings are contested or thin. Most comparable sources report short-term productivity gains (34 of 55 with a direction) and long-term maintainability costs (15 of 19). Both rows are Weak because the written rule marks a row contested when any comparable source reports the opposite direction, and both would be Moderate under a 20\% minority threshold. Short-term quality and short-term defects are also contested. Long-term productivity, long-term quality, and short-term reproducibility are Moderate. Short-term maintainability and long-term defects are Weak because their direction cannot be assessed. Agreement on the outcome fields before reconciliation was moderate for all three fields, so these labels may be best read as provisional.

Evaluation and validation practices are reported by 140 of the 159 sources that count toward RQ4, so their frequent co-occurrence with productivity gains may only reflect how common they are, and no included source tests that link. Empirical evaluations of safeguards are scarce. Prototyping is reported more often than production use, and task-domain evidence is sparse (RQ6, RQ3.3). For practitioners, the review offers a map of what the sources report and how strong that evidence is, not validated rules. The conclusions cover the coded categories of the codebook only.

Future work may study session-level dynamics, the effect of safeguards, task domains, operational definitions, and long-term maintainability. It may also extend the codebook to the four topics that were noted during coding but not coded.

\section*{Acknowledgment}
This work has been supported by FAST, the Finnish Software Engineering Doctoral Research Network, funded by the Ministry of Education and Culture, Finland.

\section*{Data availability} \label{sec:data_availability}
To support replication, we released a replication package and the package is publicly available \cite{siddeeq_2026_23963987}.

\section*{Declaration of AI Assistance}
During the preparation of this manuscript, the authors used ChatGPT to assist with grammar refinement, sentence restructuring, and formatting improvements. Following the
use of this tool, the authors carefully reviewed and revised the content and assume full responsibility for the final version of the publication.

\section*{Window 1: Selected Peer-Reviewed Literature}
\begin{enumerate}[label={[WL00]}, leftmargin=*]
    \srcitem{WL1}{WL1} Young, M., Nan, Z., \& Shen, X. (2022, May). IDE augmented with human-learning inspired natural language programming. In Proceedings of the ACM/IEEE 44th International Conference on Software Engineering: Companion Proceedings (pp. 110-114).
    \srcitem{WL2}{WL2} Yang, S., \& Yang, Y. (2024, May). FormalEval: A method for automatic evaluation of code generation via large language models. In 2024 2nd International Symposium of Electronics Design Automation (ISEDA) (pp. 660-665). IEEE.
    \srcitem{WL3}{WL3} Wang, J., \& Chen, Y. (2023, November). A review on code generation with LLMs: Application and evaluation. In 2023 IEEE International Conference on Medical Artificial Intelligence (MedAI) (pp. 284-289). IEEE.
    \srcitem{WL4}{WL4} Liang, J. T., Yang, C., \& Myers, B. A. (2024). A large-scale survey on the usability of AI programming assistants: Successes and challenges. In 2024 IEEE/ACM 46th International Conference on Software Engineering (ICSE). IEEE Computer Society, Los Alamitos, CA, USA, 605-617.
    \srcitem{WL5}{WL5} Sabouri, S., Eibl, P., Zhou, X., Ziyadi, M., Medvidovic, N., Lindemann, L., \& Chattopadhyay, S. (2025, March). Trust dynamics in AI-assisted development: Definitions, factors, and implications. In 2025 IEEE/ACM 47th International Conference on Software Engineering (ICSE) (pp. 736-736). IEEE Computer Society.
    \srcitem{WL6}{WL6} Ramler, R., Moser, M., Fischer, L., Nissl, M., \& Heinzl, R. (2024, April). Industrial experience report on AI-assisted coding in professional software development. In Proceedings of the 1st International Workshop on Large Language Models for Code (pp. 1-7).
    \srcitem{WL7}{WL7} Kumar, Y., Akinwunmi, I., \& Kruger, D. (2025, March). Evaluating the Advantage of an AI-Native IDE Cursor on Programmer Performance. In 2025 IEEE Integrated STEM Education Conference (ISEC) (pp. 1-8). IEEE.
    \srcitem{WL8}{WL8} Treude, C., \& Gerosa, M. A. (2025, April). How developers interact with AI: A taxonomy of human-AI collaboration in software engineering. In 2025 IEEE/ACM Second International Conference on AI Foundation Models and Software Engineering (Forge) (pp. 236-240). IEEE.
    \srcitem{WL9}{WL9} Jalil, S. (2025, August). The transformative influence of LLMs on software development \& developer productivity. In 2025 International Conference on Artificial Intelligence, Computer, Data Sciences and Applications (ACDSA) (pp. 1-10). IEEE.
    \srcitem{WL10}{WL10} Majdinasab, V., Bishop, M. J., Rasheed, S., Moradidakhel, A., Tahir, A., \& Khomh, F. (2024, March). Assessing the security of GitHub Copilot's generated code-a targeted replication study. In 2024 IEEE International Conference on Software Analysis, Evolution and Reengineering (SANER) (pp. 435-444). IEEE.
    \srcitem{WL11}{WL11} Feng, L., Yen, R., You, Y., Fan, M., Zhao, J., \& Lu, Z. (2024, May). Coprompt: Supporting prompt sharing and referring in collaborative natural language programming. In Proceedings of the 2024 CHI conference on human factors in computing systems (pp. 1-21).
    \srcitem{WL12}{WL12} Xu, F. F., Vasilescu, B., \& Neubig, G. (2022). In-ide code generation from natural language: Promise and challenges. ACM Transactions on Software Engineering and Methodology (TOSEM), 31(2), 1-47.
    \srcitem{WL13}{WL13} Ross, S. I., Martinez, F., Houde, S., Muller, M., \& Weisz, J. D. (2023, March). The programmer's assistant: Conversational interaction with a large language model for software development. In Proceedings of the 28th international conference on intelligent user interfaces (pp. 491-514).
    \srcitem{WL14}{WL14} Cox, D., Murray, J., \& Salter, A. (2025, April). Routine, twisty, and queer: pasts and futures of games programming pedagogy with no and low code tools. In Proceedings of the 20th International Conference on the Foundations of Digital Games (pp. 1-8).
    \srcitem{WL15}{WL15} Mitchell, J., \& Shaaban, Y. (2025, October). Position: Vibe coding needs vibe reasoning: improving vibe coding with formal verification. In Proceedings of the 1st ACM SIGPLAN International Workshop on Language Models and Programming Languages (pp. 84-90).
    \srcitem{WL16}{WL16} Klemmer, J. H., Horstmann, S. A., Patnaik, N., Ludden, C., Burton Jr, C., Powers, C., ... \& Fahl, S. (2024, December). Using AI assistants in software development: A qualitative study on security practices and concerns. In Proceedings of the 2024 on ACM SIGSAC Conference on Computer and Communications Security (pp. 2726-2740).
    \srcitem{WL17}{WL17} Donato, B., Mariani, L., Micucci, D., Riganelli, O., \& Somaschini, M. (2025, June). Multimind: A plug-in for the implementation of development tasks aided by ai assistants. In Proceedings of the 33rd ACM International Conference on the Foundations of Software Engineering (pp. 1310-1317).
    \srcitem{WL18}{WL18} Johnson, B., Bird, C., Ford, D., Al Haque, E., Forsgren, N., \& Zimmermann, T. (2025, October). Facilitating Trust in AI-assisted Software Tools. ACM Transactions on Software Engineering and Methodology.
    \srcitem{WL19}{WL19} Flores-Saviaga, C., Hanrahan, B. V., Imteyaz, K., Clarke, S., \& Savage, S. (2025, April). The impact of generative AI coding assistants on developers who are visually impaired. In Proceedings of the 2025 CHI Conference on Human Factors in Computing Systems (pp. 1-17).
    \srcitem{WL20}{WL20} Moore, J. H., \& Tatonetti, N. (2025, July). Vibe coding: a new paradigm for biomedical software development. BioData Mining, 18(1), 46.
    \srcitem{WL21}{WL21} Lyu, M. R., Ray, B., Roychoudhury, A., Tan, S. H., \& Thongtanunam, P. (2025, May). Automatic programming: Large language models and beyond. ACM Transactions on Software Engineering and Methodology, 34(5), 1-33.
    \srcitem{WL22}{WL22} Sergeyuk, A., Golubev, Y., Bryksin, T., \& Ahmed, I. (2025). Using AI-based coding assistants in practice: State of affairs, perceptions, and ways forward. \emph{Information and Software Technology}, 178, 107610. \url{https://doi.org/10.1016/j.infsof.2024.107610}
    \srcitem{WL23}{WL23} Cheng, R., Wang, R., Zimmermann, T., \& Ford, D. (2024). ``It would work for me too'': How online communities shape software developers' trust in AI-powered code generation tools. ACM Transactions on Interactive Intelligent Systems, 14(2), 1-39.
    \srcitem{WL24}{WL24} Malamas, N., Tsardoulias, E., Panayiotou, K., \& Symeonidis, A. L. (2025). Toward efficient vibe coding: An LLM-based agent for low-code software development. Journal of Computer Languages, 101367.
    \srcitem{WL25}{WL25} De Silva, D., Mills, N., Issadeen, Z., Moraliyage, H., Jennings, A., \& Manic, M. (2025, June). Generative AI Vibe Coding for Prototyping Industrial Systems. In 2025 IEEE 34th International Symposium on Industrial Electronics (ISIE) (pp. 1-6). IEEE.
    \srcitem{WL26}{WL26} Beheshti, A. (2024, July). Natural language-oriented programming (NLOP): Towards democratizing software creation. In 2024 IEEE International Conference on Software Services Engineering (SSE) (pp. 258-267). IEEE.
    \srcitem{WL27}{WL27} Meske, C., Hermanns, T., Von der Weiden, E., Loser, K. U., \& Berger, T. (2025). Vibe coding as a reconfiguration of intent mediation in software development: Definition, implications, and research agenda. IEEE Access, 13, 213242-213259.
    \srcitem{WL28}{WL29} Zaytsev, V. (2025). Compilation of feature models by expert vibing. In \emph{Joint Proceedings of the STAF 2025 Workshops (CEUR Workshop Proceedings, Vol. 4122)}. \url{https://ceur-ws.org/Vol-4122/paper22.pdf}
    \srcitem{WL29}{WL30} Cabot, J. (2025). Exploring the use of large language models in domain-specific language development: an experience report. In \emph{Joint Proceedings of the STAF 2025 Workshops (CEUR Workshop Proceedings, Vol. 4122)}. \url{https://ceur-ws.org/Vol-4122/paper6.pdf}
    \srcitem{WL30}{WL31} Fortes-Ferreira, M., Alam, M. S., \& Bazilinskyy, P. (2025, September). Vibe Coding in Practice: Building a Driving Simulator Without Expert Programming Skills. In \emph{Adjunct Proceedings of the 17th International Conference on Automotive User Interfaces and Interactive Vehicular Applications} (pp. 60-66). \url{https://doi.org/10.1145/3744335.3758482}
    \srcitem{WL31}{WL74} Huang, D., Nan, Z., Hu, X., Jin, P., Peng, S., Wen, Y., Zhang, R., Du, Z., Guo, Q., Pu, Y., \& Chen, Y. (2023). ANPL: Towards Natural Programming with Interactive Decomposition. \emph{Advances in Neural Information Processing Systems 36}, 69404-69440. \url{https://doi.org/10.52202/075280-3040}
    \srcitem{WL32}{WL75} Borg, M., Bjarnason, E., \& Hedin, F. (2025). Vibe Coding and the New Prototyping Playbook. \emph{IEEE Software}, 42(6), 12-16. \url{https://doi.org/10.1109/ms.2025.3597032}
    \srcitem{WL33}{WL76} Feldman, M. Q., \& Anderson, C. J. (2024, June). Non-Expert Programmers in the Generative AI Future. In \emph{Proceedings of the 3rd Annual Meeting of the Symposium on Human-Computer Interaction for Work} (pp. 1-19). \url{https://doi.org/10.1145/3663384.3663393}
    \srcitem{WL34}{WL77} Thiede, C., Taeumel, M., Böhme, L., \& Hirschfeld, R. (2024, October). Talking to Objects in Natural Language: Toward Semantic Tools for Exploratory Programming. In \emph{Proceedings of the 2024 ACM SIGPLAN International Symposium on New Ideas, New Paradigms, and Reflections on Programming and Software} (pp. 68-84). \url{https://doi.org/10.1145/3689492.3690049}
\end{enumerate}

\section*{Window 1: Selected Grey Literature}
\begin{enumerate}[label={[GL000]}, leftmargin=*]
    \srcitem{GL1}{GL1} Ray, P. P. (2025). \emph{A review on vibe coding: Fundamentals, state-of-the-art, challenges and future directions} [Preprint]. TechRxiv. \url{https://doi.org/10.36227/techrxiv.174681482.27435614/v1}
    \srcitem{GL2}{GL2} Maes, S. H. (2025). \emph{The gotchas of AI coding and vibe coding. It's all about support and maintenance} [Preprint]. Zenodo. \url{https://doi.org/10.5281/zenodo.15343349}
    \srcitem{GL3}{GL3} Horvat, M. (2025). \emph{What is Vibe coding and when should you use it (or not)?} [Preprint]. TechRxiv. \url{https://doi.org/10.36227/techrxiv.175459780.03758839/v1}
    \srcitem{GL4}{GL4} Karpathy, A. (2025, February). There's a new kind of coding I call ``vibe coding''\ldots [Post]. X. \url{https://x.com/karpathy/status/1886192184808149383?lang=en}
    \srcitem{GL5}{GL5} Gaspar, N. (2025, May 11). \emph{The state of vibe coding tools (May 2025).} LinkedIn. \url{https://www.linkedin.com/pulse/state-vibe-coding-tools-may-2025-nufar-gaspar-x1znf/}
    \srcitem{GL6}{GL6} Huang, T. (2025, May 19). \emph{Vibe coding fundamentals in 33 minutes} [Video]. YouTube. \url{https://youtu.be/iLCDSY2XX7E}
    \srcitem{GL7}{GL7} Sarkar, A., \& Drosos, I. (2025). \emph{Vibe coding: programming through conversation with artificial intelligence} (arXiv:2506.23253) [Preprint]. arXiv. \url{https://arxiv.org/abs/2506.23253}
    \srcitem{GL8}{GL8} Sapkota, R., Roumeliotis, K. I., \& Karkee, M. (2025). \emph{Vibe coding vs. agentic coding: Fundamentals and practical implications of agentic AI} (arXiv:2505.19443) [Preprint]. arXiv. \url{https://arxiv.org/abs/2505.19443}
    \srcitem{GL9}{GL9} Gómez, O. S. (2025). \emph{The Twilight of Software Engineering in the Age of Vibe Coding} [Preprint]. ResearchGate. \url{https://www.researchgate.net/publication/393638047}
    \srcitem{GL10}{GL10} Fawzy, A., Tahir, A., \& Blincoe, K. (2026). Vibe Coding in Practice: Motivations, Challenges, and a Future Outlook -- a Grey Literature Review. In Proceedings of the IEEE/ACM 48th International Conference on Software Engineering: Software Engineering in Practice (ICSE-SEIP '26) (pp. 212-223). ACM. \url{https://doi.org/10.1145/3786583.3786866}
    \srcitem{GL11}{GL11} Google Cloud. (2025, April). \emph{Vibe coding explained: Tools and guides.} \url{https://cloud.google.com/discover/what-is-vibe-coding}
    \srcitem{GL12}{GL12} Samsyudin, I. (2025). \emph{Vibe Coding and AI-Led Conversational Programming: Emerging Trends in Software Development} [Preprint]. SSRN. \url{https://doi.org/10.2139/ssrn.5469367}
    \srcitem{GL13}{GL13} Edwards, B. (2025, March 6). \emph{Will the future of software development run on vibes?} Ars Technica. \url{https://arstechnica.com/ai/2025/03/is-vibe-coding-with-ai-gnarly-or-reckless-maybe-some-of-both/}
    \srcitem{GL14}{GL14} Geng, F., Shah, A., Li, H., Mulla, N., Swanson, S., Soosai Raj, G., Zingaro, D., \& Porter, L. (2026). Exploring Student-AI Interactions in Vibe Coding. In Proceedings of the 28th Australasian Computing Education Conference (ACE 2026) (pp. 45-54). ACM. \url{https://doi.org/10.1145/3786228.3786236}
    \srcitem{GL15}{GL15} van Hurne, M. H. (2025). \emph{A Technical Debt-Aware Prompting Framework for Sustainable Vibe Coding: Addressing the Production Readiness Crisis in AI-Assisted Software Development} [Preprint]. TechRxiv. \url{https://doi.org/10.36227/techrxiv.175459417.76916566/v1}
    \srcitem{GL16}{GL16} Crowson, M. G., \& Celi, L. C. A. (2025). \emph{Academic vibe coding: Opportunities for accelerating research in an era of resource constraint} (arXiv:2508.00952) [Preprint]. arXiv. \url{https://arxiv.org/abs/2508.00952}
    \srcitem{GL17}{GL17} Bollikonda, T., \& Kovi, M. (2025). \emph{Intent-Driven Code Synthesis: Redefining Software Development with Transformers} [Preprint]. Preprints.org. \url{https://doi.org/10.20944/preprints202504.1732.v1}
    \srcitem{GL18}{GL18} Harkar, S. (2025). \emph{What is vibe coding?} IBM. \url{https://www.ibm.com/think/topics/vibe-coding}
    \srcitem{GL19}{GL19} Chowdhury, H., \& Mann, J. (2025, February 13). Silicon Valley's next act: Bringing ``vibe coding'' to the world. \emph{Business Insider}. \url{https://www.businessinsider.com/vibe-coding-ai-silicon-valley-andrej-karpathy-2025-2}
    \srcitem{GL20}{GL21} Wang, H., Gong, J., Zhang, H., Xu, J., \& Wang, Z. (2025, August 15). \emph{AI Agentic Programming: A Survey of Techniques, Challenges, and Opportunities} (arXiv:2508.11126) [Preprint]. arXiv. \url{https://arxiv.org/abs/2508.11126}
    \srcitem{GL21}{GL22} Chang, H.-F., Shirazi, M., Cao, L., \& Mobasser, S. K. (2025, December 30). \emph{Coding With AI: From a Reflection on Industrial Practices to Future Computer Science and Software Engineering Education} (arXiv:2512.23982) [Preprint]. arXiv. \url{https://arxiv.org/abs/2512.23982}
    \srcitem{GL22}{GL32} Huang, K. (2025, April 9). \emph{Secure vibe coding guide}. Cloud Security Alliance. \url{https://cloudsecurityalliance.org/blog/2025/04/09/secure-vibe-coding-guide}
    \srcitem{GL23}{GL33} Miller, S. (2025, September 15). \emph{Vibe Coding Examples: 15 Projects I Actually Built with AI}. \url{https://www.stephanmiller.com/the-great-vibe-coding-experiment/}
    \srcitem{GL24}{GL34} Pimenova, V., Fakhoury, S., Bird, C., Storey, M.-A., \& Endres, M. (2025, September 15). \emph{Good Vibrations? A Qualitative Study of Co-Creation, Communication, Flow, and Trust in Vibe Coding} (arXiv:2509.12491) [Preprint]. arXiv. \url{https://arxiv.org/abs/2509.12491}
    \srcitem{GL25}{GL35} Chen, N., Qiu, L. K., Wang, A. Z., Wang, Z., \& Yang, Y. (2025, June 16). \emph{Screen Reader Programmers in the Vibe Coding Era: Adaptation, Empowerment, and New Accessibility Landscape} (arXiv:2506.13270) [Preprint]. arXiv. \url{https://arxiv.org/abs/2506.13270}
    \srcitem{GL26}{GL37} Ushio, T. (2025, July 19). \emph{How I Use Vibe Coding in Production Code}. Medium. \url{https://tsuyoshiushio.medium.com/how-i-use-vibe-coding-in-production-code-f0786981ea82}
    \srcitem{GL27}{GL38} Fanny (2025, August 18). \emph{I Tested 5 Vibe Coding Tools So You Don't Have To}. UX Planet. \url{https://uxplanet.org/i-tested-5-ai-coding-tools-so-you-dont-have-to-b229d4b1a324}
    \srcitem{GL28}{GL44} Gupta, V. (2025, October 8). \emph{How to Make LLM-based Development Stable and Scalable}. LinkedIn. \url{https://www.linkedin.com/posts/vishnugupta12_guide-activity-7381652122653622272-RFF1}
    \srcitem{GL29}{GL46} Andrej Karpathy [@karpathy] (2025, February 2). There's a new kind of coding I call ``vibe coding'', where you fully give in to the vibes, embrace exponentials [Post]. X. \url{https://x.com/karpathy/status/1886192184808149383}
    \srcitem{GL30}{GL47} Andrej Karpathy [@karpathy] (2025, May 1). I attended a vibe coding hackathon recently and used the chance to build a web app (with auth, payments, deploy [Post]. X. \url{https://x.com/karpathy/status/1917961248031080455}
    \srcitem{GL31}{GL48} Willison, S. (2025, March 19). \emph{Not all AI-assisted programming is vibe coding (but vibe coding rocks)}. Substack. \url{https://simonw.substack.com/p/not-all-ai-assisted-programming-is}
    \srcitem{GL32}{GL94} Krings, K., Bohn, N. S., \& Ludwig, T. (2025, October 14). \emph{(R)evolution of Programming: Vibe Coding as a Post-Coding Paradigm} (arXiv:2510.12364) [Preprint]. arXiv. \url{https://arxiv.org/abs/2510.12364}
    \srcitem{GL33}{GL96} Meyer, J. G. (2025, October 10). \emph{Rapid Development of Omics Data Analysis Applications through Vibe Coding} (arXiv:2510.09804) [Preprint]. arXiv. \url{https://arxiv.org/abs/2510.09804}
    \srcitem{GL34}{GL97} Bamil, V. (2025, October 9). \emph{Vibe Coding: Toward an AI-Native Paradigm for Semantic and Intent-Driven Programming} (arXiv:2510.17842) [Preprint]. arXiv. \url{https://arxiv.org/abs/2510.17842}
    \srcitem{GL35}{GL98} Bursuc, S., Ehrenborg, T., Lin, S., Astefanoaei, L., Chiosa, I. E., Kukovec, J., Singh, A., Butterley, O., Bizid, A., Dougherty, Q., Zhao, M., Tan, M., \& Tegmark, M. (2025, September 26). \emph{A benchmark for vericoding: formally verified program synthesis} (arXiv:2509.22908) [Preprint]. arXiv. \url{https://arxiv.org/abs/2509.22908}
    \srcitem{GL36}{GL99} Chen, J., Huang, H., Lyu, Y., An, J., Shi, J., Yang, C., Zhang, T., Tian, H., Li, Y., Li, Z., Zhou, X., Hu, X., \& Lo, D. (2025, September 26). \emph{SecureVibeBench: Benchmarking Secure Vibe Coding of AI Agents via Reconstructing Vulnerability-Introducing Scenarios} (arXiv:2509.22097) [Preprint]. arXiv. \url{https://arxiv.org/abs/2509.22097}
    \srcitem{GL37}{GL100} Hayashi, S.-I., Morita, K., Mukunoki, D., Hoshino, T., \& Katagiri, T. (2025, September 26). \emph{VibeCodeHPC: An Agent-Based Iterative Prompting Auto-Tuner for HPC Code Generation Using LLMs} (arXiv:2510.00031) [Preprint]. arXiv. \url{https://arxiv.org/abs/2510.00031}
    \srcitem{GL38}{GL101} Woo, D. J., Guo, K., \& Yu, Y. (2025, September 9). \emph{A vibe coding learning design to enhance EFL students' talking to, through, and about AI} (arXiv:2509.08854) [Preprint]. arXiv. \url{https://arxiv.org/abs/2509.08854}
    \srcitem{GL39}{GL102} Peng, X., \& Wang, C. (2025, March 11). \emph{Code Digital Twin: A Knowledge Infrastructure for AI-Assisted Complex Software Development} (arXiv:2503.07967) [Preprint]. arXiv. \url{https://arxiv.org/abs/2503.07967}
    \srcitem{GL40}{GL104} Noah [@noahreeveshq] (2025, April 30). Vibe coding is a scam. I tried vibe coding for 2 days. I'm talking real vibe coding where I turn [Post]. X. \url{https://x.com/noahreeveshq/status/1917700874677412249}
    \srcitem{GL41}{GL105} Lex Fridman [@lexfridman] (2025, August 19). Programming with AI is insanely fun. Process is: 1. generate code 2. read \& understand code that was generated 3 [Post]. X. \url{https://x.com/lexfridman/status/1957940412179497327}
    \srcitem{GL42}{GL115} Y Combinator (2025, April 25). \emph{How To Get The Most Out Of Vibe Coding} [Video]. YouTube. \url{https://youtube.com/watch?v=BJjsfNO5JTo}
    \srcitem{GL43}{GL120} Sarkar, A., \& Drosos, I. (2025, June 29). \emph{Vibe coding: programming through conversation with artificial intelligence} (arXiv:2506.23253) [Preprint]. arXiv. \url{https://arxiv.org/abs/2506.23253}
    \srcitem{GL44}{GL121} Shields, T. (2025, April 6). \emph{My first attempt at vibe coding}. TechTarget. \url{https://www.techtarget.com/it-infrastructure/opinion/My-first-attempt-at-vibe-coding}
    \srcitem{GL45}{GL122} Goedecke, S. (2025, March 14). \emph{Refactoring to understand and ``vibe coding''}. seangoedecke.com. \url{https://www.seangoedecke.com/vibe-coding/}
    \srcitem{GL46}{GL123} Hodges, N. (2025, March 25). \emph{Vibe coding is groovy}. InfoWorld. \url{https://www.infoworld.com/article/3853805/vibe-coding-with-claude-code.html}
    \srcitem{GL47}{GL124} Elisa (2025, October 6). \emph{Vibe Coding for the Code-Illiterate: My Ongoing Journey from Idea Guy to Creator}. Elisa Digital. \url{https://digital.elisa.fi/blog/vibe-coding-for-the-code-illiterate-my-ongoing-journey-from-idea-guy-to-creator/}
    \srcitem{GL48}{GL125} Goode, L. (2025, August 20). \emph{Why Did a \$10 Billion Startup Let Me Vibe-Code for Them—and Why Did I Love It?} WIRED. \url{https://www.wired.com/story/why-did-a-10-billion-dollar-startup-let-me-vibe-code-for-them-and-why-did-i-love-it/}
    \srcitem{GL49}{GL126} Willison, S. (2025, October 6). \emph{Vibe engineering}. Simon Willison's Weblog. \url{https://simonwillison.net/2025/Oct/7/vibe-engineering/}
    \srcitem{GL50}{GL127} Asseo, I. (2025, September 11). \emph{Vibe-coding convinced me to fire my agency}. LinkedIn. \url{https://www.linkedin.com/pulse/vibe-coding-convinced-me-fire-my-agency-itai-asseo-djlpe}
    \srcitem{GL51}{GL128} MacCaw, A. (2025, June 19). \emph{How to Vibe Code as a Senior Engineer}. \url{https://blog.alexmaccaw.com/how-to-vibe-code-as-a-senior-engineer/}
    \srcitem{GL52}{GL129} Lo, V. (2025, April 6). \emph{The Truth about Vibe Coding (feat. GitHub Copilot Agent Mode)}. lo-victoria.com. \url{https://lo-victoria.com/the-truth-about-vibe-coding-feat-github-copilot-agent-mode}
    \srcitem{GL53}{GL130} Mohl, D. (2025, May 6). \emph{Better Vibe Coding - Part 1, Foundation \& Basics}. \url{https://david.coffee/vibe-coding-part-1/}
    \srcitem{GL54}{GL131} Tenner, D. (2025, September 27). \emph{5 very different ways of ``vibe coding''}. danieltenner.com. \url{https://danieltenner.com/5-very-different-ways-of-vibe-coding/}
    \srcitem{GL55}{GL132} Macneale, N. (2025, March 28). \emph{Vibe Coding with Cursor}. dolthub.com. \url{https://www.dolthub.com/blog/2025-03-29-vibin/}
    \srcitem{GL56}{GL133} Hayashi, M. (2025, June 13). \emph{I Built a Whole App by Vibe Coding with AI — And It Changed How I Think About Software}. blog.mariohayashi.com. \url{https://blog.mariohayashi.com/p/i-built-a-whole-app-by-vibe-coding}
    \srcitem{GL57}{GL143} What Team (2025, September 17). \emph{How we supercharged our WordPress coding with AI}. \url{https://what.digital/ai-assisted-wordpress-coding/}
    \srcitem{GL58}{GL144} Bowen, G. (2025, April 15). \emph{AI and the Theory Building View of programming}. gareth.nz. \url{https://gareth.nz/ai-programming-as-theory-building.html}
    \srcitem{GL59}{GL145} Malykhin, N. (2025, July 3). \emph{Can We Think Less with AI?} \url{https://www.nikmalykhin.com/p/can-we-think-less-with-ai}
    \srcitem{GL60}{GL146} Sharma, S. (2025, July 20). \emph{Adventures with Claude Code: Reflections on Building a Full-Stack System with AI Assisted Coding}. sajalsharma.com. \url{https://sajalsharma.com/posts/adventures-claude-code/}
    \srcitem{GL61}{GL147} Xu, F., Medappa, P. K., Tunc, M. M., Vroegindeweij, M., \& Fransoo, J. C. (2025, October 11). \emph{AI-Assisted Programming Decreases the Productivity of Experienced Developers by Increasing the Technical Debt and Maintenance Burden} (arXiv:2510.10165) [Preprint]. arXiv. \url{https://arxiv.org/abs/2510.10165}
    \srcitem{GL62}{GL156} Christie, A. (2025, June 2). \emph{Building better software with AI}. Attio. \url{https://attio.com/engineering/blog/building-better-software-with-ai}
    \srcitem{GL63}{GL157} Wise, N. (2025, September 20). \emph{Rucord - the vibe coded iOS app}. The Chicken Coop. \url{https://fastchicken.co.nz/2025/09/20/rucord-the-vibe-coded-ios-app/}
    \srcitem{GL64}{GL158} Robinson, J. (2025, April 23). \emph{GenAI Secret Sauce for Leveraging Cheap Code}. GenAI Secret Sauce. \url{https://genaisecretsauce.com/genai-secret-sauce-for-leveraging-cheap-code/}
    \srcitem{GL65}{GL159} Bandara, E., Gore, R., Liang, X., Rajapakse, S., Kularathne, I., Karunarathna, P., Foytik, P., Shetty, S., Mukkamala, R., Rahman, A., Hass, A., Keong, N. W., De Zoysa, K., Withanage, A., \& Loganathan, N. (2025, October 26). \emph{Agentsway -- Software Development Methodology for AI Agents-based Teams} (arXiv:2510.23664) [Preprint]. arXiv. \url{https://arxiv.org/abs/2510.23664}
    \srcitem{GL66}{GL167} Xu, S., Li, Z., Mei, K., \& Zhang, Y. (2024, May 11). \emph{AIOS Compiler: LLM as Interpreter for Natural Language Programming and Flow Programming of AI Agents} (arXiv:2405.06907) [Preprint]. arXiv. \url{https://arxiv.org/abs/2405.06907}
    \srcitem{GL67}{GL168} Syme, D. (2025, August 27). \emph{On Natural Language Programming}. dsyme.net. \url{https://dsyme.net/2025/08/27/on-natural-language-programming/}
    \srcitem{GL68}{GL177} Councilman, A., Fu, D. J., Gupta, A., Wang, C., Grove, D., Wang, Y.-X., \& Adve, V. (2025, July 17). \emph{Towards Formal Verification of LLM-Generated Code from Natural Language Prompts} (arXiv:2507.13290) [Preprint]. arXiv. \url{https://arxiv.org/abs/2507.13290}
    \srcitem{GL69}{GL181} grumpygrace.dev (2025, June 21). \emph{Vibe Coding Made Me a Better Coder}. \url{https://grumpygrace.dev/posts/vibe-coding-made-me-a-better-coder/}
    \srcitem{GL70}{GL182} Tompkins, D. (2025, April 3). \emph{Coding with Vibes}. Loosed Blog. \url{https://l-o-o-s-e-d.net/vibe-coding}
    \srcitem{GL71}{GL183} rigotti.nl (2025, March 17). \emph{My First Hour of Vibe Coding}. \url{https://rigotti.nl/en/blog/first-hour-vibe-coding/}
    \srcitem{GL72}{GL185} Sinik, M. (2025, August 7). \emph{Beyond Vibe Coding}. Backend Cafe. \url{https://backend.cafe/beyond-vibe-coding}
    \srcitem{GL73}{GL186} Boag, P. (2025, October 16). I finally tried vibe coding (AI-assisted coding) as a non-developer to see if the hype is real [Post]. LinkedIn. \url{https://www.linkedin.com/posts/paulboag_i-finally-tried-vibe-coding-ai-assisted-activity-7384594078631444480-Wp5P}
    \srcitem{GL74}{GL187} AppHelix Technologies (2025, April 8). \emph{Vibe Coding: When Engineers Stop Reading Their Own Code}. \url{https://apphelix.ai/vibe-coding/}
    \srcitem{GL75}{GL195} Memorys Tech Tips (2025, March 21). \emph{Vibe Coding: How To Build Software With AI (No Coding Experience)}. \url{https://memstechtips.com/vibe-coding-ai-powered-programming-guide/}
    \srcitem{GL76}{GL197} namanyayg (2025, September 12). \emph{Vibe Coding Is Creating Braindead Coders}. N's Blog. \url{https://nmn.gl/blog/vibe-coding-gambling}
    \srcitem{GL77}{GL200} Shadmand, S. (2023, February 12). \emph{Updated review of LLM based development}. \url{https://seanshadmand.com/2023/02/12/updated-review-of-llm-based-development/}
    \srcitem{GL78}{GL201} minimaxir (2025, May 5). As an experienced LLM user, I don't use generative LLMs often [Post]. Hacker News. \url{https://news.ycombinator.com/item?id=43897320}
    \srcitem{GL79}{GL204} Eibl, P., Sabouri, S., \& Chattopadhyay, S. (2025, August 11). \emph{Exploring the Challenges and Opportunities of AI-assisted Codebase Generation} (arXiv:2508.07966) [Preprint]. arXiv. \url{https://arxiv.org/abs/2508.07966}
\end{enumerate}

\section*{Window 2: Selected Peer-Reviewed Literature}
\begin{enumerate}[label={[WL00]}, leftmargin=*]
    \srcitem{WL35}{WL32} Gama, K., Calegario, F., Jackson, V., Nolte, A., Morais, L., \& Garcia, V. C. (2026, April). ``Can you feel the vibes?'': An exploration of novice programmer engagement with vibe coding. In \emph{Proceedings of the IEEE/ACM 48th International Conference on Software Engineering} (pp. 485-496). \url{https://doi.org/10.1145/3786580.3786992}
    \srcitem{WL36}{WL33} Agarwal, S., He, H., \& Vasilescu, B. (2026, April). AI IDEs or Autonomous Agents? Measuring the Impact of Coding Agents on Software Development. In \emph{Proceedings of the 23rd International Conference on Mining Software Repositories} (pp. 857-862). \url{https://doi.org/10.1145/3793302.3793589}
    \srcitem{WL37}{WL34} Chou, Y.-H., Jiang, B., Chen, Y. W., Weng, M., Jackson, V., Zimmermann, T., \& Jones, J. A. (2026). Building Software by Rolling the Dice: A Qualitative Study of Vibe Coding. In \emph{Proceedings of the ACM on Software Engineering} (pp. 1712-1734). \url{https://doi.org/10.1145/3797105}
    \srcitem{WL38}{WL35} Chen, V., Talwalkar, A., Brennan, R., \& Neubig, G. (2026, April). Code with Me or for Me? How Increasing AI Automation Transforms Developer Workflows. In \emph{Proceedings of the 2026 CHI Conference on Human Factors in Computing Systems} (pp. 1-19). \url{https://doi.org/10.1145/3772318.3790850}
    \srcitem{WL39}{WL36} Thorgeirsson, S., Weidmann, T. B., \& Su, Z. (2026, April). Computer Science Achievement and Writing Skills Predict Vibe Coding Proficiency. In \emph{Proceedings of the 2026 CHI Conference on Human Factors in Computing Systems} (pp. 1-17). \url{https://doi.org/10.1145/3772318.3791666}
    \srcitem{WL40}{WL37} Si, J. J., Li, D., Wang, Q., Liu, A. W., Lin, X., Zhu, Y., Zhou, X., Wang, A. Y., \& Wang, K. Y. (2026, April). Exploring the Impacts and Challenges of Vibe Coding Paradigm to Children's Programming Learning and Practices. In \emph{Proceedings of the 2026 CHI Conference on Human Factors in Computing Systems} (pp. 1-24). \url{https://doi.org/10.1145/3772318.3790366}
    \srcitem{WL41}{WL38} Qin, X., \& Gruppi, M. (2026, July). From Correctness to Consistency: Redefining Reliability for the Agentware Era. In \emph{Proceedings of the 3rd ACM International Conference on AI-Powered Software} (pp. 145-150). \url{https://doi.org/10.1145/3805760.3814904}
    \srcitem{WL42}{WL39} Wang, Y., Beigi Rizi, M., Khashei, F., \& Rubin, J. (2026). From Specifications to Implementation in the Gen-AI Era: Lessons from a Project-Based Software Engineering Course. In \emph{Proceedings of the ACM on Software Engineering} (pp. 1424-1447). \url{https://doi.org/10.1145/3797092}
    \srcitem{WL43}{WL40} Kobiella, C., Breidenstein, D., \& Schmidt, A. (2026, April). From Throw-Away to Takeaway: How GenAI and Vibe Coding Accelerate Prototyping Across Technical Skill Levels. In \emph{Proceedings of the 2026 CHI Conference on Human Factors in Computing Systems} (pp. 1-23). \url{https://doi.org/10.1145/3772318.3790757}
    \srcitem{WL44}{WL41} Meem, F. N., Zohra, F. T., Smith, J., \& Johnson, B. (2026, April). Hope or Hype? Understanding Vibe Coding through Software Practitioner Discussions. In \emph{Proceedings of the 19th International Conference on Cooperative and Human Aspects of Software Engineering} (pp. 115-125). \url{https://doi.org/10.1145/3794860.3794904}
    \srcitem{WL45}{WL42} Asdaque, S. A., Haider, I., Malik, M. U., Ghafoor, M. A., \& Bangash, A. A. (2026, April). Novice Developers Produce Larger Review Overhead for Project Maintainers while Vibe Coding. In \emph{Proceedings of the 23rd International Conference on Mining Software Repositories} (pp. 898-902). \url{https://doi.org/10.1145/3793302.3793597}
    \srcitem{WL46}{WL43} Watanabe, M., Li, H., Kashiwa, Y., Reid, B., Iida, H., \& Hassan, A. E. (2026). On the Use of Agentic Coding: An Empirical Study of Pull Requests on GitHub. \emph{ACM Transactions on Software Engineering and Methodology}, 35(10), 1-24. \url{https://doi.org/10.1145/3798166}
    \srcitem{WL47}{WL44} Alvarez, I. (2026, June). Upskilling UX Designers for AI-Native Work: A Pedagogical Framework and Empirical Evaluation. In \emph{Proceedings of the 5th Annual Symposium on Human-Computer Interaction for Work} (pp. 1-19). \url{https://doi.org/10.1145/3808045.3808059}
    \srcitem{WL48}{WL45} Zhao, G. Y., \& Tu, C. Z. (2026). Utilizing Vibe Coding Techniques to Improve Prototyping in the Information Systems Analysis and Design Class. \emph{Journal of Computing Sciences in Colleges}, 41(6), 19-28. \url{https://dl.acm.org/doi/10.5555/3812622.3812625}
    \srcitem{WL49}{WL46} Zhong, P., Vaezipoor, P., Cui, F., Kumar, V., Austin, J., Asgarian, A., Ho, T., Inder, P., Kedir, I., Li, Z., Ondo, N., Shafiq, A., Sheikh, I., Sioufi, E., Soltanieh, S., Wilde, B., Zhao, J., Carelli, R., Miller, H., \& Catasta, M. (2026, May). ViBench: A Benchmark on Vibe Coding. In \emph{Proceedings of the ACM Conference on AI and Agentic Systems} (pp. 497-513). \url{https://doi.org/10.1145/3786335.3813162}
    \srcitem{WL50}{WL47} Madan, S., Surabiyil Bindu, S., \& Potluri, V. (2026, April). Vibe Check: Accessibility Heuristics for Vibe Coding Interfaces. In \emph{Proceedings of the 23rd International Web for All Conference} (pp. 229-241). \url{https://doi.org/10.1145/3800424.3800455}
    \srcitem{WL51}{WL48} Tran, H., Nashold, L., Krishnan, R., Bigeard, A., \& Gu, A. (2026, May). Vibe Code Bench: Evaluating AI Models on End-to-End Web Application Development. In \emph{Proceedings of the ACM Conference on AI and Agentic Systems} (pp. 514-536). \url{https://doi.org/10.1145/3786335.3813180}
    \srcitem{WL52}{WL49} Tholander, J., \& Jonsson, M. (2026, April). Vibe Coding Entanglements – Repositioning Boundaries of Intention, Authorship, and Responsibility in Programming with Generative AI. In \emph{Proceedings of the 2026 CHI Conference on Human Factors in Computing Systems} (pp. 1-16). \url{https://doi.org/10.1145/3772318.3791847}
    \srcitem{WL53}{WL50} Li, J., Hou, Y., Lin, L., Zhu, R., Cao, H., \& El Ali, A. (2026, June). Vibe Coding in Product Teams: Reconfiguring AI-Assisted Workflows, Prototyping, and Collaboration. In \emph{Proceedings of the 5th Annual Symposium on Human-Computer Interaction for Work} (pp. 1-16). \url{https://doi.org/10.1145/3808045.3808062}
    \srcitem{WL54}{WL51} Seo, J., Deldari, E., \& Mentis, H. M. (2026, March). Whose Code Is It? How AI Autonomy Reshapes Ownership, Responsibility, and Disclosure in AI-Assisted Programming. In \emph{Proceedings of the 31st International Conference on Intelligent User Interfaces} (pp. 393-425). \url{https://doi.org/10.1145/3742413.3789121}
    \srcitem{WL55}{WL52} Garuslu, V., Jafarov, Z., Mövsümova, A., \& Namazov, A. (2026). How Software Engineers Engage with AI: A Pragmatic Workflow. \emph{IEEE Software}, 1-9. \url{https://doi.org/10.1109/ms.2026.3693964}
    \srcitem{WL56}{WL53} Vella, S., Ferworn, A., \& Sharieh, M. (2026, July). ATeam: Governance-Aware LLM-Assisted Software Sustaining Engineering for Enterprise Systems. In \emph{2026 6th International Conference on Electrical, Computer and Energy Technologies (ICECET)} (pp. 1-6). \url{https://doi.org/10.1109/icecet65726.2026.11633274}
    \srcitem{WL57}{WL54} Keyu, Z., Xiaotian, Q., \& Yang, Z. (2026, May). An Empirical Study on Engineering Modeling Efficiency of Vibe Coding: A Standardized Evaluation Framework for Generative AI-Assisted Programming. In \emph{2026 9th International Conference on Electronics Technology (ICET)} (pp. 499-504). \url{https://doi.org/10.1109/icet69987.2026.11659181}
    \srcitem{WL58}{WL55} Gudimetla, S., Gangula, S., Adinarayana, R., Challa, C., Vunnam, S., \& Buhyavarapu, S. (2026, April). From AI Drag to AI Boost: A Preceptorship Framework for Developing Software Engineering Expertise in the Agentic Era. In \emph{2026 International Conference on Artificial Intelligence, Systems, and Emerging Technologies (ICAISET)} (pp. 1-12). \url{https://doi.org/10.1109/icaiset66439.2026.11542057}
    \srcitem{WL59}{WL56} Hui, Z., \& Rongping, L. (2026, April). Research On The Application Strategies Of Generative Artificial Intelligence In User Interface Design. In \emph{2026 11th International Conference on Cloud Computing and Big Data Analytics (ICCCBDA)} (pp. 509-513). \url{https://doi.org/10.1109/icccbda69573.2026.11676493}
    \srcitem{WL60}{WL57} O'Brien, G., \& Eisty, N. U. (2026). Scientific Software in the Age of Vibe Coding. \emph{Computing in Science \& Engineering}, 28(1), 112-116. \url{https://doi.org/10.1109/mcse.2026.3661610}
    \srcitem{WL61}{WL58} Guriţă, A.-E., \& Vatavu, R.-D. (2026, June). Understanding Vibe Coding: A Phenomenological Investigation of Flow-Like Human-AI Collaboration. In \emph{2026 2nd International Conference on Human-AI Interaction and Experience Design (HAXD)} (pp. 87-94). \url{https://doi.org/10.1109/haxd70072.2026.11620988}
    \srcitem{WL62}{WL59} Suryanarayanan, S. (2026, June). Vibe Coding and its Quality Implication: An Empirical Study of AI-Native Development Practices on Software Reliability. In \emph{2026 4th International Conference on Inventive Computing and Informatics (ICICI)} (pp. 1868-1875). \url{https://doi.org/10.1109/icici68773.2026.11581171}
    \srcitem{WL63}{WL60} Kusper, G., \& Szabó, C. (2025, November). Vibe Coding in Education. In \emph{2025 International Conference on Emerging eLearning Technologies and Applications (ICETA)} (pp. 506-511). \url{https://doi.org/10.1109/iceta67772.2025.11280285}
    \srcitem{WL64}{WL61} Kim, D.-K., Baudry, B., \& Voas, J. (2026). Vibe Coding in Practice: Human Software Engineers as the Anchor of Large Language Model-Assisted Development. \emph{Computer}, 59(8), 94-97. \url{https://doi.org/10.1109/mc.2026.3662544}
    \srcitem{WL65}{WL62} Neferu, M., Marian, C. V., \& Mitrea, D. A. (2026, July). Vibe Coding: A New Paradigm in Software Development Assisted by Artificial Intelligence. In \emph{2026 18th International Conference on Electronics, Computers and Artificial Intelligence (ECAI)} (pp. 1-8). \url{https://doi.org/10.1109/ecai69016.2026.11613779}
    \srcitem{WL66}{WL63} Mesnard, F., Marianne, T., Payet, É., \& Vanhoof, W. (2026). Case study: solving P-99 with LPTP and an LLM. \emph{Electronic Proceedings in Theoretical Computer Science}, 450, 209-222. \url{https://doi.org/10.4204/eptcs.450.17}
    \srcitem{WL67}{WL64} Horvat, M., Ursić, I., \& Krmpotić, K. (2026). Comparative Evaluation of Gemini and DeepSeek for LLM-Generated Code Quality and Architectural Robustness in Backend Software Engineering. \emph{Electronics}, 15(13), 2805. \url{https://doi.org/10.3390/electronics15132805}
    \srcitem{WL68}{WL65} Al-Khalifa, H. (2026). From Code-Centric to Concept-Centric: Teaching NLP with LLM-Assisted ``Vibe Coding''. In \emph{Proceedings of the Seventh Workshop on Teaching Natural Language Processing (TeachNLP 2026)} (pp. 11-18). \url{https://doi.org/10.18653/v1/2026.teachingnlp-1.3}
    \srcitem{WL69}{WL66} Pellas, N. (2026). From Prompt to Play: Examining Computational Thinking Through Vibe Coding in Game Making for Pre-Service Teacher Education. \emph{Multimodal Technologies and Interaction}, 10(5), 57. \url{https://doi.org/10.3390/mti10050057}
    \srcitem{WL70}{WL67} Kusper, G., Szabó, J., Szabó, M., Kovács, Á., Tajti, T., Szabó, C., Perháč, J., Horváth, M., \& Sobota, B. (2026). Introducing TAIPO, an AI-based product owner assistant for vibe coding. \emph{Annales Mathematicae et Informaticae}, 63, 73-87. \url{https://doi.org/10.33039/ami.2026.06.005}
    \srcitem{WL71}{WL68} Nold, C. (2026, June). Plundergeist: Exploring generative AI through vibe coding as reflective design practice. In \emph{Proceedings of DRS}. \url{https://doi.org/10.21606/drs.2026.261}
    \srcitem{WL72}{WL69} Te'Eni, D., \& Katz, A. (2026, July). Reliable Vibe Coding: The Human-AI Context Gap in Software Development. \emph{Frontiers in Artificial Intelligence and Applications}. \url{https://doi.org/10.3233/faia260505}
    \srcitem{WL73}{WL70} Tran, H.-V., Dang, T.-T.-T., Nguyen, D.-A., \& Hung, P. N. (2026). VC-AWG: A Vibe Coding–based method for Automated Website Generation from heterogeneous artifacts. \emph{Journal of Computer Languages}, 88, 101403. \url{https://doi.org/10.1016/j.cola.2026.101403}
    \srcitem{WL74}{WL71} Hohn, C., \& Loydl, K. (2026, April). Vibe Coding: intention instead of implementation: How UX professionals build, evaluate, and hand over functional prototypes with AI. \emph{i-com}. \url{https://doi.org/10.1515/icom-2026-0018}
    \srcitem{WL75}{WL72} Emirtekin, E., Türk, M., \& Kırcaburun, K. (2026, August). The impact of vibe coding on students' programming self-efficacy and attitudes toward artificial intelligence. \emph{Scientific Reports}. \url{https://doi.org/10.1038/s41598-026-67799-x}
    \srcitem{WL76}{WL73} Reymus, M., Tichy, A., Diegritz, C., Weisbrod, B., Schwendicke, F., \& Werner, N. (2026). AI-assisted software development in digital dentistry: A technical innovation report with three open-source applications. \emph{Journal of Dentistry}, 175, 106864. \url{https://doi.org/10.1016/j.jdent.2026.106864}
\end{enumerate}

\section*{Window 2: Selected Grey Literature}
\begin{enumerate}[label={[GL000]}, leftmargin=*]
    \srcitem{GL80}{GL23} Mandl, P. [Peter], \& Mandl, P. [Paul] (2026, June 13). \emph{AI-driven Software Development: A Pragmatic Path to Agentic Development Processes} (arXiv:2606.15283) [Preprint]. arXiv. \url{https://arxiv.org/abs/2606.15283}
    \srcitem{GL81}{GL24} Bui, N. D. Q. (2026, March 5). \emph{Building Effective AI Coding Agents for the Terminal: Scaffolding, Harness, Context Engineering, and Lessons Learned} (arXiv:2603.05344) [Preprint]. arXiv. \url{https://arxiv.org/abs/2603.05344}
    \srcitem{GL82}{GL25} Fang, Z., Zhang, Y., Tang, N., McMillan, C., Li, T. J.-J., \& Huang, Y. (2026, July 6). \emph{From Conversation to Contribution: Characterizing Coding Agent in Open-Source Software} (arXiv:2607.05677) [Preprint]. arXiv. \url{https://arxiv.org/abs/2607.05677}
    \srcitem{GL83}{GL26} Thuong, P. T. (2026, July 7). \emph{From Prompt to Production: A Case Study of Human-AI Collaborative Software Development Using Claude Code Multi-Agent Orchestration}. Research Square. \url{https://www.researchsquare.com/article/rs-10253296/latest}
    \srcitem{GL84}{GL27} Tanaka, H., Igaki, H., Shimari, K., Honda, K., \& Fukuyasu, N. (2026, August 31). \emph{Practical Implementation Report on Introducing Spec-Driven Development Using AI Agents in Software Development PBL} (arXiv:2608.30572) [Preprint]. arXiv. \url{https://arxiv.org/abs/2608.30572}
    \srcitem{GL85}{GL28} Farrag, S. E. (2026, May 1). \emph{The Productivity-Reliability Paradox: Specification-Driven Governance for AI-Augmented Software Development} (arXiv:2605.01160) [Preprint]. arXiv. \url{https://arxiv.org/abs/2605.01160}
    \srcitem{GL86}{GL29} Fernández-y-Fernández, C. A., \& Aguilar-Cisneros, J. R. (2026, August 12). \emph{The Role Specialization Model (RSM): Coordinating LLM-Based Tools in Agentic Software Development - An Exploratory Case Study} (arXiv:2608.12311) [Preprint]. arXiv. \url{https://arxiv.org/abs/2608.12311}
    \srcitem{GL87}{GL30} Michels, D. L., Ghazaleh, M. A., Lazzari, F., Kassem, N., \& Klein, J. (2026, August 20). \emph{Vibe Coding: Practice, Performance, Productivity, and Risk -A State-of-the-Art Review} (arXiv:2608.20446) [Preprint]. arXiv. \url{https://arxiv.org/abs/2608.20446}
    \srcitem{GL88}{GL31} Ong, A. Y., Livingstone, I., Kilduff, C., Sevgi, M., Merle, D. A., Ruffell, E., Keane, P. A., \& Antaki, F. (2026, April 24). \emph{Vibe coding for clinicians: democratising bespoke software development for digital health innovation} (arXiv:2604.22604) [Preprint]. arXiv. \url{https://arxiv.org/abs/2604.22604}
    \srcitem{GL89}{GL36} Grauer, Y. (2026, June 22). \emph{Read this before you vibe-code another app}. The Verge. \url{https://www.theverge.com/ai-artificial-intelligence/950844/vibe-coding-security-risks-apps}
    \srcitem{GL90}{GL39} Zhao, S., Wang, D., Zhang, K., Luo, J., Li, Z., \& Li, L. (2025, December 2). \emph{Is Vibe Coding Safe? Benchmarking Vulnerability of Agent-Generated Code in Real-World Tasks} (arXiv:2512.03262) [Preprint]. arXiv. \url{https://arxiv.org/abs/2512.03262}
    \srcitem{GL91}{GL40} Uddin, S. M. J. (2026, April 14). \emph{Is Vibe Coding the Future? An Empirical Assessment of LLM Generated Codes for Construction Safety} (arXiv:2604.12311) [Preprint]. arXiv. \url{https://arxiv.org/abs/2604.12311}
    \srcitem{GL92}{GL41} r/vibecoding (2026, May 7). Vibe Coding Security Issues [Post]. Reddit. \url{https://www.reddit.com/r/vibecoding/comments/1t65xhh/what_are_security_flaws_that_you_keep_seeing_on/}
    \srcitem{GL93}{GL42} Freitag, T. (2026, April 7). \emph{Vibe coding tools compared: Cursor vs. Lovable vs. Kiro vs. Claude Code vs. Trae (2026)}. \url{https://till-freitag.com/en/blog/vibe-coding-tools-comparison}
    \srcitem{GL94}{GL43} da Mota, G., \& Gama, K. (2026, August 17). \emph{Comparing the Quality of Code Generated by Vibe Coding Tools} (arXiv:2608.16302) [Preprint]. arXiv. \url{https://arxiv.org/abs/2608.16302}
    \srcitem{GL95}{GL45} Mathias, T. G. (2026, January 28). \emph{Getting Started with Vibe Coding in Four Steps}. LinkedIn. \url{https://www.linkedin.com/pulse/getting-started-vibe-coding-four-steps-tela-gallagher-mathias-qrhde}
    \srcitem{GL96}{GL49} Osmani, A. (2025, November 30). \emph{Vibe coding is not the same as AI-Assisted engineering}. Medium. \url{https://medium.com/%40addyosmani/vibe-coding-is-not-the-same-as-ai-assisted-engineering-3f81088d5b98}
    \srcitem{GL97}{GL50} Deng, J., Fan, Z., \& Meng, R. (2026, June 22). \emph{Understanding the (In)Security of Vibe-Coded Applications} (arXiv:2606.23130) [Preprint]. arXiv. \url{https://arxiv.org/abs/2606.23130}
    \srcitem{GL98}{GL51} Aribe, S. G., \& Labastida, L. J. S. (2026, September 9). \emph{The Vibe Shift in Software Engineering: Evaluating AI-Led Conversational Programming for Performance, Cognition, and Responsible Adoption} (arXiv:2609.09560) [Preprint]. arXiv. \url{https://arxiv.org/abs/2609.09560}
    \srcitem{GL99}{GL52} Diaz, J., Gayoso, J., Cimminio, A., \& Perez, J. (2026, August 31). \emph{Spec-Driven Development for Agentic Software Engineering: Harnessing Human-Agent Teamwork} (arXiv:2609.00252) [Preprint]. arXiv. \url{https://arxiv.org/abs/2609.00252}
    \srcitem{GL100}{GL53} Zeng, Y., Tu, R., Xiang, Z., Feng, L., Li, G., \& Liu, C. H. (2026, August 30). \emph{Towards Effective Generation of Interactive Visualizations with Vibe Coding: An Empirical Study} (arXiv:2608.29550) [Preprint]. arXiv. \url{https://arxiv.org/abs/2608.29550}
    \srcitem{GL101}{GL54} Vir, R., Chilton, L., Zhang, Z., \& Wu, E. (2026, August 28). \emph{FlowCheck: Helping End-Users Specify and Verify Intent in Vibe-Coded Web Apps} (arXiv:2608.28880) [Preprint]. arXiv. \url{https://arxiv.org/abs/2608.28880}
    \srcitem{GL102}{GL55} Andročec, D. (2026, August 21). \emph{Vibe Coding and Web Application Security: A Twin-Prompt Study} (arXiv:2608.20963) [Preprint]. arXiv. \url{https://arxiv.org/abs/2608.20963}
    \srcitem{GL103}{GL56} Ren, Y., Bai, Y., Wang, X., Chen, L., \& Li, D. (2026, July 30). \emph{An Empirical Study of Coordination Mode as the First-Class Citizen in From-Scratch Multi-Agent Coding} (arXiv:2607.27877) [Preprint]. arXiv. \url{https://arxiv.org/abs/2607.27877}
    \srcitem{GL104}{GL57} Mock, M., \& Russo, B. (2026, July 24). \emph{Vibe Coding: An Experiment with Test-Driven Development} (arXiv:2607.22406) [Preprint]. arXiv. \url{https://arxiv.org/abs/2607.22406}
    \srcitem{GL105}{GL58} Yuan, Z., Lu, W., Wu, H., Jin, D., \& Wu, C. (2026, July 2). \emph{Guiding Human Validation of LLM-Generated Code via Verifiable Literate Programming} (arXiv:2607.02333) [Preprint]. arXiv. \url{https://arxiv.org/abs/2607.02333}
    \srcitem{GL106}{GL59} Sun, Z., Wen, X., Wang, F., Liu, C., Lai, Y., Hurter, C., \& Wang, Y. (2026, June 18). \emph{Vibe Coding for Visualization Implementation: An Empirical Study of Practices and Challenges} (arXiv:2606.19703) [Preprint]. arXiv. \url{https://arxiv.org/abs/2606.19703}
    \srcitem{GL107}{GL60} Sobal, V., Yang, S., Zhang, Y., Xia, W., \& Soatto, S. (2026, June 17). \emph{StaminaBench: Stress-Testing Coding Agents over 100 Interaction Turns} (arXiv:2606.19613) [Preprint]. arXiv. \url{https://arxiv.org/abs/2606.19613}
    \srcitem{GL108}{GL61} Tërnava, X. (2026, June 17). \emph{Where Did the Variability Go? From Vibe Coding to Product Lines by Regeneration} (arXiv:2606.19042) [Preprint]. arXiv. \url{https://arxiv.org/abs/2606.19042}
    \srcitem{GL109}{GL62} Barbour, C. (2026, June 15). \emph{Vibe Coding Ate My Homework: An evaluation of AI approaches to greenfield software engineering and programming} (arXiv:2606.18293) [Preprint]. arXiv. \url{https://arxiv.org/abs/2606.18293}
    \srcitem{GL110}{GL63} Larsen, O. A., \& Moghaddam, M. T. (2026, June 11). \emph{Mining Architectural Quality Under Agentic AI Adoption: A Causal Study of Java Repositories} (arXiv:2606.13298) [Preprint]. arXiv. \url{https://arxiv.org/abs/2606.13298}
    \srcitem{GL111}{GL64} Song, Y., Choi, S., Kim, J., Kim, Y., Weisberg, L., \& Moon, J. (2026, June 9). \emph{A Guiding Framework for K-12 Teachers in Creating AI-powered Learning Technologies through Vibe Coding} (arXiv:2607.05406) [Preprint]. arXiv. \url{https://arxiv.org/abs/2607.05406}
    \srcitem{GL112}{GL65} Taveter, H., \& Lepp, M. (2026, May 26). \emph{From Idea to Classroom in Days: Using ``Vibe Coding'' to Create a Programming Process Visualizer from IDE Activity Logs} (arXiv:2607.24757) [Preprint]. arXiv. \url{https://arxiv.org/abs/2607.24757}
    \srcitem{GL113}{GL66} Fawzy, A., Tahir, A., \& Blincoe, K. (2026, May 23). \emph{From Prompting to Verification: How Experience Shapes Vibe Coding Practices} (arXiv:2605.24521) [Preprint]. arXiv. \url{https://arxiv.org/abs/2605.24521}
    \srcitem{GL114}{GL67} van Bokhorst, L., \& Suilen, K. (2026, May 8). \emph{Vibe coding before the trend} (arXiv:2605.07751) [Preprint]. arXiv. \url{https://arxiv.org/abs/2605.07751}
    \srcitem{GL115}{GL68} Zigler, A. (2026, May 6). \emph{Mise en Place for Agentic Coding: Deliberate Preparation as Context Engineering Methodology} (arXiv:2605.05400) [Preprint]. arXiv. \url{https://arxiv.org/abs/2605.05400}
    \srcitem{GL116}{GL69} Saxena, S., Trivedi, N., \& Jyothi, V. (2026, May 6). \emph{SWE-WebDevBench: Evaluating Coding Agent Application Platforms as Virtual Software Agencies} (arXiv:2605.04637) [Preprint]. arXiv. \url{https://arxiv.org/abs/2605.04637}
    \srcitem{GL117}{GL70} Li, J., Ren, Y., Wang, W., Zheng, C., Wei, X.-Y., Li, Q., \& Bian, Y. (2026, May 4). \emph{MolViBench: Evaluating LLMs on Molecular Vibe Coding} (arXiv:2605.02351) [Preprint]. arXiv. \url{https://arxiv.org/abs/2605.02351}
    \srcitem{GL118}{GL71} Rinja, D., Oliveira, E. A., López-Pernas, S., Saqr, M., Specht, M., \& Misiejuk, K. (2026, April 29). \emph{Unpacking Vibe Coding: Help-Seeking Processes in Student-AI Interactions While Programming} (arXiv:2604.27134) [Preprint]. arXiv. \url{https://arxiv.org/abs/2604.27134}
    \srcitem{GL119}{GL72} Chen, A. J., Cao, Y., Shao, M., Karri, R., \& Shafique, M. (2026, April 24). \emph{Code for All: Educational Applications of the ``Vibe Coding'' Hackathon in Programming Education across All Skill Levels} (arXiv:2604.22747) [Preprint]. arXiv. \url{https://arxiv.org/abs/2604.22747}
    \srcitem{GL120}{GL73} Lipsanen, P., Rannikko, L., Christophe, F., Kalliokoski, K., Stirbu, V., \& Mikkonen, T. (2026, April 22). \emph{Shift-Up: A Framework for Software Engineering Guardrails in AI-native Software Development -- Initial Findings} (arXiv:2604.20436) [Preprint]. arXiv. \url{https://arxiv.org/abs/2604.20436}
    \srcitem{GL121}{GL74} Ma, J., Wang, S., Kung, J. H., \& Chilton, L. B. (2026, April 17). \emph{ZORO: Active Controls to Support Rule Following in Vibe Coding} (arXiv:2604.15625) [Preprint]. arXiv. \url{https://arxiv.org/abs/2604.15625}
    \srcitem{GL122}{GL75} Töpfer, M., Plášil, F., Bureš, T., \& Hnětynka, P. (2026, April 16). \emph{Vibe-Coding: Feedback-Based Automated Verification with no Human Code Inspection, a Feasibility Study} (arXiv:2604.14867) [Preprint]. arXiv. \url{https://arxiv.org/abs/2604.14867}
    \srcitem{GL123}{GL76} Lin, Y., Zhao, L., \& Shi, Y. (2026, April 10). \emph{Contract-Coding: Towards Repo-Level Generation via Structured Symbolic Paradigm} (arXiv:2604.13100) [Preprint]. arXiv. \url{https://arxiv.org/abs/2604.13100}
    \srcitem{GL124}{GL77} Ren, S., Shen, X., Zhou, Y., Dennis, Ng, \& Raj, A. (2026, April 6). \emph{Chinese Language Is Not More Efficient Than English in Vibe Coding: A Preliminary Study on Token Cost and Problem-Solving Rate} (arXiv:2604.14210) [Preprint]. arXiv. \url{https://arxiv.org/abs/2604.14210}
    \srcitem{GL125}{GL78} Hughes, J., Collier, B., \& Thomas, D. R. (2026, March 31). \emph{Stand-Alone Complex or Vibercrime? Exploring the adoption and innovation of GenAI tools, coding assistants, and agents within cybercrime ecosystems} (arXiv:2603.29545) [Preprint]. arXiv. \url{https://arxiv.org/abs/2603.29545}
    \srcitem{GL126}{GL79} Lyu, W., Xiao, Y., Zhang, Y., \& Sun, Y. (2026, March 26). \emph{TheBotCompany: Self-Organizing Multi-agent Systems for Continuous Software Development} (arXiv:2603.25928) [Preprint]. arXiv. \url{https://arxiv.org/abs/2603.25928}
    \srcitem{GL127}{GL80} Du, R., Hersh, B., Li, D., Numan, N., Qian, X., Chen, Y., Zhou, Z., Ren, J., Chen, X., Bettridge, R. T., Chen, X., Toh, S., Faruqi, F., \& Kim, D. (2026, March 25). \emph{Vibe Coding XR: Accelerating AI + XR Prototyping with XR Blocks and Gemini} (arXiv:2603.24591) [Preprint]. arXiv. \url{https://arxiv.org/abs/2603.24591}
    \srcitem{GL128}{GL81} Bansal, S., Fangkai, J., Zhou, Y., Xu, A., Joty, S., \& Yavuz, S. (2026, March 16). \emph{VIBEPASS: Can Vibe Coders Really Pass the Vibe Check?} (arXiv:2603.15921) [Preprint]. arXiv. \url{https://arxiv.org/abs/2603.15921}
    \srcitem{GL129}{GL82} Wang, S. (2026, March 16). \emph{VibeContract: The Missing Quality Assurance Piece in Vibe Coding} (arXiv:2603.15691) [Preprint]. arXiv. \url{https://arxiv.org/abs/2603.15691}
    \srcitem{GL130}{GL83} Shin, D., Gao, A., Pang, R. Y., Lee, J., Reinecke, K., \& Tseng, E. (2026, March 13). \emph{Interrogating Design Homogenization in Web Vibe Coding} (arXiv:2603.13036) [Preprint]. arXiv. \url{https://arxiv.org/abs/2603.13036}
    \srcitem{GL131}{GL85} Töpfer, M., Plášil, F., Bureš, T., \& Hnětynka, P. (2026, February 20). \emph{Feedback-based Automated Verification in Vibe Coding of CAS Adaptation Built on Constraint Logic} (arXiv:2602.18607) [Preprint]. arXiv. \url{https://arxiv.org/abs/2602.18607}
    \srcitem{GL132}{GL86} Thang, M., Wu, L., Behrouzi, S., Rostami, M., Lintelo, J. T., Picek, S., \& Sadeghi, A.-R. (2026, February 11). \emph{GoodVibe: Security-by-Vibe for LLM-Based Code Generation} (arXiv:2602.10778) [Preprint]. arXiv. \url{https://arxiv.org/abs/2602.10778}
    \srcitem{GL133}{GL87} Hu, H., Marjieh, R., Collins, K. M., Li, C., Griffiths, T. L., Sucholutsky, I., \& Jacoby, N. (2026, February 11). \emph{Why Human Guidance Matters in Collaborative Vibe Coding} (arXiv:2602.10473) [Preprint]. arXiv. \url{https://arxiv.org/abs/2602.10473}
    \srcitem{GL134}{GL88} Shi, J., Sun, W., Chen, Z., Wu, Z., Chen, X., Jin, Z., \& Liu, Y. (2026, January 23). \emph{REprompt: Prompt Generation for Intelligent Software Development Guided by Requirements Engineering} (arXiv:2601.16507) [Preprint]. arXiv. \url{https://arxiv.org/abs/2601.16507}
    \srcitem{GL135}{GL89} Hou, Z. (2026, January 8). \emph{Vibe Coding an LLM-powered Theorem Prover} (arXiv:2601.04653) [Preprint]. arXiv. \url{https://arxiv.org/abs/2601.04653}
    \srcitem{GL136}{GL90} Aiersilan, A. (2026, January 2). \emph{The Vibe-Check Protocol: Quantifying Cognitive Offloading in AI Programming} (arXiv:2601.02410) [Preprint]. arXiv. \url{https://arxiv.org/abs/2601.02410}
    \srcitem{GL137}{GL92} Li, J., Jin, Z., Zhang, H., Zhang, K., Qian, J., \& Zhao, T. (2025, November 19). \emph{A Viable Paradigm of Software Automation: Iterative End-to-End Automated Software Development} (arXiv:2511.15293) [Preprint]. arXiv. \url{https://arxiv.org/abs/2511.15293}
    \srcitem{GL138}{GL93} Cito, J., \& Bork, D. (2025, November 4). \emph{Lost in Code Generation: Reimagining the Role of Software Models in AI-driven Software Engineering} (arXiv:2511.02475) [Preprint]. arXiv. \url{https://arxiv.org/abs/2511.02475}
    \srcitem{GL139}{GL103} AI Engineer (2026, April 24). \emph{Full Walkthrough: Workflow for AI Coding — Matt Pocock} [Video]. YouTube. \url{https://www.youtube.com/watch?v=-QFHIoCo-Ko&pp=ygUZIkFJLWFzc2lzdGVkIHByb2dyYW1taW5nIg%3D%3D}
    \srcitem{GL140}{GL106} Andrej Karpathy [@karpathy] (2025, December 28). I was inspired by this so I wanted to see if Claude Code can get into my Lutron home automation [Post]. X. \url{https://x.com/karpathy/status/2005067301511630926}
    \srcitem{GL141}{GL107} AI Search [@aisearchio] (2026, September 5). If you're vibe coding games or 3D assets with GPT 6 Astra, by default it'll still output pretty basic, sloppy-looking [Post]. X. \url{https://x.com/aisearchio/status/2096331480763113795}
    \srcitem{GL142}{GL108} Prajwal Tomar [@PrajwalTomar\_] (2026, September 3). The most expensive habit in vibe coding right now is shipping the second the app works. A 20+ year dev [Post]. X. \url{https://x.com/PrajwalTomar_/status/2095490245785735195}
    \srcitem{GL143}{GL113} Dudzik, J. W. (2026, September 9). \emph{Spec Driven Development, vibe coding, and your obligation to say ``slowdown''}. LinkedIn. \url{https://linkedin.com/pulse/spec-driven-development-vibe-coding-your-obligation-jonathan-wesley-bpule}
    \srcitem{GL144}{GL117} Sequoia Capital (2026, April 29). \emph{Andrej Karpathy: From Vibe Coding to Agentic Engineering w/ Stephanie Zhan} [Video]. YouTube. \url{https://youtube.com/watch?v=96jN2OCOfLs}
    \srcitem{GL145}{GL118} Lex Fridman (2026, August 26). \emph{DHH: Future of Programming, AI, Agentic Engineering, Vibe Coding \& Linux} [Video]. YouTube. \url{https://youtube.com/watch?v=NYFGCESmikA}
    \srcitem{GL146}{GL119} Chris Raroque (2026, March 16). \emph{Vibe Coding Has A Security Problem (And How To Fix It)} [Video]. YouTube. \url{https://youtube.com/watch?v=tK4NQtzfZbM}
    \srcitem{GL147}{GL134} r/vibecoding (2026, May 17). Vibe coding feels amazing until an experienced developer [Post]. Reddit. \url{https://www.reddit.com/r/vibecoding/comments/1tfpdvt/vibe_coding_feels_amazing_until_an_experienced/}
    \srcitem{GL148}{GL135} Simic, I. (2026, April 29). \emph{Vibe Coding Mistakes you can easily avoid}. ShiftMag. \url{https://shiftmag.dev/vibe-coding-mistakes-how-to-avoid-them-9255/}
    \srcitem{GL149}{GL136} Thomas, R. (2026, January 28). \emph{Breaking the Spell of Vibe Coding}. fast.ai. \url{https://www.fast.ai/posts/2026-01-28-dark-flow/}
    \srcitem{GL150}{GL137} Colin, C. (2026, May 18). \emph{I'm a Normie. Can Normies Really Vibe Code?} WIRED. \url{https://www.wired.com/story/normie-vibe-code/}
    \srcitem{GL151}{GL138} DW Innovation (2025, December 15). \emph{Vibe Coding: How Useful Is It For Newsroom Projects?} \url{https://innovation.dw.com/articles/vibe-coding-for-newsroom-projects}
    \srcitem{GL152}{GL139} Yoskovitz, B. (2026, January 6). \emph{Vibe Coding Without System Design is a Trap}. Focused Chaos. \url{https://www.focusedchaos.co/p/vibe-coding-without-system-design-is-a-trap}
    \srcitem{GL153}{GL140} Kundapur, N. (2026, March 30). \emph{Vibe Coding, Viable Coding: PM's Perspective}. \url{https://niran-kundapur.com/2026/03/29/vibe-coding-viable-coding-pms-perspective/}
    \srcitem{GL154}{GL141} Liu, Y. (2025, December 22). \emph{No 44. Vibe Coding Workflow for Non-Programmers}. UX Planet. \url{https://uxplanet.org/no-44-vibe-coding-workflow-for-non-programmers-9c695f01e7b7}
    \srcitem{GL155}{GL142} Amditis, J. (2025, December 1). \emph{Vibe Coding for Newsrooms}. Generative AI in the Newsroom. \url{https://generative-ai-newsroom.com/vibe-coding-for-newsrooms-6848b17dac99}
    \srcitem{GL156}{GL148} Wilson, G. (2026, May 20). \emph{Twelve ways to be wrong about AI-assisted coding}. The Third Bit. \url{https://third-bit.com/2026/05/20/twelve-ways-to-be-wrong/}
    \srcitem{GL157}{GL149} Dhameja, G. (2026, May 24). \emph{AI-Assisted Coding: Field Notes}. \url{https://www.gautamdhameja.com/ai-assisted-coding-notes/}
    \srcitem{GL158}{GL150} Chepurin, I., \& Turner, T. (2026, May 19). \emph{AI-assisted engineers are burning out, is this fine?} Evil Martians. \url{https://evilmartians.com/chronicles/ai-assisted-engineers-are-burning-out-is-this-fine}
    \srcitem{GL159}{GL151} Remy (2026, January 11). \emph{What Is OpenSpec? Workflow, Files, and When to Use It}. Redreamality. \url{https://redreamality.com/garden/notes/openspec-guide/}
    \srcitem{GL160}{GL152} Yang, E. Z. (2026, March 10). \emph{AI-assisted programming for spmd\_types}. ezyang's blog. \url{https://blog.ezyang.com/2026/03/ai-assisted-programming-for-spmd-types/}
    \srcitem{GL161}{GL153} Tang, N., Chen, C., Fang, Z., Xu, G., Dhakal, M., Shi, Y., McMillan, C., Huang, Y., \& Li, T. J.-J. (2026, April 1). \emph{Programming by Chat: A Large-Scale Behavioral Analysis of 11,579 Real-World AI-Assisted IDE Sessions} (arXiv:2604.00436) [Preprint]. arXiv. \url{https://arxiv.org/abs/2604.00436}
    \srcitem{GL162}{GL154} Piglei (2026, March 23). \emph{AI Coding: A Concise Guide for Software Engineers}. \url{https://www.piglei.com/articles/en-ai-coding-a-concise-guide-for-swe/}
    \srcitem{GL163}{GL155} Banes, C. (2025, December 29). \emph{The Disappearing Middle: How AI Coding Tools Are Breaking Software Apprenticeship}. \url{https://chrisbanes.me/posts/disappearing-middle-ai-software-apprenticeship/}
    \srcitem{GL164}{GL160} Bailey, E. (2026, June 15). \emph{The case for an accessibility designer vibe coding when all his coworkers are also vibe coding}. ericwbailey. \url{https://ericwbailey.website/published/the-case-for-an-accessibility-designer-vibe-coding-when-all-his-coworkers-are-also-vibe-coding/}
    \srcitem{GL165}{GL161} Uptime Consulting (2025, December 8). \emph{Coding Agents Need Deterministic Feedback: A Case for Phoenix}. \url{https://www.uptimeconsulting.no/coding-agents-need-deterministic-feedback-a-case-for-phoenix/}
    \srcitem{GL166}{GL162} Prashish (2026, January 17). \emph{Software for One: Powered by LLMs}. prashish.xyz. \url{https://prashish.xyz/ai/software-for-one-powered-by-llms/}
    \srcitem{GL167}{GL163} Fiene (2026, March 17). \emph{Beyond Vibe-Coding: A Disciplined Workflow for AI-Assisted Software Development with Claude Code}. \#InnoBlog. \url{https://blog.innogames.com/beyond-vibe-coding-a-disciplined-workflow-for-ai-assisted-software-development-with-claude-code/}
    \srcitem{GL168}{GL164} Turkovic, I. (2026, January 17). \emph{Why Ruby Might Be the Most AI-Friendly Language Nobody's Talking About}. Signal Through the Noise. \url{https://www.ivanturkovic.com/2026/01/17/ruby-token-efficiency-llm-ai-friendly-language/}
    \srcitem{GL169}{GL165} Corbet, J. (2026, June 15). \emph{Development statistics for the 7.1 kernel}. LWN.net. \url{https://lwn.net/Articles/1077425/}
    \srcitem{GL170}{GL166} Abdrakhmanov, D. (2026, July 6). \emph{Porting fmodel to Dart}. DClimber — Business Software, Event Modeling, and Delivery. \url{https://dclimber.com/posts/fmodel-dart/}
    \srcitem{GL171}{GL172} Li, X., Yu, S., Shen, L., Zhang, J., Zheng, Y., \& Xie, Y. (2026, March 19). \emph{AutORAN: LLM-driven Natural Language Programming for Agile xApp Development} (arXiv:2603.18604) [Preprint]. arXiv. \url{https://arxiv.org/abs/2603.18604}
    \srcitem{GL172}{GL173} Servile, V. (2026, July 23). \emph{Should we still design code for humans?} Thoughtworks. \url{https://www.thoughtworks.com/insights/blog/programming-languages/should-still-design-code-humans}
    \srcitem{GL173}{GL174} Venkatasubramanian, R. K. (2026, June 30). \emph{The fallacy of using natural language to code}. ragzzy.com. \url{https://ragzzy.com/blog/llm_natural_language/}
    \srcitem{GL174}{GL175} Syme, D. (2026, January 27). \emph{Generative AI and Changing Inputs}. dsyme.net. \url{https://dsyme.net/2026/01/27/generative-ai-and-changing-inputs/}
    \srcitem{GL175}{GL176} Liberty, A. (2026, March 7). \emph{Towards Understandable Software}. \url{https://gracefulliberty.com/articles/towards-understandable-software/}
    \srcitem{GL176}{GL178} Simpson, K. (2026, February 12). \emph{LLMs vs Code: Can Natural Language Outperform Programming?} LinkedIn. \url{https://www.linkedin.com/posts/getify_those-of-you-who-believe-natural-language-activity-7427746649160032256-NmPP}
    \srcitem{GL177}{GL179} Schleicher, D. (2026, January 4). \emph{How Creating a Ubiquitous Language Ensures AI Builds What You Actually Want}. \url{https://www.danielschleicher.com/software/engineering,/ai,/spec-driven/development/2026/01/04/removing-ambiguity-with-spec-driven-development.html}
    \srcitem{GL178}{GL180} Doug (2026, May 19). \emph{Why English will never be a programming language}. Doug Slater. \url{https://www.slater.dev/2026/05/why-english-will-never-be-a-programming-language/}
    \srcitem{GL179}{GL184} AI-Assisted Software Development (2026). \emph{Vibe Coding Is FAAFO}. \url{https://ai-assisted-software-development.com/vibe-coding-is-faafo/}
    \srcitem{GL180}{GL188} Phung, T. (2026, May 9). \emph{Vibe Coding Interview Guide: Ace AI-Assisted Coding Assessments}. DEV Community. \url{https://dev.to/truongpx396/vibe-coding-interview-guide-ace-ai-assisted-coding-assessments-1gbh}
    \srcitem{GL181}{GL189} Drotárová, Z. (2026, September 8). \emph{We Need to Stop Calling Everything Vibe Coding}. Unite.AI. \url{https://www.unite.ai/vibe-coding-vs-agentic-engineering-ai-software-development/}
    \srcitem{GL182}{GL190} Naveen (2026, June 19). \emph{Vibe Coding Dies the Moment It Meets a Legacy System}. HackerNoon. \url{https://hackernoon.com/vibe-coding-dies-the-moment-it-meets-a-legacy-system}
    \srcitem{GL183}{GL191} Chung, J.-W. (2025, December 19). \emph{Vibe Coding is Speculative Decoding}. \url{https://jaewonchung.me/thoughts/Vibe-coding-is-speculative-decoding/}
    \srcitem{GL184}{GL193} Aging Coder (2026, January 6). \emph{Book review: Vibe coding}. \url{https://agingcoder.com/posts/book-review-vibe-coding/}
    \srcitem{GL185}{GL194} Circelli, R. (2026, May 20). \emph{Want to Vibe Code? Use ChatGPT Instead of Claude}. PCMag UK. \url{https://uk.pcmag.com/ai/165073/want-to-vibe-code-use-chatgpt-instead-of-claude}
    \srcitem{GL186}{GL196} Bboysoul (2026, March 16). \emph{Vibe Coding with GitHub Copilot}. Bboysoul's Blog. \url{https://www.bboy.app/2026/03/16/vibe-coding-with-github-copilot/}
    \srcitem{GL187}{GL198} Cress, L. (2025, November 6). \emph{'Vibe coding' named word of the year by Collins Dictionary}. bbc.com. \url{https://www.bbc.com/news/articles/cpd2y053nleo}
    \srcitem{GL188}{GL199} Hickson, I. (2026, June 12). \emph{llmdevsilo: Using LLMs to develop code securely}. Hixie's Natural Log. \url{https://loge.hixie.ch/?start=1781308547&count=1}
    \srcitem{GL189}{GL202} Pendergast, G. (2026, June 3). \emph{Your New Job Description}. \url{https://pento.net/2026/06/03/your-new-job-description/}
    \srcitem{GL190}{GL203} Ferreyra, A. (2026, July 21). \emph{Engineering AI for the Future of Health Care}. Harvard T.H. Chan School of Public Health. \url{https://hsph.harvard.edu/advanced-learning-academy/news/the-hard-part-isnt-vibe-coding-its-getting-ai-safely-into-healthcare-workflows/}
\end{enumerate}

\bibliographystyle{elsarticle-num}
\bibliography{References}

\end{sloppypar}
\end{document}